\documentclass[prb, twocolumn, reprint, superscriptaddress, aps, longbibliography, floatfix]{revtex4-2}

\usepackage[utf8]{inputenc}
\usepackage[T1]{fontenc}

\usepackage{graphicx}
\usepackage{amsmath}
\usepackage{amssymb}
\usepackage{afterpage}
\usepackage{amsfonts}
\usepackage{braket}
\usepackage{float}
\usepackage{bm}
\usepackage{siunitx}
\usepackage{lipsum}
\usepackage[]{xcolor}
\usepackage[english]{babel}
\usepackage{hyperref}
\usepackage{physics}
\hypersetup{
	colorlinks,
	linkcolor={blue!90!black},
	citecolor={blue!90!black},
	urlcolor=	{blue!90!black}
}
\usepackage[]{placeins}
\usepackage{makecell}

\usepackage{orcidlink}

\usepackage[]{newtxtext} 
\usepackage[subscriptcorrection,nosymbolsc,smallerops,bigdelims]{newtxmath} 
\DeclareMathAlphabet{\mathcal}{OMS}{cmsy}{m}{n} 
\DeclareMathAlphabet{\mathbcal}{OMS}{cmsy}{b}{n} 

\begin{document}
	
	\title{Theory of resonance Raman profiles in transition metal dichalcogenide monolayers:\\Interference effects due to inter-valley phonon scattering}
	
	\author{Daniel Groll\,\orcidlink{0000-0001-5588-7073}}
	\email{daniel.groll@uni-muenster.de}
	\affiliation{Institute of Solid State Theory, University of M\"unster, 48149 M\"unster, Germany}
	
	\author{Daniel Wigger\,\orcidlink{0000-0002-4190-8803}}
	\affiliation{Department of Physics, University of M\"unster, 48149 M\"unster, Germany}
	
	\author{Tilmann Kuhn\,\orcidlink{0000-0001-7449-9287}}
	\affiliation{Institute of Solid State Theory, University of M\"unster, 48149 M\"unster, Germany}
	
	
	\begin{abstract}
		The optical properties of monolayer transition metal dichalcogenides (TMDCs) are strongly impacted by exciton-phonon coupling, especially inter-valley scattering in the case of tungsten-based materials. Together with strain-tunability of the energetic K-Q valley separation this material class therefore provides a platform for semiconductor optomechanics. Here we investigate the influence of inter-valley scattering and strain on resonance Raman scattering in monolayer TMDCs using a microscopic model based on an effective deformation potential coupling for the exciton-phonon interaction. Considering exemplarily the resonance Raman profile of the A$_1$-mode in hBN-encapsulated monolayer WSe$_2$, we find that inter-valley scattering between the electronic K- and Q-valley leads to destructive interference between a first order and a third order Raman process. As a result we obtain a suppression of the incoming resonance in WSe$_2$, consistent with recently published experimental data.
	\end{abstract}
	
	\date{\today}
	
	\maketitle
	
	\section{Introduction}
	Monolayers of transition metal dichalcogenides (TMDCs) exhibit a large variety of remarkable properties making them an interesting platform for both technological applications and fundamental research~\cite{lv2015transition,mueller2018exciton,wang2018colloquium}. In the context of semiconductor optics their large excitonic binding energies together with coupled spin and valley physics allow for valley-selective optical control even at room temperature~\cite{wang2016control,mueller2018exciton,wang2018colloquium,dogadov2026dissecting}.
	
	In addition to direct (bright) excitons, these materials host a multitude of momentum-dark excitons whose energies can be similar to the bright ones~\cite{deilmann2019finite}. This implies that exciton-phonon scattering between bright and momentum-dark excitons can significantly impact the properties of monolayer TMDCs, e.g., their photoluminescence and homogeneous exciton linewidth~\cite{selig2016excitonic,selig2018dark,brem2019intrinsic,brem2020phonon}. Furthermore the energy separation between bright and momentum-dark excitons can be controlled via strain, making monolayer TMDCs an excellent platform for applications of strain-tuning~\cite{niehues2018strain,peng2020strain, niehues2024excitons}.
	
	To better understand the interplay between valley-dependent exciton physics and phonons, resonance Raman scattering provides an excellent optical probe, combining sensitivity on excitonic resonances with the possibility to spectrally isolate single-phonon creation (Stokes) or annihilation (anti-Stokes) processes. Resonance Raman scattering is often described theoretically via simple effective few-level models~\cite{yu2010fundamentals,mcdonnell2018probing,mcdonnell2020observation,sahu2025resonance,lambers2026resonant} or fully fledged ab initio calculations~\cite{guo2015double,miranda2017quantum, carvalho2017intervalley,taghizadeh2020library, reichardt2020nonadiabatic,bergmann2026enhancement}. Here we bridge the gap between these two approaches and provide a microscopic model of resonance Raman scattering in monolayer TMDCs based on a widely established theory of exciton-phonon scattering using effective deformation potentials~\cite{jin2014intrinsic}. This approach has been shown to be extraordinarily successful in explaining experimental data of various optical experiments~\cite{selig2016excitonic,selig2018dark,niehues2018strain,brem2019intrinsic,brem2020phonon,dogadov2026dissecting}, however to the best of our knowledge it has not been applied to resonance Raman scattering previously. 
	
	We start in Sec.~\ref{sec:2} with a derivation of the dipole scattering spectrum of a generic coupled exciton-phonon system by modeling the idealized spectrometer signal in the far field. We provide a representation of the scattering spectrum in terms of Green's functions, which allows us to use a diagrammatic approach in the calculation of the resonance Raman scattering amplitudes.
	
	Before considering the actual resonance Raman signal, we present a detailed study of the exciton-phonon model in Sec.~\ref{sec:exc_phon_model}. We discuss the properties of the phonon-induced exciton self-energies of the bright and momentum-dark excitons obtained in a self-consistent Born approximation. We consider the example of an hBN-encapsulated monolayer of WSe$_2$, providing a strain-dependent study of the exciton self-energies.
		
	In Sec.~\ref{sec:stokes} we then derive the resonance Raman scattering amplitude of a single-phonon Stokes process in up to at least third order with respect to the exciton-phonon coupling. We explicitly consider the resonance Raman profile of the A$_1$-mode in a backscattering configuration. Our result depends on two effective parameters: (i) Strain, modeled as a relative energy shift of the electronic Q-valley, and (ii) the strength of the A$_1$-phonon emission from the electronic Q-valley relative to the emission from the K-valley.
	
	Finally in Sec.~\ref{sec:RRP} we provide a parameter study of our two-parameter model for the case of hBN-encapsulated monolayer WSe$_2$. We discuss the impact of these two parameters on the resonance Raman profile of the A$_1$-mode, highlighting the role of phonon-mediated inter-valley scattering, and compare the results with existing experimental data.
	
	\textit{Note that throughout the paper we work in units such that $\hbar=1$.}
	\section{Modeling dipole scattering spectra using Green's functions}\label{sec:2}
	\subsection{Time-integrated spectrum in the far field}\label{sec:avg_spec}
	We consider detection of light emitted by a sample placed at the origin $\vb{r}=\vb{0}$, using a spectrometer in the far field at $\vb{r}=\vb{r}_{\mathrm{D}}$ with a polarization filter placed in front that selects for a specific polarization vector $\bm{\epsilon}_{\mathrm{D}}\bot\vb{r}_{\mathrm{D}}$. In this case the time-integrated spectrum reads
	\begin{equation}\label{eq:spectrum}
		I(\omega)=\expval{\qty[\bm{\epsilon}_{\mathrm{D}}^*\cdot \vb{E}^{(+)}(\vb{r}_{\mathrm{D}},\omega)]^{\dagger}\qty[\bm{\epsilon}^*_{\mathrm{D}}\cdot \vb{E}^{(+)}(\vb{r}_{\mathrm{D}},\omega)]}
	\end{equation}
	and can be derived from the detected time-dependent spectrum~\cite{eberly1977time}, considering an ideal spectrometer with perfect resolution. Note that the presence of a spectrometer with finite resolution would lead to an additional convolution of this ideal spectrum, e.g., with a Lorentzian of width $\Gamma$ describing the spectrometer resolution (see App.~\ref{app:avg_spec}). Here, $\vb{E}^{(\pm)}(\vb{r},\omega)$ denote the Fourier transforms of the positive/negative frequency components of the electric field operator, obeying $\vb{E}^{(-)}(\vb{r},t)=\vb{E}^{(+)}(\vb{r},t)^{\dagger}$ in the Heisenberg picture time domain~\cite{glauber1963coherent, cohen2022photons}.
	
	Assuming that stray light from the sample excitation is filtered out and does not reach the spectrometer, as well as that the relevant optical modes are not (thermally) occupied without emission or scattering of light by the sample, the electric field in the far field of the sample, relevant to the detected spectrum in Eq.~\eqref{eq:spectrum}, stems from its dipole radiation~\cite{griffiths2005introduction,stokes2016quantum,cohen2022photons}, leading to the following operator relation in the Heisenberg picture
	\begin{equation}\label{eq:dipole_rad}
		\bm{\epsilon}_{\mathrm{D}}^*\cdot\vb{E}^{(+)}(\vb{r}_{\mathrm{D}},t)=-\frac{1}{4\pi\epsilon_0 c^2 r_{\mathrm{D}}}\bm{\epsilon}_{\mathrm{D}}^*\cdot\ddot{\vb{d}}^{(+)}(t-r_{\mathrm{D}}/c)\,.
	\end{equation}
	Here, $c$ denotes the speed of light in vacuum and $\epsilon_0$ the vacuum permittivity, while $\vb{d}^{(\pm)}(t)$ denote the positive/negative frequency components of the dipole operator which are related by $\vb{d}^{(-)}(t)=\vb{d}^{(+)}(t)^{\dagger}$. These account for downwards/upwards electronic transitions in the optical domain, i.e., where the final state has less/more energy than the initial state and these states are separated by optical frequencies. The relation for the negative frequency components is obtained by conjugating Eq.~\eqref{eq:dipole_rad}. Note that in Eq.~\eqref{eq:dipole_rad} a rotating wave approximation is implicitly assumed.
	
	As derived in App.~\ref{app:avg_spec}, we can now express the time-integrated spectrum in Eq.~\eqref{eq:spectrum} in terms of the two-time dipole correlation function
	\begin{equation}\label{eq:def_C}
		C(t_1,t_2)=\expval{\qty[\bm{\epsilon}_{\mathrm{D}}\cdot\vb{d}^{(-)}(t_1)]\qty[\bm{\epsilon}^*_{\mathrm{D}}\cdot\vb{d}^{(+)}(t_2)]}\,
	\end{equation}
	as
	\begin{align}
		I(\omega)&=\int\limits_{-\infty}^{\infty}\dd t\frac{\omega^4}{32\pi^4\epsilon_0^2 c^4 r_{\mathrm{D}}^2}\text{Re}\qty[\int\limits_{0}^{\infty}\dd\tau\,C(t+\tau,t) e^{-i\omega\tau}]\,.\label{eq:I_id_final}
	\end{align}
\subsection{Light scattering spectra for monochromatic laser driving}
In the following we consider optical driving of the sample in dipole approximation with the Hamiltonian~\cite{cohen2022photons}
\begin{equation}\label{eq:H_l}
	H_l(t)=-\vb{d}\cdot\vb{E}_l(t)
\end{equation}
describing the coupling of the material to a classical laser field $\vb{E}_l$ at the position $\vb{r}=\vb{0}$ of the sample. The field is assumed to consist of a single coherent monochromatic mode with coherent amplitude $\mathcal{E}$, frequency $\omega_l$, traveling in direction $\vb{k}_l$ with polarization vector $\bm{\epsilon}_l\bot\vb{k}_l$, i.e., it describes a continuous wave (cw) laser. Then its positive frequency component is given by~\cite{glauber1963coherent,cohen2022photons}
\begin{equation}
	\vb{E}_l^{(+)}(t)=i\sqrt{\frac{\omega_l}{2\epsilon_0 V_0}}\bm{\epsilon}_le^{-i\omega_lt}\mathcal{E}\,
\end{equation}
with the normalization volume $V_0$. In the following we model a situation where the intensity of the exciting laser is fixed
\begin{equation}
	\vb{E}_l^{(-)}(t)\cdot\vb{E}_l^{(+)}(t)\sim \omega_l|\mathcal{E}|^2=\text{const.}
\end{equation}
which implies that the coherent amplitude $\mathcal{E}$ and therefore also the average number of photons varies with frequency. In an experimental setting this can be ensured by separating some part of the laser field off via a beam splitter and measuring the intensity via a photodetector to potentially adjust the input power to guarantee constant excitation intensity.

Expanding the correlation function from Eq.~\eqref{eq:def_C} in powers of the optical driving strength $\mathcal{E}$ we obtain in the linear response regime after applying the rotating wave approximation and adiabatic switching-on of the cw laser (see App.~\ref{app:lin_response})
\begin{align}
	C(t+\tau,t)&=C(\tau,0)\equiv C(\tau)\notag\\
	&={\frac{\omega_l|\mathcal{E}|^2}{2\epsilon_0  V_0}}e^{i\omega_l\tau}\expval{e^{iH\tau}P^{\dagger}e^{-iH\tau}P}\label{eq:C_stat}
\end{align}
with the polarizability operator~\cite{page1981separation}
\begin{align}
	P=\int\limits_{0}^{\infty}\dd t\,&e^{i(\omega_l+i0^+)t}\qty(\bm{\epsilon}^*_{\mathrm{D}}\cdot\vb{d}^{(+)})e^{-iHt}\qty(\bm{\epsilon}_l\cdot\vb{d}^{(-)})e^{iHt}\,,\label{eq:polariz}
\end{align}	
where the Hamiltonian $H$ describes the free dynamics of the sample, i.e., the dynamics in absence of external optical driving, and we assume an initial thermal state at $t_0\rightarrow-\infty$ making the correlation function $C$ stationary. Note that this renders Eq.~\eqref{eq:I_id_final} ill-defined since the integrand becomes independent of $t$. However we can of course switch from a time-integrated to a time-averaged spectrum, which is finally given by
\begin{align}\label{eq:I_id_final_S}
	I(\omega)=\xi(\omega)S(\omega-\omega_l)
\end{align}
with the overall Rayleigh/Raman scattering frequency dependence
\begin{equation}\label{eq:phase_space}
	\xi(\omega)\sim\omega^4
\end{equation}
and the scattering spectrum
\begin{equation}\label{eq:def_S}
	S(\Omega)=2\text{Re}\qty(\int\limits_{0}^{\infty}\dd\tau\,\expval{e^{iH\tau}P^{\dagger}e^{-iH\tau}P}e^{-i\Omega\tau})\,,
\end{equation}
where the expectation value is taken with respect to a thermal state induced by the Hamiltonian $H$, corresponding to the state of the sample before optical excitation. According to Eq.~\eqref{eq:I_id_final_S}, the parts of the scattering spectrum $S(\Omega)$ with a certain fixed frequency $\Omega$ describe a frequency shift between the incoming light $\omega_l$ and the outgoing light $\omega=\omega_l+\Omega$. Thus it contains Stokes scattering for $\Omega<0$, Rayleigh scattering for $\Omega=0$ and anti-Stokes scattering for $\Omega>0$~\cite{yu2010fundamentals}.

\subsection{Green's function representation of the scattering spectrum}\label{sec:green}
In the following we will evaluate the scattering spectrum in Eq.~\eqref{eq:def_S} further. To do so, we have to specify the structure of the Hamiltonian $H$ describing the sample properties. Focusing on the optical properties of spatially homogeneous semiconductors, we consider a generic exciton-phonon Hamiltonian~\cite{krummheuer2002theory, selig2018dark, brem2020phonon, groll2026phonon}
\begin{equation}\label{eq:H_generic}
	H=H_{\mathrm{e}}+H_{\mathrm{p}}+H_{\mathrm{ep}}\,.
\end{equation}
Here
\begin{equation}
	H_{\mathrm{e}}=\sum_{\alpha\vb{K}} E_{\alpha}(\vb{K})X_{\alpha}^{\dagger}(\vb{K})X_{\alpha}^{}(\vb{K})
\end{equation}
describes the free dynamics of excitons with energies $E_{\alpha}(\vb{K})$, where the center of mass momentum is given by $\vb{K}$ and further quantum numbers are denoted collectively by $\alpha$. We consider a neutral semiconductor in the linear response regime, consistent with the derivation of Eq.~\eqref{eq:C_stat}, implying small exciton densities and an equal number of electrons and holes. This implies that we can treat the exciton annihilation (creation) operators $X_{\alpha}^{(\dagger)}(\vb{K})$ as bosonic operators~\cite{katsch2018theory}.

The free dynamics of the phonons are furthermore described by the Hamiltonian
\begin{equation}
	H_{\mathrm{p}}=\sum_{j\vb{q}}\Omega_{j}(\vb{q})b_{j}^{\dagger}(\vb{q})b_{j}^{}(\vb{q})
\end{equation}
for phonons in branch $j$ with momentum $\vb{q}$ and dispersion relation $\Omega_{j}(\vb{q})$. Here, the phonon annihilation (creation) operators $b_{j}^{(\dagger)}(\vb{q})$ fulfill the usual bosonic commutation relations.

Finally, the exciton-phonon coupling is given by
\begin{equation}\label{eq:H_ep_general}
	H_{\mathrm{ep}}=\sum_{j\vb{q}}V_{j}(\vb{q})\qty[b_{j}^{}(\vb{q})+b_{j}^{\dagger}(-\vb{q})]
\end{equation}
with $V_j(\vb{q})$ being an operator acting on the Hilbert space of excitons, which describes scattering between exciton states induced by absorption of a phonon in branch $j$ with momentum $\vb{q}$ or emission of a phonon in branch $j$ with momentum $-\vb{q}$. Crucially, this scattering process conserves the number of excitons, i.e.,
\begin{equation}
	\comm{V_{j}(\vb{q})}{N_x}=0\,,
\end{equation}
where the exciton number operator is given by
\begin{equation}
	N_x=\sum_{\alpha\vb{K}} X_{\alpha}^{\dagger}(\vb{K})X_{\alpha}^{}(\vb{K})\,.
\end{equation}
This property holds due to the large energy mismatch between excitons $E_{\alpha}(\vb{K})\sim$1~eV and phonons $\Omega_j(\vb{q})\sim 10$~meV  in semiconductors~\cite{yu2010fundamentals}.

This implies that the full Hamiltonian in Eq.~\eqref{eq:H_generic} also conserves the total number of excitons due to $\comm{N_x}{H}=0$. Furthermore, the total number of phonons is conserved in absence of exciton-phonon scattering $	H_{\mathrm{ep}}=0$. In the following we will be interested in excitonic dipole transitions driven by the exciton-light coupling Hamiltonian in Eq.~\eqref{eq:H_l}, such that the positive frequency component of the dipole operator removes excitons from the system, while the negative frequency component adds excitons
\begin{equation}
	\comm{N_x}{\vb{d}^{(+)}}=-\vb{d}^{(+)}\,,\qquad \comm{N_x}{\vb{d}^{(-)}}=\vb{d}^{(-)}\,.\label{eq:dipole_number}
\end{equation}

We now approximate the initial thermal state of the sample with respect to the full Hamiltonian $H$ as 
\begin{equation}\label{eq:rho}
	\rho(-\infty)=\frac{e^{-\beta H}}{\Tr\qty(e^{-\beta H})}\approx \ket{0_x}\bra{0_x}\frac{e^{-\beta H_{\mathrm{p}}}}{Z_{\mathrm{p}}}
\end{equation}
with $Z_{\mathrm{p}}=\Tr\qty(e^{-\beta H_{\mathrm{p}}})$, $\beta=\qty(k_{\mathrm{B}}T)^{-1}$, $T$ being the temperature and $k_{\mathrm{B}}$ being Boltzmann's constant. Here, $\ket{0_x}$ is the state without any excitons, i.e., $N_x\ket{0_x}=0$. This approximation is justified since excitons in semiconductors have energies on the order of 1~eV and are therefore not thermally occupied at cryogenic temperatures or even at room temperature. Note that since the state without any excitons fulfills
\begin{equation}
	H\ket{0_x}=H_{\mathrm{p}}\ket{0_x}\,,
\end{equation}
the approximated density matrix in Eq.~\eqref{eq:rho} is still stationary with $\comm{H}{\rho(-\infty)}=0$. Writing an arbitrary multi-mode Fock eigenstate of the phonon Hamiltonian as $\ket{\qty{n}}$ with
\begin{equation}\label{eq:E_n}
	H_{\mathrm{p}}\ket{\qty{n}}=E_{\qty{n}}\ket{\qty{n}}=\sum_{j\vb{q}}\Omega_{j}(\vb{q})n_{j}(\vb{q})\ket{\qty{n}}
\end{equation}
and $n_{j}(\vb{q})$ being the number of phonons in mode $(j,\vb{q})$, we can rewrite the scattering spectrum in Eq.~\eqref{eq:def_S} as
\begin{align}\label{eq:def_S_2}
	S(\Omega)=&2\text{Re}\Bigg(\int\limits_{0}^{\infty}\dd\tau\,e^{-i\Omega\tau}Z_{\mathrm{p}}^{-1}\sum_{\qty{n}}e^{-\beta E_{\qty{n}}}\notag\\
	&\qquad\times\bra{\qty{n}}\bra{0_x}e^{iH\tau}P^{\dagger}e^{-iH\tau}P\ket{0_x}\ket{\qty{n}}\Bigg)\notag\\
	=&\sum_{\qty{m},\qty{n}}2\pi Z_{\mathrm{p}}^{-1}e^{-\beta E_{\qty{n}}}\delta(E_{\qty{n}}-E_{\qty{m}}-\Omega)\notag\\
	&\qquad\times\qty|\bra{\qty{m}}\bra{0_x}P\ket{0_x}\ket{\qty{n}}|^2\,,
\end{align}
where we inserted the identity $1=\sum_{\qty{m}}\ket{\qty{m}}\bra{\qty{m}}$ of the phonon subspace to the right of $P^{\dagger}$ and made use of the fact that $P\ket{0_x}\sim\ket{0_x}$, i.e., the polarizability operator in Eq.~\eqref{eq:polariz} does not change the number of excitons due to the considered structure of the dipole operator given in Eq.~\eqref{eq:dipole_number}. Here we find the light-induced scattering amplitude
\begin{align}\label{eq:scattering_amplitude}
	&\frac{1}{i}\bra{\qty{m}}\bra{0_x}P\ket{0_x}\ket{\qty{n}}\\
	&=\frac{1}{i}\bra{\qty{m}}\bra{0_x}\qty(\bm{\epsilon}^*_{\mathrm{D}}\cdot\vb{d}^{(+)})\int\limits_{0}^{\infty}\dd t\,e^{i(\omega_l+i0^+)t}e^{-iHt}e^{iE_{\qty{n}}t}\notag\\
	&\qquad\qquad\times\qty(\bm{\epsilon}_l\cdot\vb{d}^{(-)})\ket{0_x}\ket{\qty{n}}\notag\\
	&=\bra{\qty{m}}\bra{0_x}\qty(\bm{\epsilon}^*_{\mathrm{D}}\cdot\vb{d}^{(+)})G(\omega_l+E_{\qty{n}})\qty(\bm{\epsilon}_l\cdot\vb{d}^{(-)})\ket{0_x}\ket{\qty{n}}\notag
\end{align}
by performing the integral stemming from the definition of the polarizability operator $P$ in Eq.~\eqref{eq:polariz} using $H\ket{0_x}\ket{\qty{n}}=E_{\qty{n}}\ket{0_x}\ket{\qty{n}}$. The result is written in terms of the retarded Green's function (operator)~\cite{fujimura1979theoretical, taylor2002quantum, grosso2013solid}
\begin{equation}\label{eq:def_G}
	G(E)=\frac{1}{E-H+i0^+}\,.
\end{equation}
Courtesy of the Dyson series we can represent this fully interacting retarded Green's function as
\begin{equation}\label{eq:Dyson}
	G(E)=G_0(E)+G_0(E)VG(E)\,,\qquad V=H-H_0
\end{equation}
with the free retarded Green's function
\begin{equation}
	G_0(E)=\frac{1}{E-H_0+i0^+}\,,
\end{equation}
leading to
\begin{equation}
	G(E)=G_0(E)\sum_{n=0}^{\infty}\qty[VG_0(E)]^n\,.\label{eq:G_series}
\end{equation}
For our purposes we choose the free Hamiltonian to be
\begin{equation}
	H_0=H_{\mathrm{e}}+H_{\mathrm{p}}
\end{equation}
describing the free dynamics of excitons and phonons, as well as the interaction Hamiltonian to be
\begin{equation}
	V=H_{\mathrm{ep}}\,.
\end{equation}
The presence of the Green's function in Eq.~\eqref{eq:scattering_amplitude} suggests that we apply diagrammatic techniques to evaluate the scattering spectrum in Eq.~\eqref{eq:def_S_2}. 

To determine the light-induced scattering amplitude in Eq.~\eqref{eq:scattering_amplitude}, we need to consider the properties of the retarded Green's function $G(E)$ on the subspace containing a single exciton, since the Hamiltonian $H$ and thus also $G(E)$ conserves the number of excitons, while
\begin{equation}
	N_x\vb{d}^{(-)}\ket{0_x}=\vb{d}^{(-)}\ket{0_x}
\end{equation}
courtesy of Eq.~\eqref{eq:dipole_number}. This means that the dipole operator $\vb{d}^{(-)}$ in Eq.~\eqref{eq:scattering_amplitude} creates an exciton due to the interaction of the sample with the laser. The subsequent propagation of the system is then described by the retarded Green's function $G(E)$ and the exciton is finally annihilated by $\vb{d}^{(+)}$, leading to the optical detection of the emitted light. This interpretation explains the projection of $\vb{d}^{(-)}$ onto the polarization vector of the laser $\bm{\epsilon}_l$, as well as the projection of $\vb{d}^{(+)}$ onto the polarization vector of the detector $\bm{\epsilon}_{\mathrm{D}}$ in Eq.~\eqref{eq:scattering_amplitude}.

For the sake of consistency of the presented material, in App.~\ref{app:diagrams} we derive the necessary Feynman rules used in the diagrammatic evaluation of scattering amplitudes. We summarize here only some crucial definitions which are relevant to the following discussions~\cite{fetter2003quantum, haug2008quantum}.

We are generally interested in matrix elements of the retarded Green's function between single exciton states, i.e.,
\begin{equation}
	\bra{\qty{n}}\bra{\vb{K}',\alpha'}G(E)\ket{\vb{K},\alpha}\ket{\qty{m}}\,.
\end{equation}
Here, $\ket{\vb{K},\alpha}=X^{\dagger}_{\alpha}(\vb{K})\ket{0_x}$ is an arbitrary single exciton state with momentum $\vb{K}$ and additional quantum numbers denoted by $\alpha$. Considering the series expansion in Eq.~\eqref{eq:G_series}, we find that such a matrix element consists of sums of different contributions that themselves consist of products of matrix elements of the free Green's function $G_0$ and the interaction $V$.

Starting with the free Green's function $G_0$, this leads us to the definition of the free (single) exciton propagator
\begin{align}
	&\bra{\qty{0}}\bra{\vb{K},\alpha}G_0(E)\ket{\vb{K},\alpha}\ket{\qty{0}}\notag\\
	&\qquad=\mathcal{G}^{(0)}_{\alpha}(E,\vb{K})=\frac{1}{E-E_{\alpha}(\vb{K})+i0^+}\label{eq:propagator_free}
\end{align}
whose pole determines the exciton energy. Furthermore we define the exciton-phonon interaction vertex 
\begin{equation}\label{eq:vertex_def}
	\mathcal{V}^{\vb{K}\alpha}_{\vb{K}'\alpha'}(j,\vb{q})=\bra{\vb{K}',\alpha'}V_j(\vb{q})\ket{\vb{K},\alpha}\,.
\end{equation}
Scattering with phonons generally modifies the exciton energies, i.e., the poles of exciton propagators such as in Eq.~\eqref{eq:propagator_free}. This effect is described by defining the (interacting) exciton propagator (at vanishing temperature)
\begin{equation}\label{eq:propagator_interacting}
	\mathcal{G}_{\alpha}(E,\vb{K})=\bra{\qty{0}}\bra{\vb{K},\alpha}G(E)\ket{\vb{K},\alpha}\ket{\qty{0}}\,, 
\end{equation}
which can be written as 
\begin{equation}\label{eq:G_full}
	\mathcal{G}_{\alpha}(E,\vb{K})=\frac{1}{E-E_{\alpha}(\vb{K})-\Sigma_{\alpha}(E,\vb{K})+i0^+}\,.
\end{equation}
with $\Sigma_{\alpha}(E,\vb{K})$ denoting the exciton self-energy. The exciton energies are thus renormalized by the real part of the self-energy and the resonances, i.e., poles of the Green's function, are broadened by its imaginary part. These effects describe polaron shifts and phonon-induced dephasing of the exciton, respectively. In the following we work in the self-consistent Born approximation, such that the exciton self-energy reads
\begin{equation}
	\Sigma_{\alpha}(E,\vb{K})=\sum_{j\vb{q}\alpha'\vb{K}'}\qty|\mathcal{V}_{\vb{K}\alpha}^{\vb{K}'\alpha'}(j,\vb{q})|^2 \mathcal{G}_{\alpha'}(E-\Omega_j(\vb{q}),\vb{K}')\,.\label{eq:self_energy_new_new}
\end{equation}

\section{Strain-dependent resonance Raman profiles in monolayer TMDCs}\label{sec:results}
\subsection{Details of the exciton-phonon model}\label{sec:exc_phon_model}
In the following we present a short discussion on the considered models for the excitons and their coupling to phonons. The corresponding Hamiltonian for a generic multi-valley semiconductor is derived in App.~\ref{app:Hamilton} and the exciton and phonon parameters are summarized in App.~\ref{app:model_params}.
\subsubsection{Exciton landscape}
We consider a neutral TMDC monolayer in the linear response regime implying small exciton densities and an equal number of electrons and holes. Starting from a multi-band model for electrons and holes, each parabolic band corresponding to a certain valley being treated in the effective mass approximation, in App.~\ref{app:Hamilton} we derive the Hamiltonian describing the excitons formed by an electron in valley $\lambda_{\mathrm{e}}$ and a hole in valley $\lambda_{\mathrm{h}}$~\cite{selig2016excitonic, selig2018dark, brem2019intrinsic, brem2020phonon}. The exciton dispersion $E^{(n)}_{\lambda_{\mathrm{e}}\lambda_{\mathrm{h}}}(\vb{Q})$ is given by
\begin{equation}
	E^{(n)}_{\lambda_{\mathrm{e}}\lambda_{\mathrm{h}}}(\vb{Q})=E_{\lambda_{\mathrm{e}}}^{\mathrm{e}}-E_{\lambda_{\mathrm{h}}}^{\mathrm{h}}+\frac{\vb{Q}^2}{2M_{\lambda_{\mathrm{e}}\lambda_{\mathrm{h}}}}+B^{(n)}_{\lambda_{\mathrm{e}}\lambda_{\mathrm{h}}}\,.\label{eq:disp_ex}
\end{equation}
Here, $M_{\lambda_{\mathrm{e}}\lambda_{\mathrm{h}}}$ is the total mass of the electron hole pair, $E_{\lambda_{\mathrm{e}}}^{\mathrm{e}}$ is the energy of the electronic conduction band minimum of valley $\lambda_{\mathrm{e}}$, while $E_{\lambda_{\mathrm{h}}}^{\mathrm{h}}$ is the energy of the valence band maximum of valley $\lambda_{\mathrm{h}}$. The exciton binding energy $B^{(n)}_{\lambda_{\mathrm{e}}\lambda_{\mathrm{h}}}$ is determined by the Wannier equation [Eq.~\eqref{eq:Wannier_2d} in App.~\ref{app:Hamilton}] for an electron-hole pair in the corresponding valleys and the label $n=$1s, 2p, ... denotes the quantum number corresponding to the relative motion of electron and hole, while the total center of mass momentum of the exciton is given by
\begin{equation}
	\overline{\vb{Q}}_{\lambda_{\mathrm{e}}\lambda_{\mathrm{h}}}=\vb{Q}+\vb{k}_{\lambda_{\mathrm{e}}}-\vb{k}_{\lambda_{\mathrm{h}}}\,.\label{eq:momentum_exc}
\end{equation}
Here, $\vb{k}_{\lambda_{\mathrm{e}}}$ denotes the momentum of the electron in the conduction band minimum of valley $\lambda_{\mathrm{e}}$ and $-\vb{k}_{\lambda_{\mathrm{h}}}$ denotes the momentum of the hole in the valence band maximum of valley $\lambda_{\mathrm{h}}$, while $\vb{Q}=\vb{0}$ corresponds to the minimum of the exciton dispersion. In terms of the previous notation we can label the excitons uniquely by $(\vb{K},\alpha)\rightarrow(\vb{Q},\lambda_{\mathrm{e}},\lambda_{\mathrm{h}},n)$. To model the substrate- or encapsulation-induced screening we solve Poisson's equation for a three-layer geometry, as described in App.~\ref{app:screening}. Additional details on the numerical solution of the Wannier equation can be found in App.~\ref{app:Wannier_numerical}. 

We will focus on the case of co-polarized detection $\bm{\epsilon}_l=\bm{\epsilon}_{\mathrm{D}}$ and nearly resonant excitation of the 1s exciton in the K-valley with a laser having suitable circular polarization~\cite{xiao2012coupled, wang2018colloquium}, such that we can write the dipole operator as
\begin{equation}
	\bm{\epsilon}_{l}\cdot \vb{d}^{(-)}=\bm{\epsilon}_{\mathrm{D}}\cdot \vb{d}^{(-)}\approx d_{\mathrm{1s}} X^{(n=1\mathrm{s})\dagger}_{\lambda_{\mathrm{e}}=\mathrm{K},\lambda_{\mathrm{h}}=\mathrm{K}}(\vb{Q}=\vb{0})\label{eq:excitation_dipole}
\end{equation}
with $d_{\mathrm{1s}}$ being the dipole moment (oscillator strength) of the bright 1s exciton in the K-valley. Note that we assume here perpendicular illumination of the monolayer, such that only excitons with vanishing in-plane center of mass momentum can be excited, i.e., $\vb{Q}=\vb{0}$ for direct excitons [see Eq.~\eqref{eq:momentum_exc}]. Furthermore, $\bm{\epsilon}_l=\bm{\epsilon}_{\mathrm{D}}$ implies that the wavevectors of incident and scattered light have to be antiparallel, i.e., we assume a backscattering configuration~\cite{yu2010fundamentals,chen2015helicity, drapcho2017apparent, miranda2017quantum}.

Due to the quasi-resonant excitation of the 1s exciton in the K-valley, in the following we suppress the index $n$ describing the quantum number of the relative motion of electron and hole, focusing on the interaction of 1s excitons with phonons. This is justified due to the sufficient spectral separation of the 1s excitons from the higher excited exciton states in TMDCs~\cite{chernikov2014exciton, brem2019intrinsic}. 

\begin{figure}
	\centering
	\includegraphics[width=\linewidth]{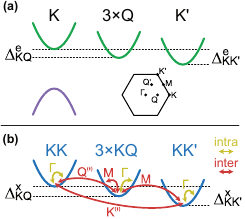}
	\caption{Reduced exciton landscape for monolayer TMDCs under optical excitation conditions creating electron-hole pairs in the K-valley. (a) We consider a hole in the K-valley (purple parabola) and five valleys for the electron (green parabolas): the K- and K'-valley, as well as three degenerate Q-valleys. Their energetic separation without the influence of Coulomb interaction $\Delta^{\mathrm{e}}_{\mathrm{K}\lambda_{\mathrm{e}}}=E^{\mathrm{e}}_{\mathrm{K}}-E^{\mathrm{e}}_{\mathrm{\lambda_{\mathrm{e}}}}$ depends on the chosen material and is furthermore impacted by strain. Inset: First Brillouin zone with relevant high symmetry points. (b) Resulting direct (KK) and indirect (KK$'$ and \textit{three} KQ) excitons (blue parabolas) with phonon intra-valley (yellow arrows) and inter-valley scattering processes (red arrows). The arrows are labeled according to the region around high symmetry points of the first BZ, from which the phonons mediating the exciton scattering stem. Compared to (a), the energetic separation $\Delta^{\mathrm{x}}_{\mathrm{K}\lambda_{\mathrm{e}}}=\Delta^{\mathrm{e}}_{\mathrm{K}\lambda_{\mathrm{e}}}+B_{\mathrm{K}\mathrm{K}}^{(1\mathrm{s})}-B_{\lambda_{\mathrm{e}}\mathrm{K}}^{(1\mathrm{s})}$  is modified by the relative exciton binding energy.}
	\label{fig:electronbands}
\end{figure}

This leads us to the following approximation for the exciton landscape of monolayer TMDCs, as shown in Fig.~\ref{fig:electronbands}~\cite{selig2016excitonic, selig2018dark, brem2019intrinsic,brem2020phonon}. We include the bright exciton with electron and hole both at the K-point. In addition we consider four different valley combinations leading to momentum-dark excitons that are close in energy to the bright exciton: These are the three equivalent (degenerate) excitons with the hole at the K point and the electron at one of the three equivalent Q points (also often referred to as $\Lambda$~\cite{jin2014intrinsic,selig2016excitonic,selig2018dark,deilmann2019finite}), as well as the hole at the K point and the electron at the K' point. The corresponding three non-equivalent exciton parabolas are schematically depicted in Fig.~\ref{fig:electronbands}~(b). Here, the KQ exciton is triply degenerate. Due to the large energetic separation between the spin-split electron bands at the Q-point~\cite{kormanyos2015k}, the KQ$^\prime$ exciton lies energetically sufficiently far above the excitons presented in (b) and is thus not considered in our model. 

Note that the relative energetic position $\Delta^{\mathrm{x}}_{\mathrm{K}\lambda_{\mathrm{e}}}$ of the minima depends strongly on the considered TMDC monolayer material and is furthermore impacted by screening (through the exciton binding energy), as well as, importantly for the following discussion, by strain.
\subsubsection{Exciton-phonon coupling}\label{sec:exciton-phonon_coupling}
Starting from the electron-hole picture, in App.~\ref{app:Hamilton} we derive the coupling between excitons and phonons, which has the general form assumed in Eq.~\eqref{eq:H_ep_general}. Using the quantum numbers $(\vb{Q},\lambda_{\mathrm{e}},\lambda_{\mathrm{h}})$ as discussed above, this leads to the exciton-phonon interaction vertices (see App.~\ref{sec:def_pot})
\begin{equation}
	\mathcal{V}^{\vb{Q}\lambda_{\mathrm{e}}\lambda_{\mathrm{h}}}_{\vb{Q}'\lambda_{\mathrm{e}}'\lambda_{\mathrm{h}}'}(j,\vb{q})=\bra{\vb{Q}'\lambda_{\mathrm{e}}'\lambda_{\mathrm{h}}'}V_{j}(\vb{q})\ket{\vb{Q}\lambda_{\mathrm{e}}\lambda_{\mathrm{h}}}\,.\label{eq:vertex_main_new}
\end{equation}
Intra-valley scattering is described by
\begin{align}\label{eq:vertex_intra_main_new}
	\mathcal{V}^{\vb{Q}\lambda_{\mathrm{e}}\lambda_{\mathrm{h}}}_{\vb{Q}'\lambda_{\mathrm{e}}\lambda_{\mathrm{h}}}(j,\vb{q})=\delta_{\vb{Q}',\vb{Q}+\vb{q}}	g_{j\lambda_{\mathrm{e}}\lambda_{\mathrm{h}}}^{(x)}(\vb{q})
\end{align}
with the intra-valley exciton-phonon coupling constant
\begin{align}
	g_{j\lambda_{\mathrm{e}}\lambda_{\mathrm{h}}}^{(x)}(\vb{q})&=g_{j\lambda_{\mathrm{e}}\lambda_{\mathrm{e}}}^{(\mathrm{e})}(\vb{q})F_{\lambda_{\mathrm{e}}\lambda_{\mathrm{h}};\lambda_{\mathrm{e}}\lambda_{\mathrm{h}}}\qty(-\mu^{\mathrm{h}}_{\lambda_{\mathrm{e}}\lambda_{\mathrm{h}}}\vb{q})\notag\\&\qquad-g_{j\lambda_{\mathrm{h}}\lambda_{\mathrm{h}}}^{(\mathrm{h})}(\vb{q})F_{\lambda_{\mathrm{e}}\lambda_{\mathrm{h}};\lambda_{\mathrm{e}}\lambda_{\mathrm{h}}}\qty(\mu^{\mathrm{e}}_{\lambda_{\mathrm{e}}\lambda_{\mathrm{h}}}\vb{q})\,.\label{eq:g_x_new}
\end{align}
Here and in the following $g^{(\mathrm{e/h} )}_{j\lambda\lambda'}(\vb{k})$ denote the scattering matrix elements for electrons/holes from valley $\lambda'$ to valley $\lambda$ via a phonon in mode $j$ with wavevector $\vb{k}$.

The exciton landscape in Fig.~\ref{fig:electronbands} contains a single valley for the holes, such that only electron-phonon inter-valley scattering is relevant here. The corresponding matrix elements are given by (with $\lambda_{\mathrm{e}'}\neq\lambda_{\mathrm{e}}$)
\begin{align}
	&\eval{\mathcal{V}^{\vb{Q}\lambda_{\mathrm{e}}\lambda_{\mathrm{h}}}_{\vb{Q}'\lambda_{\mathrm{e}}'\lambda_{\mathrm{h}}}(j,\vb{q}+\vb{k}_{\lambda_{\mathrm{e}}'}-\vb{k}_{\lambda_{\mathrm{e}}})}_{\lambda_{\mathrm{e}}'\neq \lambda_{\mathrm{e}}}\!\!\!=\delta_{\vb{Q}',\vb{Q}+\vb{q}}g_{j\lambda_{\mathrm{e}}'\lambda_{\mathrm{e}}}^{(\mathrm{e})}(\vb{q}+\vb{k}_{\lambda_{\mathrm{e}}'}-\vb{k}_{\lambda_{\mathrm{e}}})\notag\\
	&\qquad\times	F_{\lambda_{\mathrm{e}}'\lambda_{\mathrm{h}};\lambda_{\mathrm{e}}\lambda_{\mathrm{h}}}\qty[-\mu^{\mathrm{h}}_{\lambda_{\mathrm{e}}'\lambda_{\mathrm{h}}}(\vb{Q}+\vb{q})+\mu^{\mathrm{h}}_{\lambda_{\mathrm{e}}\lambda_{\mathrm{h}}}\vb{Q}]\,.\label{eq:vertex_electron_main_new}
\end{align}
Here the momentum argument of the phonons is shifted such that $\vb{q}$ reflects the phonon momentum relative to the momentum difference of electrons between the final and initial valleys. For the sake of completeness, the general form of the hole-phonon inter-valley scattering matrix elements is presented in App.~\ref{sec:def_pot}.


Crucially, electron-phonon and hole-phonon scattering are modified by the exciton form factor~\cite{krummheuer2002theory,lengers2020theory,groll2026phonon} [see also Eq.~\eqref{eq:form_factor} in App.~\ref{app:Hamilton}]
\begin{equation}\label{eq:form_factor_main}
	F^{}_{\lambda_{\mathrm{e}}\lambda_{\mathrm{h}};\lambda_{\mathrm{e}}'\lambda_{\mathrm{h}}'}(\vb{q})=\sum_{\vb{K}}\phi_{\lambda_{\mathrm{e}}\lambda_{\mathrm{h}}}^{*}\qty(\vb{K}-\vb{q})\phi_{\lambda_{\mathrm{e}}'\lambda_{\mathrm{h}}'}^{}\qty(\vb{K})\,,
\end{equation}
where $\phi_{\lambda_{\mathrm{e}}\lambda_{\mathrm{h}}}(\vb{k})$ is the 1s exciton wavefunction in momentum space, i.e., the solution of the Wannier equation corresponding to the binding energy $B^{(\mathrm{1s})}_{\lambda_{\mathrm{e}}\lambda_{\mathrm{h}}}$. Since, assuming an isotropic effective mass, 1s exciton wavefunctions are isotropic in momentum space,  the form factors only depend on the absolute value of their argument 
\begin{equation}\label{eq:form_factor_isotropy}
	F^{}_{\lambda_{\mathrm{e}}\lambda_{\mathrm{h}};\lambda_{\mathrm{e}}'\lambda_{\mathrm{h}}'}(\vb{q})=F^{}_{\lambda_{\mathrm{e}}\lambda_{\mathrm{h}};\lambda_{\mathrm{e}}'\lambda_{\mathrm{h}}'}(q)
\end{equation}
with $q=|\vb{q}|$. These form factors present a limitation on the relevant momentum transfers in an exciton-phonon scattering process.

To model the electron-phonon and hole-phonon coupling constants, we use established effective deformation potential parameters obtained from ab initio calculations (see App.~\ref{sec:def_pot})~\cite{jin2014intrinsic}. These provide momentum-space averaged coupling constants for the different inter-valley ($\lambda_{\mathrm{e}}\neq\lambda_{\mathrm{e}'}$ and $\lambda_{\mathrm{h}}=\lambda_{\mathrm{h}}'$) and intra-valley ($\lambda_{\mathrm{e}}=\lambda_{\mathrm{e}'}$ and $\lambda_{\mathrm{h}}=\lambda_{\mathrm{h}}'$) scattering processes, depicted by red and yellow arrows in Fig.~\ref{fig:electronbands}(b), respectively. These parameters have been very successful in explaining experimentally observed features such as exciton linewidths in various TMDCs as well as the impact of inter-valley scattering on optical signals~\cite{selig2016excitonic,niehues2018strain,selig2018dark,brem2019intrinsic, brem2020phonon,dogadov2026dissecting}. However it is important to note that their application relies on averaging the frequencies of the different optical and acoustic branches around the relevant high symmetry points into single effective optical and acoustic branches, respectively. These are assumed to be dispersion-less in the vicinity of these high symmetry points except for the acoustic branch at the $\Gamma$-point, which is treated within the Debye model. These details will become important again when discussing Raman scattering amplitudes. 

The relevant high symmetry points for the phonon modes inducing inter-valley and intra-valley scattering can be found labeling the arrows in Fig.~\ref{fig:electronbands}(b). Since phonon mode frequencies do not depend on the sign of their associated wavevector, in the following we will suppress the prime $'$ when talking about the phononic properties for linguistic simplicity, e.g., 'scattering with phonons at the Q-point' means either scattering with phonons at the Q- or Q$'$-point, depending on the direction of the wavevector in the scattering process.
\subsubsection{Notes on the self-energy calculation}\label{sec:self_energy_details}
To obtain the self-energy in Eq.~\eqref{eq:self_energy_new_new}, we can separate it into contributions stemming from scattering with phonons from the different high symmetry points in the first Brillouin zone, which are modeled according to the previous discussion (see also App.~\ref{sec:def_pot}). The explicit self-energy calculations are given in App.~\ref{app:self_energy}. 

For the contributions stemming from acoustic phonons around the $\Gamma$ point, which are always thermally occupied for any non-vanishing temperature due to their linear dispersion relation, we derive an effectively temperature-dependent self-energy in the quasi-particle approximation in App.~\ref{app:master_eq}, employing an open quantum system master equation approach in their derivation~\cite{breuer2002theory}. There we rely on the fact that the Raman signal contained in Eq.~\eqref{eq:def_S} is described entirely by a fully time-ordered correlation function~\cite{groll2023read} (see App.~\ref{app:isolation}). This master equation approach effectively removes the explicit appearance of these low frequency acoustic phonons from any scattering diagrams that we calculate in the following, while yielding temperature-dependent self-energy contributions due to scattering with a thermal acoustic phonon bath, calculated explicitly in App.~\ref{app:self_energy_low}. This approach is furthermore justified by the comparably small impact of these modes on exciton linewidths in TMDCs at cryogenic temperatures~\cite{selig2016excitonic, brem2019intrinsic}.

This allows us to use the zero-temperature Green's function formalism presented in Sec.~\ref{sec:green} and App.~\ref{app:diagrams}~\cite{fetter2003quantum, haug2008quantum}, while still retaining the temperature-dependent impact due to low frequency acoustic modes. Typical high-frequency phonon modes around high symmetry points in TMDCs, i.e., optical phonons around the $\Gamma$-point and phonon modes around other high symmetry points, have energies on the order of 10~meV~\cite{jin2014intrinsic}. For these high-frequency phonon modes, we can assume that they are thermally unoccupied for typical cryogenic temperatures encountered in a laboratory setting. Their contributions to the self-energy in Eq.~\eqref{eq:self_energy_new_new} are processed further in App.~\ref{app:self_energy_high}. All following results have been obtained for a temperature of $T=4$~K.

For the self-consistent numerical calculation of the exciton self-energies we employ the quasi-particle approximation (see App.~\ref{app:self_consistent})~\cite{fetter2003quantum, haug2008quantum} by fixing the energy argument in Eq.~\eqref{eq:self_energy_new_new} and replacing
\begin{equation}
	\Sigma_{\lambda_{\mathrm{e}}\lambda_{\mathrm{h}}}(E,\vb{Q})\rightarrow	\Sigma_{\lambda_{\mathrm{e}}\lambda_{\mathrm{h}}}(\vb{Q})=	\Sigma_{\lambda_{\mathrm{e}}\lambda_{\mathrm{h}}}(E=E_{\lambda_{\mathrm{e}}\lambda_{\mathrm{h}}}(\vb{Q}),\vb{Q})\,.
\end{equation}
As discussed in App.~\ref{app:self_consistent}, we furthermore absorb the polaron shifts, i.e., the real part of the self-energy, into the exciton energies shown in Eq.~\eqref{eq:disp_ex}, which we keep fixed.

Finally, we include radiative decay of the bright exciton in the K-valley by calculating the corresponding decay rate $\gamma_{\mathrm{r}}$ microscopically using band parameters obtained from a $\vb{k}\cdot\vb{p}$ treatment, as well as the numerically determined exciton wavefunction~\cite{xiao2012coupled,brem2019intrinsic}. Details can be found in App.~\ref{app:form_factors}. This radiative decay modifies the exciton self-energy for exciton states within the light cone, i.e., in good approximation only for the bright exciton at the K-point according to
\begin{equation}
	\Sigma_{\lambda_{\mathrm{e}}\lambda_{\mathrm{h}}}(\vb{Q})\rightarrow \Sigma_{\lambda_{\mathrm{e}}\lambda_{\mathrm{h}}}(\vb{Q})-\frac{i}{2}\gamma_{\mathrm{r}}\delta_{\lambda_{\mathrm{e}}\mathrm{K}}\delta_{\lambda_{\mathrm{h}}\mathrm{K}}\delta_{\vb{Q}\vb{0}}\,.
\end{equation}
\subsubsection{Properties of the exciton self-energy}
\begin{figure}
	\centering
	\includegraphics[width=\linewidth]{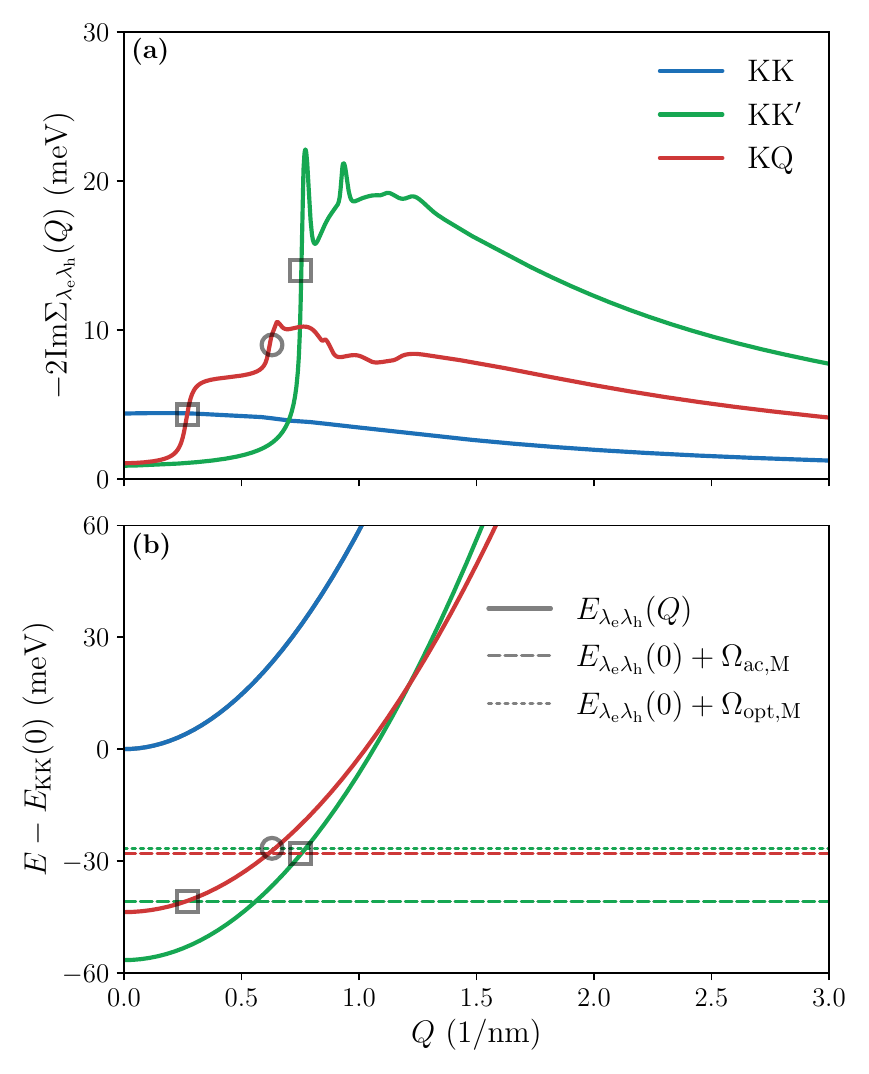}
	\caption{Impact of phonons on excitons in hBN-encapsulated monolayer WSe$_2$ at $T=4$~K. (a) Phonon-induced decay rates for the $\lambda_{\mathrm{h}}\lambda_{\mathrm{e}}$-excitons with momentum $Q=|\vb{Q}|$ relative to the minimum of the corresponding dispersion relation. (b) Momentum-dependent exciton energies (solid) with colors denoting the excitons species according to the labels in (a). Additional horizontal lines mark representative exciton energies with vanishing momentum $Q=0$ relative to the minimum of the corresponding dispersion relation, offset by the acoustic phonon energy $\Omega_{\mathrm{ac},\mathrm{M}}$ (dashed) and the optical phonon energy $\Omega_{\mathrm{opt},\mathrm{M}}$ (dotted) at the M-point. Distinct features in (a) and (b) are marked by squares (onset of acoustic phonon inter-valley scattering via phonons at the M point) and circles (onset of optical phonon inter-valley scattering via phonons at the M point).}
	\label{fig:rates}
\end{figure}
To give a concrete example for the calculated exciton self-energies, in Fig.~\ref{fig:rates}(a) we plot the phonon-induced decay rates for hBN-encapsulated monolayer WSe$_2$, i.e., $-2\Im\Sigma_{\lambda_{\mathrm{e}}\lambda_{\mathrm{h}}}(Q)$ with vanishing radiative decay $\gamma_{\mathrm{r}}=0$, for the three different exciton species labeled by the hole and electron valley quantum numbers: KK (blue), KK$'$ (green) and KQ (red). Note that the self-energy is isotropic, i.e., depends only on $Q=|\vb{Q}|$ due to the assumed isotropy of the exciton energies in Eq.~\eqref{eq:disp_ex}, as well as of the exciton form factors in Eq.~\eqref{eq:form_factor_isotropy}.

In general we find that the phonon-induced decay rates start with an almost constant value around $Q=0$ for all three exciton species, followed by an intricate intermediate $Q$-dependent behavior, whose details depend on the specific kind of exciton, as well as a general decaying behavior for large $Q$. The latter feature stems from the exciton form factor in Eq.~\eqref{eq:form_factor_main}, which weights all exciton-phonon coupling matrix elements and limits the relevant momentum transfers to $q\lesssim a_{\mathrm{B}}^{-1}$, where $a_{\mathrm{B}}$ is the typical exciton Bohr radius, which is here on the order of 1~nm (see Tab.~\ref{tab:wannier} in App.~\ref{app:model_params}). 

To better understand the intermediate behavior of the decay rates, in Fig.~\ref{fig:rates}(b) we show the exciton dispersion relations (solid) with colors corresponding to the exciton species in (a). In addition we show some representative lines corresponding to the respective exciton energies with vanishing momentum $Q=0$ relative to the minimum of the corresponding dispersion relation, offset by the acoustic phonon energy $\Omega_{\mathrm{ac},\mathrm{M}}$ (dashed) and the optical phonon energy $\Omega_{\mathrm{opt},\mathrm{M}}$ (dotted) at the M-point.

In the case of the KK$'$ exciton (green), there is a strong increase of the phonon-induced decay rate around $Q\approx 0.75/$nm, marked by a square in Fig.~\ref{fig:rates}(a). This originates from the onset of inter-valley scattering to the KQ exciton parabola via acoustic M-phonon emission [red dashed line crosses green solid line, marked by square in (b)]. This process dominates the phonon-induced decay properties of the KK$'$ exciton due to the comparably large deformation potential coupling constant (see Tab.~\ref{tab:phon} in App.~\ref{app:model_params}) and the fact that there are in total three KQ exciton parabolas implying a large final state phase space. This then leads to the strong increase in the KK$'$ decay rate from roughly 1~meV at $Q=0$ to above 20~meV around $Q=0.75/$nm in (a).

Similarly, the sudden increase of the decay rate of the KQ exciton (red, marked by square) in Fig.~\ref{fig:rates}(a) around $Q=0.3/$nm stems from the onset of inter-valley scattering from the KQ exciton to the KK$'$ exciton via emission of acoustic M-phonons [green dashed line crossing red solid line, marked by square in (b)]. This increase is roughly 1/3 of the increase in the case of the KK$'$ exciton (green) discussed previously, reflecting the fact that there is only one KK$'$ exciton final state parabola for inter-valley scattering from the KQ exciton, while there are three final state KQ exciton parabolas for scattering from the KK$'$ exciton. The further increase of the phonon-induced KQ exciton decay rate around $Q=0.6/$nm [marked by circle in (a)] originates from the onset of inter-valley scattering from the KQ to the KK$'$ exciton via optical phonons at the M-point [green dotted line crossing red solid line, marked by circle in (b)].

Finally we note that due to the applied self-consistent Born approximation, energy conservation requirements for the scattering processes are less strict than if we would have applied the normal Born approximation. Therefore the onsets of different scattering processes at certain $Q$-values in Fig.~\ref{fig:rates}(a) are not sudden, i.e., in the form of Heaviside functions, but smeared out.

\subsubsection{Impact of strain on phonon-induced decay rates}
As we have seen in the previous discussion on phonon-induced decay rates in the context of Fig.~\ref{fig:rates}, there is a strong dependence on the exciton momentum $Q$. Certain scattering channels are 'turned on', as depicted by significant increases in the scattering rates in (a), only when the $Q$-dependent exciton energy crosses a certain threshold, as denoted by the solid lines in (b) crossing certain horizontal dashed or dotted lines. The strong dependence on the exciton momentum $Q$ is therefore rather a strong dependence on the relative exciton energies, whose energetic separation has to be large enough to allow for scattering processes induced by emission of certain high frequency phonons.

It is well known that the relative energetic position of the KQ exciton parabola with respect to the KK and KK$'$ excitons is very sensitive to external influences, especially strain, while the relative energetic position between KK and KK$'$ excitons is approximately protected by time-reversal symmetry~\cite{niehues2018strain,deilmann2019finite}. While all three exciton species shift with strain, only the \textit{relative} energetic position of the KQ exciton is affected. To capture this effect we introduce an additional phenomenological detuning $\delta_{\mathrm{KQ}}$ of the KQ exciton parabola 
\begin{align}\label{eq:def_detuning}
	&E_{\lambda_{\mathrm{e}}=\mathrm{Q},\lambda_{\mathrm{h}}=\mathrm{K}}(\vb{Q})-E_{\mathrm{KK}}(0)\notag\\
	&\qquad\rightarrow E_{\lambda_{\mathrm{e}}=\mathrm{Q},\lambda_{\mathrm{h}}=\mathrm{K}}(\vb{Q})-E_{\mathrm{KK}}(0)+\delta_{\mathrm{KQ}}\,.
\end{align}
The pristine unstrained situation depicted in Fig.~\ref{fig:rates} is given by $\delta_{\mathrm{KQ}}=0$.

\begin{figure}
	\centering
	\includegraphics[width=\linewidth]{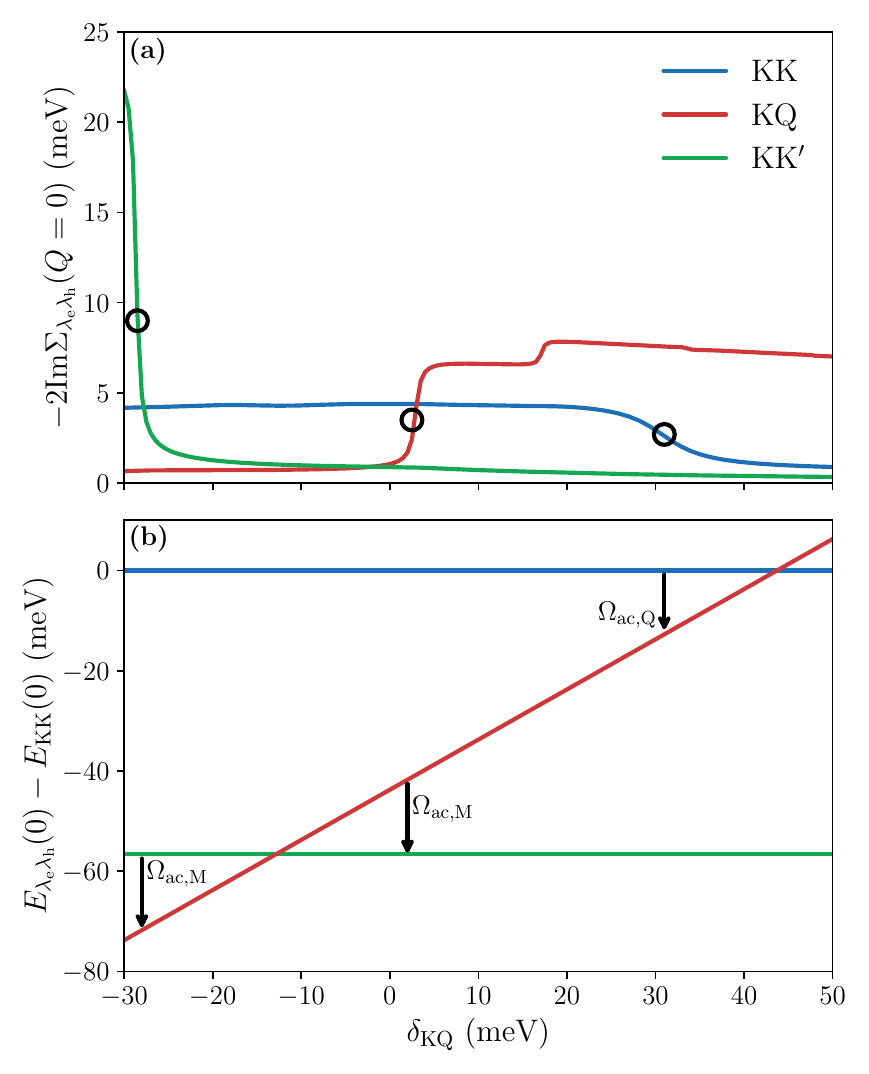}
	\caption{Impact of strain on phonon-induced decay rates in hBN-encapsulated monolayer WSe$_2$ at $T=4$~K. (a) Phonon-induced decay rates for the $\lambda_{\mathrm{h}}\lambda_{\mathrm{e}}$-excitons with momentum $Q=0$ relative to the minimum of the corresponding dispersion relation as a function of the KQ-exciton detuning $\delta_{\mathrm{KQ}}$ [Eq.~\eqref{eq:def_detuning}]. (b) Exciton energies corresponding to the decay rates in (a). Downward arrows having a length corresponding to the phonon frequency labels denote detuning values where the corresponding scattering channels are turned on/off [circles in (a)] according to energy conservation requirements.}
	\label{fig:strainrates}
\end{figure}

To better understand the impact of strain on the phonon-induced decay rates, in Fig.~\ref{fig:strainrates}(a) we show them for the lowest-lying exciton from each species, i.e., $-2\Im\Sigma_{\lambda_{\mathrm{e}}\lambda_{\mathrm{h}}}(Q=0)$ for vanishing radiative decay $\gamma_{\mathrm{r}}=0$, in hBN-encapsulated monolayer WSe$_2$ as a function of the detuning $\delta_{\mathrm{KQ}}$. To analyze the detuning-dependent behavior of these rates, in (b) we plot the corresponding exciton energies and denote detuning values at which specific phonon emission-induced scattering channels are turned on/off by downward arrows having a length corresponding to the respective phonon frequency. The impacted decay rates corresponding to these on/off-sets of scattering channels are marked by circles in (a), positioned at the same detuning value.

We find that the KK$'$ exciton decay rate (green) increases dramatically for detuning values below $\delta_{\mathrm{KQ}}<-25$~meV, since the KQ exciton then lies far enough below the KK$'$ exciton to allow acoustic M-phonon-induced inter-valley scattering from the KK$'$ to the KQ excitons. Conversely the KQ exciton decay rate (red) increases drastically for $\delta_{\mathrm{KQ}}>2$~meV since then the KQ exciton lies sufficiently far above the KK$'$ exciton to allow for the reverse inter-valley scattering process via acoustic M-phonons.

The decay rate of the bright KK exciton (blue) is rather stable for a wide range of detuning values, however it decreases by more than 50\% for $\delta_{\mathrm{KQ}}\gtrsim 30$~meV, since the inter-valley scattering channel from the KK to the KQ excitons via emission of acoustic Q-phonons is then turned off.

\subsection{Modeling a Stokes process in back-scattering geometry with circularly co-polarized detection}\label{sec:stokes}
\subsubsection{Determination of the scattering amplitude}
In this section we derive the scattering amplitude for the simplest non-trivial Raman process, which is light-scattering-induced creation of a single phonon, i.e., a Stokes process~\cite{yu2010fundamentals}. Such a process appears as a peak at $S(\Omega=-\Omega_{\mathrm{R}})$ in the scattering spectrum in Eq.~\eqref{eq:def_S_2}. Assuming that we can uniquely associate the frequency shift $-\Omega_{\mathrm{R}}$ with a certain phonon mode with branch $j_{\mathrm{R}}$ and momentum $\vb{q}_{\mathrm{R}}$, as well as working at cryogenic temperatures, the relevant scattering amplitude in Eq.~\eqref{eq:scattering_amplitude} is proportional to
\begin{equation}\label{eq:A_R}
	A_{\mathrm{R}}(\omega_l)=\bra{\vb{0}\mathrm{K}\mathrm{K}}\bra{\qty{0}}b_{j_{\mathrm{R}}}(\vb{q}_{\mathrm{R}})G(\omega_l)\ket{\qty{0}}\ket{\vb{0}\mathrm{K}\mathrm{K}}\,.
\end{equation}
Here, $\ket{\qty{0}}$ is the phonon vacuum of the high frequency phonon modes (see discussion in Sec.~\ref{sec:self_energy_details}) and $\ket{\vb{0}\mathrm{K}\mathrm{K}}=X^{\dagger}_{\mathrm{K}\mathrm{K}}(\vb{0})\ket{0_x}$ describes the bright exciton that is created by nearly resonant perpendicular illumination of the sample, as well as annihilated by light emission into the backscattering direction, according to Eq.~\eqref{eq:excitation_dipole}. Since the interaction Hamiltonian in Eq.~\eqref{eq:H_ep_general} conserves the total momentum of the exciton-phonon system (see also App.~\ref{sec:def_pot}), the only non-vanishing contribution to this scattering amplitude is given by the case $\vb{q}_{\mathrm{R}}=\vb{0}$. Note that this is only strictly true for perpendicular illumination of the sample and detection in the backscattering configuration. Otherwise the light can induce a small initial in-plane momentum, which would be modeled via a modification of Eq.~\eqref{eq:excitation_dipole}. Also, defects in the sample, which we do not take into account in our model, can in principle lead to momentum-non-conserving processes and $\vb{q}_{\mathrm{R}}\neq \vb{0}$~\cite{martin1971breakdown,menendez1985interference, drapcho2017apparent}.

\begin{figure}
	\centering
	\includegraphics[width=\linewidth]{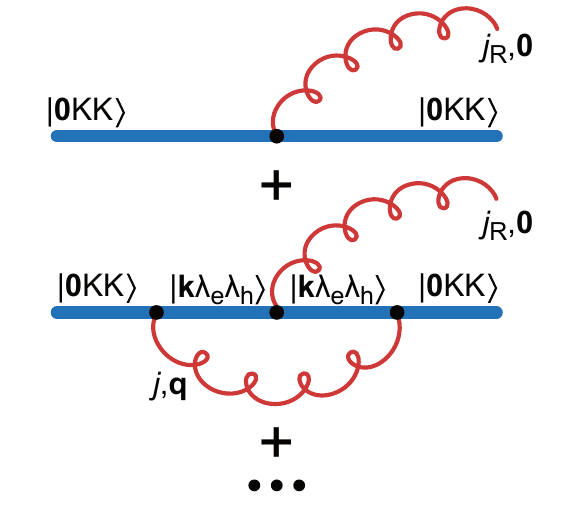}
	\caption{Diagrammatic representation of the scattering amplitude $A_{\mathrm{R}}(\omega_l)$ for the Stokes process [Eq.~\eqref{eq:A_R}]. The partially resummed diagrams up to at least third order in the exciton-phonon coupling are shown, which yield Eq.~\eqref{eq:A_R_final}, when employing the Feynman rules presented in App.~\ref{app:diagrams}.}
	\label{fig:stokes}
\end{figure}

Up to at least third order in the exciton-phonon vertices, we can represent Eq.~\eqref{eq:A_R} diagrammatically as in Fig.~\ref{fig:stokes}. While the complete Feynman rules are given in App.~\ref{app:diagrams}, this diagram can be understood by considering the following summary: In each of the two diagrams, the initial state is on the left and the final state is on the right. Exciton-phonon interaction vertices [Eq.~\eqref{eq:vertex_main_new}] are depicted as black dots. Thick blue lines between vertices and or initial/final states with labels $(\lambda_{\mathrm{e}}\lambda_{\mathrm{h}},\vb{k})$ represent the interacting single exciton propagator $\mathcal{G}_{\lambda_{\mathrm{e}}\lambda_{\mathrm{h}}}(E,\vb{k})$ [Eq.~\eqref{eq:G_full}], whose pole is furthermore shifted by the sum of all phonon energies of the phonons that are explicitly present in the diagram (red wavy lines). Note that the upper diagram in Fig.~\ref{fig:stokes} already implicitly contains phonon loops before and after the Raman phonon emission vertex, similar to the phonon loop in the lower diagram. However these loops are accounted for by the exciton self-energy and thus do not appear explicitly.


The scattering amplitude for the single-phonon Stokes process up to at least the third order in the exciton-phonon coupling is now given by the sum of the two diagrams in Fig.~\ref{fig:stokes}, leading to
%
%
%
%

\begin{align}\label{eq:A_R_final}
	&A_{\mathrm{R}}(\omega_l)=\mathcal{G}_{\mathrm{K}\mathrm{K}}(\omega_l,\vb{0})\mathcal{G}_{\mathrm{K}\mathrm{K}}(\omega_l-\Omega_{\mathrm{R}},\vb{0})\\
	&\qquad\times\Big[\mathcal{V}^{\vb{0}\mathrm{K}\mathrm{K}}_{\vb{0}\mathrm{K}\mathrm{K}}(j_{\mathrm{R}},\vb{0})^*+\sum_{j\vb{q}\lambda_{\mathrm{e}}\lambda_{\mathrm{h}}}\mathcal{V}^{-\vb{q}\lambda_{\mathrm{e}}\lambda_{\mathrm{h}}}_{-\vb{q}\lambda_{\mathrm{e}}\lambda_{\mathrm{h}}}(j_{\mathrm{R}},\vb{0})^*\qty|\mathcal{V}_{\vb{0}\mathrm{K}\mathrm{K}}^{-\vb{q}\lambda_{\mathrm{e}}\lambda_{\mathrm{h}}}(j,\vb{q})|^2\notag\\
	&\qquad\times \mathcal{G}_{\lambda_{\mathrm{e}}\lambda_{\mathrm{h}}}\qty(\omega_l-\Omega_{j}(\vb{q}),-\vb{q})\mathcal{G}_{\lambda_{\mathrm{e}}\lambda_{\mathrm{h}}}\qty(\omega_l-\Omega_{j}(\vb{q})-\Omega_{\mathrm{R}},-\vb{q})\Big]\notag\\
	&\qquad +...\notag
\end{align}
Here we made explicit use of momentum conservation at each vertex, such that in the second diagram the only non-vanishing contributions are for $\vb{k}=-\vb{q}$. The corresponding Stokes scattering spectrum in Eq.~\eqref{eq:def_S_2} is finally given by
\begin{equation}
	S(\Omega=-\Omega_{\mathrm{R}};\omega_l)\sim \qty|A_{\mathrm{R}}(\omega_l)|^2
\end{equation}
and depends parametrically on the carrier frequency $\omega_l$ of the exciting cw-laser. The time-averaged spectrum from Eq.~\eqref{eq:I_id_final_S} at the Stokes peak $\omega=\omega_l-\Omega_{\mathrm{R}}$ is further modified by the prefactor from Eq.~\eqref{eq:phase_space}, leading to
\begin{equation}\label{eq:I_final}
	I(\omega_l-\Omega_{\mathrm{R}})\sim (\omega_l-\Omega_{\mathrm{R}})^4\qty|A_{\mathrm{R}}(\omega_l)|^2\,.
\end{equation}
\subsubsection{Important details of the Raman phonon - exciton coupling}
Considering the result in Eq.~\eqref{eq:A_R_final}, emission of the Raman phonon in intra-valley scattering is determined by vertices of the form $\mathcal{V}_{\vb{k}\lambda_{\mathrm{e}}\lambda_{\mathrm{h}}}^{\vb{k}\lambda_{\mathrm{e}}\lambda_{\mathrm{h}}}(j_{\mathrm{R}},\vb{0})$, which can generally be written as [see Eqs.~\eqref{eq:vertex_intra_main_new} and \eqref{eq:g_x_new}]
\begin{equation}\label{eq:vertex_raman}
	\mathcal{V}_{\vb{k}\lambda_{\mathrm{e}}\lambda_{\mathrm{h}}}^{\vb{k}\lambda_{\mathrm{e}}\lambda_{\mathrm{h}}}(j_{\mathrm{R}},\vb{0})=g_{j_{\mathrm{R}}\lambda_{\mathrm{e}}\lambda_{\mathrm{h}}}^{(x)}(\vb{0})=g_{j_{\mathrm{R}}\lambda_{\mathrm{e}}\lambda_{\mathrm{e}}}^{(\mathrm{e})}(\vb{0})-g_{j_{\mathrm{R}}\lambda_{\mathrm{h}}\lambda_{\mathrm{h}}}^{(\mathrm{h})}(\vb{0})\,
\end{equation}
with $g_{j_{\mathrm{R}}\lambda_{\mathrm{e/h}}\lambda_{\mathrm{e/h}}}^{(\mathrm{e/h})}(\vb{0})$ being the electron/hole-phonon coupling constant for intra-valley scattering in valley $\lambda_{\mathrm{e/h}}$ via phonons in branch $j_{\mathrm{R}}$ with vanishing momentum. Here, $g_{j_{\mathrm{R}}\lambda_{\mathrm{e}}\lambda_{\mathrm{h}}}^{(x)}(\vb{0})$ is a shorthand for the resulting exciton-phonon coupling constant.  Note that hermiticity of the electron- and hole-phonon coupling Hamiltonians requires that the right-hand-side in Eq.~\eqref{eq:vertex_raman} is real for zone-center phonons with momentum $\vb{q}=\vb{0}$. Furthermore the form factors present in Eq.~\eqref{eq:g_x_new} do not appear here since due to the normalization of the exciton wavefunctions they satisfy $F_{\lambda_{\mathrm{e}}\lambda_{\mathrm{h}};\lambda_{\mathrm{e}}\lambda_{\mathrm{h}}}(0)=1$.

In monolayer TMDCs there are three high frequency phonon branches with significant electron-phonon coupling strengths (i.e. we omit the weakly coupling E$''$ and A$_2$$''$ modes from the following discussion)~\cite{kaasbjerg2012phonon,jin2014intrinsic}. Labeled by the symmetry of their zone-center phonon modes, these are the homopolar A$_1$-mode (out-of-plane optical vibration) and the two quasi-degenerate E$'$-modes [in-plane LO and transverse optical (TO) mode]. However for circularly co-polarized detection of the Raman signal in a back-scattering configuration, only the homopolar A$_1$-mode can contribute in Eq.~\eqref{eq:A_R_final}, i.e., $j_{\mathrm{R}}=\mathrm{A}_1$~\cite{ribeiro2014group,chen2015helicity,drapcho2017apparent}. 

This is consistent with the fact that the E$'$ LO mode couples to excitons in TMDCs mainly via the Fröhlich coupling, which is identical for conduction and valence band when interacting with zone-center phonons~\cite{kaasbjerg2012phonon, sohier2016two}, such that its contribution would exactly cancel in the Raman phonon emission vertex in Eq.~\eqref{eq:vertex_raman}. Furthermore, the E$'$ TO mode couples via a first order deformation potential, which vanishes for zone-center phonons~\cite{kaasbjerg2012phonon}, such that the only non-vanishing contribution in Eq.~\eqref{eq:vertex_raman} is for $j_{\mathrm{R}}=\mathrm{A}_1$.

As discussed previously in Sec.~\ref{sec:exciton-phonon_coupling}, for the calculation of the exciton self-energy we use established ab-initio parameters for the exciton-phonon coupling in the form of momentum-averaged effective deformation potential parameters~\cite{jin2014intrinsic}. These effective parameters furthermore combine the relevant optical modes, i.e., A$_1$ and E$'$ (LO/TO), into a single effective optical mode. This is a reasonable assumption if one is interested only in excitonic properties, since phonon-induced dephasing leads to sufficient broadening, such that we cannot spectrally separate the impact of the different optical modes on the excitons. However this is not a valid approximation anymore when considering the scattering spectrum in Eq.~\eqref{eq:def_S_2}, where the phonon energies appear in a $\delta$-function, which is only potentially broadened by the spectrometer resolution in the detected spectrum~(see App.~\ref{app:avg_spec}). Furthermore the effective ab-initio parameters are momentum-averaged in the vicinity of high symmetry points and not momentum-resolved, such that we cannot infer the strength of the Raman phonon emission vertex in Eq.~\eqref{eq:vertex_raman} from these parameters. 

This leads to a situation, where we are able to calculate the impact of exciton-phonon scattering on the exciton lifetimes~\cite{selig2016excitonic, brem2019intrinsic}, as discussed in Sec.~\ref{sec:self_energy_details}, but cannot deduce the relative strength of the Raman (A$_1$) phonon emission for the different excitons, i.e., we cannot determine 
\begin{equation}
	\mathcal{V}_{\vb{k}\lambda_{\mathrm{e}}\mathrm{K}}^{\vb{k}\lambda_{\mathrm{e}}\mathrm{K}}(\mathrm{A}_1,\vb{0})=g^{(x)}_{\mathrm{A}_1\lambda_{\mathrm{e}}\mathrm{K}}(\vb{0})
\end{equation}
from the considered ab initio parameters, which are however needed to calculate the scattering amplitude in Eq.~\eqref{eq:A_R_final}.

Having five electron valleys in our model (see Fig.~\ref{fig:electronbands}), this amounts to five different coupling constants. However the three Q-valleys are identical. Furthermore electron-phonon intra-valley scattering in the K- and K'-valley can be assumed identical, such that only two parameters remain: the intra-valley A$_1$-phonon coupling constant for the KQ and the KK$^{(')}$ excitons. In the following we will investigate resonance Raman profiles, i.e., we are interested in the parametric dependence of Eq.~\eqref{eq:I_final} and thus of Eq.~\eqref{eq:A_R_final} with respect to the laser frequency $\omega_l$. We are not concerned with overall prefactors and can thus factor out one of the coupling constants from Eq.~\eqref{eq:A_R_final}, leaving only the ratio
\begin{equation}
	\bar{g}^{\mathrm{Q}}_{\mathrm{K}}=\frac{g^{(x)}_{\mathrm{A}_1\mathrm{Q}\mathrm{K}}(\vb{0})}{g^{(x)}_{\mathrm{A}_1\mathrm{K}\mathrm{K}}(\vb{0})}
\end{equation}
as a free real-valued parameter in the calculation of Eq.~\eqref{eq:A_R_final}, which can be rewritten as
\begin{align}\label{eq:A_R_final_final}
	&A_{\mathrm{A}_1}(\omega_l)\sim\mathcal{G}_{\mathrm{K}\mathrm{K}}(\omega_l,\vb{0})\mathcal{G}_{\mathrm{K}\mathrm{K}}(\omega_l-\Omega_{\mathrm{A}_1},\vb{0})\\
	&\qquad\times\Big[1+\sum_{j\vb{q}\lambda_{\mathrm{e}}}\qty(\delta_{\lambda_{\mathrm{e}}\mathrm{K}}+\delta_{\lambda_{\mathrm{e}}\mathrm{K}'}+\delta_{\lambda_{\mathrm{e}}\mathrm{Q}}\bar{g}^{\mathrm{Q}}_{\mathrm{K}})\qty|\mathcal{V}_{\vb{0}\mathrm{K}\mathrm{K}}^{-\vb{q}\lambda_{\mathrm{e}}\mathrm{K}}(j,\vb{q})|^2\notag\\
	&\qquad\times \mathcal{G}_{\lambda_{\mathrm{e}}\mathrm{K}}\qty(\omega_l-\Omega_{j}(\vb{q}),-\vb{q})\mathcal{G}_{\lambda_{\mathrm{e}}\mathrm{K}}\qty(\omega_l-\Omega_{j}(\vb{q})-\Omega_{\mathrm{A}_1},-\vb{q})\Big]\notag\,.
\end{align}
Note that the sum over $\lambda_{\mathrm{e}}$ involves three equivalent Q-valleys, all denoted by $\lambda_{\mathrm{e}}=$\ Q for notational simplicity. This leads to three identical contributions $\sim \bar{g}^{\mathrm{Q}}_{\mathrm{K}}$. For the remaining vertices present in this scattering amplitude we still employ the effective momentum-averaged deformation potentials since they are summed over all phonon branches $j$ and phonon momenta~$\vb{q}$.
\subsection{Impact of intra- and inter-valley scattering on resonance Raman profiles}\label{sec:RRP}
\subsubsection{Model calculations in monolayer WSe$_2$}
\begin{figure}[b]
	\centering
	\includegraphics[width=\linewidth]{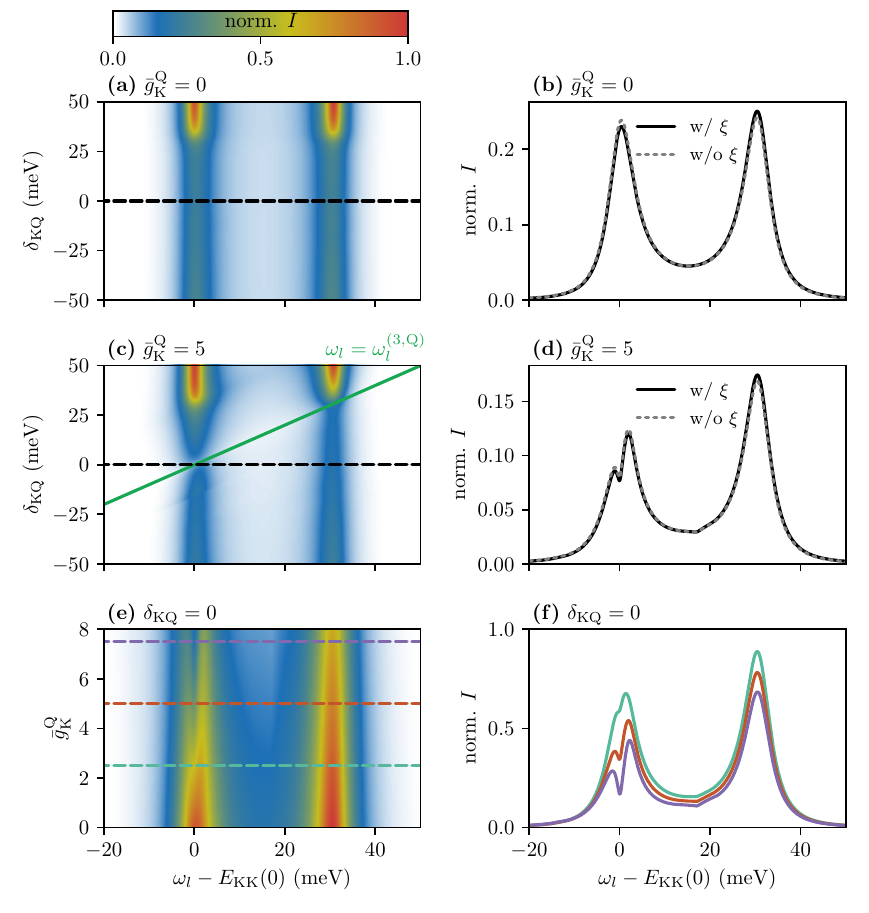}
	\caption{Impact of strain and inter-valley scattering on resonance Raman profiles $I(\omega_l-\Omega_{\mathrm{A}_1})$ of hBN-encapsulated monolayer WSe$_2$ at $T=4$~K. (a) Strain-dependent resonance Raman profile for suppressed A$_1$-phonon emission from the Q-valley $\bar{g}_{\mathrm{K}}^{\mathrm{Q}}=0$. (b) Cut in (a) as marked by horizontal dashed line, showing the unstrained case. Full simulation including the factor $\xi(\omega_l-\Omega_{\mathrm{A}_1})$ from Eq.~\eqref{eq:phase_space} (solid) vs. neglected factor $\xi(\omega_l-\Omega_{\mathrm{A}_1})\rightarrow \xi(E_{\mathrm{KK}}(0)-\Omega_{\mathrm{A}_1}/2)$ (dashed). (c,d) same as (a,b) for non-vanishing A$_1$-phonon emission from the Q-valley $\bar{g}_{\mathrm{K}}^{\mathrm{Q}}=5$. The green line in (c) marks the energetic position of the third order resonance $\omega_l=E_{\mathrm{KQ}}(0)+\Omega_{\mathrm{A}_1}+\Omega_{\mathrm{ac},\mathrm{Q}}\equiv \omega_l^{(3,\mathrm{Q})}$. (e,f) Resonance Raman profiles for the unstrained case $\delta_{\mathrm{KQ}}=0$ as a function of the relative A$_1$-phonon emission strength from the Q-valley $\bar{g}^{\mathrm{Q}}_{\mathrm{K}}$. Lines in (f) correspond to identically colored horizontal cuts in (e).}
	\label{fig:resonance}
\end{figure}

In the following we discuss resonance Raman profiles given by Eq.~\eqref{eq:I_final} with the scattering amplitude from Eq.~\eqref{eq:A_R_final_final}. The results for hBN-encapsulated monolayer WSe$_2$ are shown in Fig.~\ref{fig:resonance}. As discussed previously, we have two free parameters in our model: (i) The relative strength of A$_1$-phonon emission from the Q-valley $\bar{g}^{\mathrm{Q}}_{\mathrm{K}}$ contributing in third order in Eq.~\eqref{eq:A_R_final_final}, and (ii) the strain-induced detuning of the Q-valley $\delta_{\mathrm{KQ}}$ introduced in Eq.~\eqref{eq:def_detuning}.

Figure~\ref{fig:resonance} shows the dependence of the resonance Raman profile on both parameters. Starting in (a) with a suppressed A$_1$-phonon emission from the Q-valley $\bar{g}_{\mathrm{K}}^{\mathrm{Q}}=0$, we find a quasi-symmetric two-peak structure for all considered strain-induced detunings $\delta_{\mathrm{KQ}}$. For sufficiently large detunings $\delta_{\mathrm{KQ}}\gtrsim 30$~meV the resonance Raman signal increases drastically. This stems from the decrease of the phonon-induced decay rate of the KK exciton due to the removal of the acoustic Q-phonon assisted inter-valley scattering channel for such large detunings, as discussed in the context of Fig.~\ref{fig:strainrates}.

The two peaks present in Fig.~\ref{fig:resonance}(a) are the so-called incoming and outgoing resonances at $\omega_l=E_{\mathrm{KK}}(0)$ and $\omega_l=E_{\mathrm{KK}}(0)+\Omega_{\mathrm{A}_1}$, respectively~\cite{yu2010fundamentals}. Formally they appear as the propagators $\mathcal{G}_{\mathrm{KK}}(\omega_l,\vb{0})$ and $\mathcal{G}_{\mathrm{KK}}(\omega_l-\Omega_{\mathrm{A}_1},\vb{0})$ in the first line of Eq.~\eqref{eq:A_R_final_final}, respectively. The incoming resonance corresponds to direct excitation of the optically active KK exciton, followed by the emission of the Raman phonon and a photon with reduced energy, while the outgoing resonance stems from phonon-assisted excitation of the KK exciton. Using language that is common in the context of solid state single photon emitters coupled to phonons, the incoming resonance corresponds to direct excitation of the zero-phonon line and emission on the first phonon sideband (below the zero-phonon line) while the outgoing resonance corresponds to excitation on the first phonon sideband (above the zero-phonon line) and emission on the zero-phonon line.~\cite{krummheuer2002theory, wigger2019phonon, preuss2022resonant}.

To discuss the symmetry properties of the resonance Raman profiles, in Fig.~\ref{fig:resonance}(b) we show the unstrained case $\delta_{\mathrm{KQ}}=0$ as marked by the dashed horizontal line in (a). The full simulation via Eq.~\eqref{eq:I_final} (solid) contains the factor $\xi(\omega_l-\Omega_{\mathrm{A}_1})\sim (\omega_l-\Omega_{\mathrm{A}_1})^4$, which naturally leads to slightly asymmetric peaks [note that for WSe$_2$ we use $E_{\mathrm{KK}}(0)=1.74$~eV, see App.~\ref{app:model_params}]. When we remove this influence by replacing the factor with a constant $\xi(\omega_l-\Omega_{\mathrm{A}_1})\rightarrow \xi(E_{\mathrm{KK}}(0)-\Omega_{\mathrm{A}_1}/2)$ (dashed) we find a virtually symmetric double peak structure. Considering the scattering amplitude in Eq.~\eqref{eq:A_R_final_final}, the first order contribution is completely symmetric under reflection at $\omega_l=E_{\mathrm{KK}}(0)+\Omega_{\mathrm{A}_1}/2$. Any third order scattering from the bright KK exciton into the KK- and KK$'$-exciton parabolas accompanied by A$_1$-phonon emission from these intermediate states (second contribution in Fig.~\ref{fig:stokes}) would likely lead to a deviation from the perfectly symmetric first order contribution, indicating that these third order processes are negligible.

Turning on the A$_1$-phonon emission from the Q-valley qualitatively changes the resonance Raman profiles, as shown in Fig.~\ref{fig:resonance} (c,d) for $\bar{g}_{\mathrm{K}}^{\mathrm{Q}}=5$. The detuning-dependent profiles in (c) exhibit a clear suppression of the incoming resonance around the unstrained case $\delta_{\mathrm{KQ}}=0$. Considering the scattering amplitude in Eq.~\eqref{eq:A_R_final_final}, the additional contributions due to A$_1$-phonon emission from the Q-valley are weighted by the propagator products
\begin{equation}
	\mathcal{G}_{\mathrm{QK}}\qty(\omega_l-\Omega_{j}(\vb{q}),-\vb{q})\mathcal{G}_{\mathrm{QK}}\qty(\omega_l-\Omega_{j}(\vb{q})-\Omega_{\mathrm{A}_1},-\vb{q})\,.\label{eq:third_order_resonance_G}
\end{equation}
Since scattering between the K- and Q-valley is mediated by Q-phonons, we only need to consider $j=\mathrm{ac/opt},\mathrm{Q}$. Here, the acoustic phonons at the Q-point in WSe$_2$ have an average frequency of $\Omega_{\mathrm{ac},\mathrm{Q}}\approx 13$~meV, while the optical phonons have an average frequency of $\Omega_{\mathrm{opt},\mathrm{Q}}\approx 30$~meV (see Tab.~\ref{tab:phon} in App.~\ref{app:model_params}). Furthermore the energetic distance between the lowest-lying KK exciton and the lowest-lying KQ exciton in the unstrained case is given by $E_{\mathrm{KK}}(0)-E_{\mathrm{KQ}}(0)\approx 43$~meV [see Fig.~\ref{fig:strainrates}(b)], while the frequency of the A$_1$-phonon is $\Omega_{\mathrm{A}_1}=30.8$~meV. This implies that in the unstrained case we have
\begin{equation}\label{eq:third-order-resonance}
	\delta_{\mathrm{KQ}}=0:\qquad E_{\mathrm{KK}}(0)\approx E_{\mathrm{KQ}}(0)+\Omega_{\mathrm{A}_1}+\Omega_{\mathrm{ac},\mathrm{Q}}\equiv \omega_l^{(3,\mathrm{Q})}
\end{equation}
in hBN-encapsulated monolayer WSe$_2$. As a guide to the eye, in Fig.~\ref{fig:resonance} (c), we show the strain-dependent right-hand-side of the previous equation as a green line. The interference between first order and third order phonon scattering processes in Eq.~\eqref{eq:A_R_final_final} at the resonance from Eq.~\eqref{eq:third-order-resonance} is destructive, leading to a suppression of the incoming resonance peak at $\omega_l=E_{\mathrm{KK}}(0)$. However the same qualitative effect cannot be seen at a larger detuning of $\delta_{\mathrm{KQ}}\approx 30$~meV, where the green line crosses the outgoing resonance. For such large detuning values the KQ exciton has a much larger linewidth compared to the unstrained case [see Fig.~\ref{fig:strainrates}(a, red)], reducing its impact on the resonance Raman amplitude in Eq.~\eqref{eq:A_R_final_final} stemming from the propagator product in Eq.~\eqref{eq:third_order_resonance_G}.

%

We conclude this discussion on the impact of inter-valley scattering on resonance Raman profiles in hBN-encapsulated WSe$_2$ by showing the dependence of these profiles  on the relative strength of the A$_1$-phonon emission from the Q-valley $\bar{g}_{\mathrm{K}}^{\mathrm{Q}}$ in Fig.~\ref{fig:resonance}(e) in the unstrained case $\delta_{\mathrm{KQ}}=0$. We generally find that, due to the suppression of the incoming resonance observed in (c), the double peak structure becomes increasingly asymmetric when increasing the influence of the Q-valley via $\bar{g}_{\mathrm{K}}^{\mathrm{Q}}$. It is to note however that for sufficiently large $\bar{g}_{\mathrm{K}}^{\mathrm{Q}}$ the third order description in Eq.~\eqref{eq:A_R_final_final} will become insufficient and even higher order diagrams have to be included. Nonetheless the qualitative trend towards more asymmetric resonance Raman profiles with a suppressed incoming resonance is clearly visible. For a more quantitative picture in (f) we show cuts in (e) as denoted by the respective colors. Most notably, the incoming resonance splits into two sub-peaks for a sufficiently strong impact of the Q-valley (red and violet curve). This can be seen as a simple geometric feature of the different lines shown in (c): The first order incoming resonance is a broadened vertical line at $\omega_l=E_{\mathrm{KK}}(0)$, while the third order resonance $\omega_l=\omega_l^{(3,\mathrm{Q})}$ that leads to destructive interference (green line) crosses in a diagonal fashion. Therefore the incoming resonance bends towards the left when approaching $\delta_{\mathrm{KQ}}=0$ from above, while it bends to the right, when approaching from below. For a sufficiently strong destructive interference these two bends can effectively be separated and lead to a double peak substructure of the incoming resonance around $\omega_l=E_{\mathrm{KK}}(0)$ for the unstrained case $\delta_{\mathrm{KQ}}=0$.
\subsubsection{Comparison with existing experimental data}
\begin{figure}[t]
	\centering
	\includegraphics[width=\linewidth]{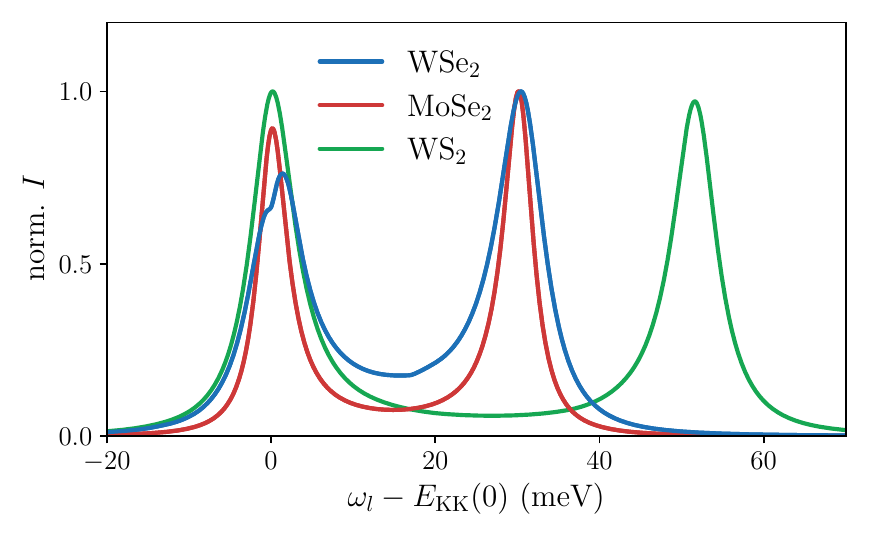}
	\caption{Resonance Raman profiles $I(\omega_l-\Omega_{\mathrm{A}_1})$ of different unstrained $(\delta_{\mathrm{KQ}}=0)$ hBN-encapsulated TMDC monolayers at $T=4$~K for $\bar{g}_{\mathrm{K}}^{\mathrm{Q}}=2.5$ in all three cases.}
	\label{fig:compare}
\end{figure}

To support the following discussion, in Fig.~\ref{fig:compare} we show simulations of resonance Raman profiles for unstrained hBN-encapsulated monolayers of WSe$_2$ (blue), MoSe$_2$ (red) and WS$_2$ (green) at $T=4$~K. For each presented result we consider $\bar{g}_{\mathrm{K}}^{\mathrm{Q}}=2.5$. Note however that this effective parameter can in principle be different for different materials. Parameter scans for WSe$_2$ have already been discussed in the context of Fig.~\ref{fig:resonance}. Additional parameter scans for MoSe$_2$ and WS$_2$ can be found in App.~\ref{app:rrp_add}.

Recently published measurements on resonance Raman scattering in WSe$_2$~\cite{lambers2026resonant} show asymmetric resonance Raman profiles of the A$_1$-peak similar to the ones in Fig.~\ref{fig:resonance} (f) and Fig.~\ref{fig:compare} (blue line), however with a large quantitative variation between the two presented samples. The authors attribute the asymmetry to inter-valley scattering via Q-phonons, which is supported by the microscopic modeling presented here. The variation between samples can furthermore be explained by the influence of strain, since the resonance Raman profiles in WSe$_2$ are particularly sensitive to strain around the pristine case of $\delta_{\mathrm{KQ}}=0$ due to the interference with the third order resonance [see Fig.~\ref{fig:resonance} (c)]. Note that our effective model in principle allows for negative values of $\bar{g}_{\mathrm{K}}^{\mathrm{Q}}$, which however would lead to an increase of the incoming resonance, instead of a suppression, in Fig.~\ref{fig:resonance}, which is not consistent with the data in Ref.~\cite{lambers2026resonant} and is therefore ruled out.

Similar resonance Raman profiles for the A$_1$-peak in WSe$_2$ around the 1s-exciton resonance have been published in Ref.~\cite{mcdonnell2020observation}. While the presented datapoints constitute symmetric resonance Raman profiles, especially the uncertainty at the out-going resonance is exceptionally large compared with the uncertainty of all other datapoints [Fig.~2 (b) therein]. The presented data in Ref.~\cite{mcdonnell2020observation} is therefore consistent with our model calculations in Figs.~\ref{fig:resonance} and \ref{fig:compare} (blue line). 

Furthermore the authors in Ref.~\cite{mcdonnell2020observation} present symmetric resonance Raman profiles for the A$_1$-peak in MoSe$_2$, which is also supported by our modeling (red line in Fig.~\ref{fig:compare}; Fig.~\ref{fig:resonance_MoSe2} in App.~\ref{app:rrp_add}). Note that the slight asymmetry in Fig.~\ref{fig:compare} (red) stems from the factor $\xi(\omega_l-\Omega_{\mathrm{A}_1})\sim (\omega_l-\Omega_{\mathrm{A}_1})^4$ in Eq.~\eqref{eq:I_final}. In MoSe$_2$ both the KQ exciton and the KK' exciton lie above the KK exciton (Tabs.~\ref{tab:bands} and~\ref{tab:wannier} in App.~\ref{app:model_params}), such that the inter-valley scattering terms in the Raman scattering amplitude in Eq.~\eqref{eq:A_R_final_final} are effectively removed. This leads to symmetric resonance Raman profiles for not too strong coupling of the 1s exciton with optical phonons at the $\Gamma$-point, such that the symmetric first order term in Eq.~\eqref{eq:A_R_final_final} dominates. Note that this symmetric situation can only be obtained when choosing the relative signs of the optical phonon intra-valley scattering deformation potential correctly~\cite{brem2020microscopic} (see Tab.~\ref{tab:phon} in App.~\ref{app:model_params}), since otherwise the third order intra-valley term in Eq.~\eqref{eq:A_R_final_final} becomes too strong, which is inconsistent with the experimental data.

Further data on resonance Raman profiles for the A$_1$-peak in WS$_2$ from Ref.~\cite{mcdonnell2018probing} shows an approximate symmetry between incoming and outgoing resonance. This is largely supported by our model (green line in Fig.~\ref{fig:compare}; Fig.~\ref{fig:resonance_WS2} in App.~\ref{app:rrp_add}), however we find that slightly asymmetric resonance Raman profiles can emerge depending on strain and the specific value of the parameter $\bar{g}_{\mathrm{K}}^{\mathrm{Q}}$, which needs to be on the order of $\bar{g}_{\mathrm{K}}^{\mathrm{Q}}\lesssim2$ to produce symmetric profiles for the unstrained monolayer. Note that the resonance Raman profiles in Ref.~\cite{mcdonnell2018probing} show an additional shoulder below the incoming resonance, which is attributed to trions. We cannot reproduce this feature in our model, which assumes a neutral monolayer with small charge densities, i.e., considers only the influence of excitons in the low-density limit~\cite{katsch2018theory}.

In total we find that our modeling of the resonance Raman profile of the A$_1$-peak in different TMDC monolayers is consistent with experimental data~\cite{mcdonnell2018probing,mcdonnell2020observation,lambers2026resonant}. The situation in WSe$_2$ however is unique in the sense that the interference with the third order resonance shown in Fig.~\ref{fig:resonance} (c) is present in the case of a pristine, unstrained monolayer, leading to a significant suppression of the incoming resonance. The strength of this suppression depends on strain and the specific value of the relative A$_1$-phonon emission strength from the electronic Q-valley compared to the electronic K-valley $\bar{g}_{\mathrm{K}}^{\mathrm{Q}}$, which thus becomes an effective parameter that is directly measurable in strain-dependent resonance Raman scattering experiments.
\section{Conclusion}
We have studied resonance Raman profiles of the A$_1$-mode in monolayer TMDCs, starting with a derivation of the scattering spectrum based on modeling the actual spectrometer signal in the far field. Considering monochromatic laser driving we have established a representation of the scattering spectrum in terms of Green's functions, allowing for a diagrammatic evaluation of the resonance Raman scattering process and inclusion of phonon-induced exciton linewidths, i.e., exciton self-energies, in terms of a self-consistent Born approximation. We have discussed the dependence of the various exciton self-energies of the direct (bright) and the indirect (momentum-dark) excitons on strain in the form of a relative energy shift of the electronic Q-valley.

While we consider established ab initio parameters for the exciton-phonon coupling in the form of effective momentum-averaged deformation potentials~\cite{jin2014intrinsic}, the calculated resonance Raman profiles contain two free parameters: (i) Strain as described above, and (ii) the strength of the A$_1$-phonon emission from the electronic Q-valley relative to the emission from the K-valley $\bar{g}_{\mathrm{K}}^{\mathrm{Q}}$. In a case study of hBN-encapsulated monolayer WSe$_2$ we have discussed the impact of these two parameters on the resonance Raman profile of the A$_1$-mode and found that the profile exhibits a suppressed incoming resonance, i.e., asymmetric peaks, due to electron inter-valley scattering between the K-valley and the Q-valley. The suppression of the incoming resonance is essentially facilitated by destructive interference between the first order Raman process, i.e., involving only the emission of the A$_1$-phonon from the electronic K-valley, and a third order Raman process where the A$_1$-phonon is emitted from the electronic Q-valley. This effect can potentially lead to a double peak substructure of the incoming resonance. 

Finally we discuss our theoretical results in the context of existing experimental data~\cite{mcdonnell2020observation,lambers2026resonant} on resonance Raman scattering in WSe$_2$ and find that our simulated resonance profiles agree qualitatively well with these measurements. As a check for our model we furthermore provide simulations of resonance Raman profiles for MoSe$_2$ and WS$_2$ (App.~\ref{app:rrp_add}), which also agree qualitatively with published data~\cite{mcdonnell2018probing,mcdonnell2020observation}.

While our model calculation depends on the two free parameters described previously, we conclude that the relative strength of the A$_1$-phonon emission $\bar{g}_{\mathrm{K}}^{\mathrm{Q}}$ can in principle be directly measured in resonance Raman scattering when performing a systematic strain-tuning study of monolayer TMDCs.

\begin{acknowledgments}
	We thank Hendrik Lambers and Ursula Wurstbauer for fruitful discussions.
\end{acknowledgments}

	\appendix
	
	\section{Detailed derivation of the scattering spectrum}

	\subsection{Time-integrated spectrum and dipole correlation function}\label{app:avg_spec}
		The time-dependent spectrum detected by the spectrometer setup described in Sec.~\ref{sec:avg_spec} is given by~\cite{eberly1977time,groll2025fundamentals}
	\begin{align}\label{eq:I_t}
		I_{\mathrm{D}}(\omega,t)&=\int\limits_{-\infty}^{\infty}\dd t_1\,\int\limits_{-\infty}^{\infty}\dd t_2\, r_{\omega}^*(t_1)r_\omega(t_2)\\
		&\quad\times\expval{\qty[\bm{\epsilon}_{\mathrm{D}}\cdot \vb{E}^{(-)}(\vb{r}_{\mathrm{D}},t-t_1)]\qty[\bm{\epsilon}^*_{\mathrm{D}}\cdot \vb{E}^{(+)}(\vb{r}_{\mathrm{D}},t-t_2)]}\notag\,.
	\end{align}
	 Here, $r_\omega(t)$ is the linear, time-invariant and causal spectrometer response. Considering a spectrometer with Lorentzian broadening $\Gamma$~\cite{eberly1977time}
	\begin{equation}
		r_\omega(t)=\Gamma\Theta(t)e^{-i(\omega-i\Gamma) t}
	\end{equation}
	with $\Theta$ being the Heaviside function, the time-integrated spectrum then reads
	\begin{align}
		\bar{I}_{\mathrm{D}}(\omega)&=\int\limits_{-\infty}^{\infty}\dd t\, I_{\mathrm{D}}(\omega,t)\notag\\
		&= \int\limits_{-\infty}^{\infty}\dd\omega'\,\frac{2\pi\Gamma^2}{\Gamma^2+(\omega-\omega')^2}I(\omega')\,.\label{eq:I_int}
	\end{align}
	Here the ideal ($\Gamma\rightarrow 0$) time-integrated spectrum
	\begin{equation}
		I(\omega)=\expval{\qty[\bm{\epsilon}_{\mathrm{D}}^*\cdot \vb{E}^{(+)}(\vb{r}_{\mathrm{D}},\omega)]^{\dagger}\qty[\bm{\epsilon}^*_{\mathrm{D}}\cdot \vb{E}^{(+)}(\vb{r}_{\mathrm{D}},\omega)]}
	\end{equation}
	is broadened by the spectrometer and we made use of the Fourier representation of Heisenberg picture operators, which are denoted by an explicit time dependence throughout the current subsection
	\begin{subequations}\label{eq:op_fourier}
		\begin{align}
			A(t)&=\int\limits_{-\infty}^{\infty}\dd\omega\,A(\omega)e^{-i\omega t} \,,\\ A(\omega)&=\frac{1}{2\pi}\int\limits_{-\infty}^{\infty}\dd t\, A(t)e^{i\omega t}\,.\label{eq:A_omega}
		\end{align}
	\end{subequations}
	Using Eq.~\eqref{eq:A_omega} together with Eq.~\eqref{eq:dipole_rad} from the main text, we obtain Eq.~\eqref{eq:I_id_final} after time-ordering by using $C(t_1,t_2)^*=C(t_2,t_1)$ for the two-time dipole correlation function from Eq.~\eqref{eq:def_C}. 

\subsection{Two-time dipole correlation function in linear response}\label{app:lin_response}
To evaluate the correlation function in Eq.~\eqref{eq:def_C}, we need to determine the dynamics of the dipole operator. To this aim, we can write the total time evolution operator of the optically driven system in terms of a power series with respect to $H_l(t)$ from Eq.~\eqref{eq:H_l} yielding
\begin{align}
	U_{\mathrm{tot}}(t,t_0)&=U(t,t_0)U_{lI}(t,t_0)\notag\\
	&=U(t,t_0)\qty[1-i\int\limits_{t_0}^t\dd\tau\,H_{lI}(\tau)+...]
\end{align}
with the free time evolution operator of the sample
\begin{equation}\label{eq:U_free}
	U(t,t_0)=\exp[-iH(t-t_0)]
\end{equation}
and interaction picture operators defined by
\begin{equation}
	A_I(t)=U^{\dagger}(t,t_0)A_S(t)U(t,t_0)
\end{equation}
with $A_S(t)$ being a Schrödinger picture operator with potentially explicit time dependence. The optically driven dynamics of the considered system, as described by $U_{\mathrm{tot}}(t,t_0)$, are governed by the total Hamiltonian $H_{\mathrm{tot}}(t)=H+H_l(t)$. With this we obtain the Heisenberg picture dynamics of the dipole operator
\begin{align}
	\bm{\epsilon}^*_{\mathrm{D}}\cdot\vb{d}^{(+)}(t)&=U_{\mathrm{tot}}^{\dagger}(t,t_0)\bm{\epsilon}^*_{\mathrm{D}}\cdot\vb{d}^{(+)}U_{\mathrm{tot}}(t,t_0)\label{eq:dyn_dip}\\
	&\approx\bm{\epsilon}^*_{\mathrm{D}}\cdot\vb{d}_I^{(+)}(t)-i\int\limits_{t_0}^t\dd\tau\,\comm{\bm{\epsilon}^*_{\mathrm{D}}\cdot\vb{d}_I^{(+)}(t)}{H_{lI}(\tau)}\notag
\end{align}
neglecting terms on the order of $\mathcal{O}(\mathcal{E}^2)$.
Note that the positive frequency component of the dipole operator in the interaction picture $\vb{d}_I^{(+)}(t)$ is nothing else than the positive frequency component present in Eq.~\eqref{eq:def_C} for vanishing laser driving $\mathcal{E}=0$, since Heisenberg and interaction picture coincide for $H_l(t)\rightarrow 0$. For any eigenstates of the free time evolution $H\ket{E}=E\ket{E}$ we can write 
\begin{equation}
	\bra{E}\vb{d}^{(+)}_I(t)\ket{E'}= e^{i(E-E')(t-t_0)}\bra{E}\vb{d}^{(+)}\ket{E'}\,.
\end{equation}
In the following we are interested in dipole transitions in the optical range, which are furthermore in close resonance with the monochromatic laser $E'-E\sim\omega_l\sim 1$~eV, and we neglect all other dipole transitions which amounts to performing a rotating wave approximation~\cite{cohen1998atom, groll2025fundamentals}. It is then reasonable to assume that for not too strong light-matter coupling the definition of the positive frequency component of the dipole operator is equivalent to
\begin{equation}
	\bra{E}\vb{d}^{(+)}\ket{E'}=\left\lbrace\begin{matrix}\bra{E}\vb{d}^{(+)}\ket{E'}\,,&E'-E\approx\omega_l\\0\,,&\text{else}\end{matrix}\right.
\end{equation}
for the considered model of the sample. With this definition the positive frequency component $\vb{d}^{(+)}$ is a superposition of operators leading to a downwards transition in energy, i.e., de-excitation of the material, while $\vb{d}^{(-)}$ describes upwards transitions, i.e., excitation of the material. Since the transition frequency for such a transition lies in the optical range around $\omega_l$, excitation of the sample is only possible by absorption of photons in this optical range. This implies that if the photonic environment of the sample is initially in the vacuum state (valid for $k_{\mathrm{B}}T\ll\omega_l$), and if the material is initially close to its ground state (thermal excitations of modes with frequencies $\omega\ll\omega_l$ are allowed), then we have
\begin{equation}
	\vb{d}^{(+)}_I(t)\rho(t_0)=\rho(t_0)\vb{d}^{(-)}_I(t+\tau)=0
\end{equation}
for the initial density matrix $\rho(t_0)$. It follows that the correlation function $C(t_1,t_2)$ in Eq.~\eqref{eq:def_C} vanishes in absence of external laser driving, i.e., for $H_l\rightarrow 0$, since the dynamics of the dipole operator is then given by $\vb{d}^{(+)}(t_2)=\vb{d}_I^{(+)}(t_2)$ [see Eq.~\eqref{eq:dyn_dip}]. Since no photons can be absorbed from the vacuum, the material remains close to its ground state until time $t_2$, when the action of the operator $\vb{d}^{(+)}$ would enforce a downwards transition in energy, which is not possible, leading to the vanishing of the correlation function. Including the laser driving via Eq.~\eqref{eq:H_l} allows for the excitation of the material prior to time $t_2$, such that it can be de-excited via $\vb{d}^{(+)}(t_2)$. Considering a similar argument for the action of $\vb{d}^{(-)}(t_1)$ to the left in Eq.~\eqref{eq:def_C}, the correlation function is of at least second order in $H_l(t)$. Using Eq.~\eqref{eq:dyn_dip}, we can write the correlation function from Eq.~\eqref{eq:def_C} as
\begin{align}
	C(t+\tau,t)&=\frac{\omega_l|\mathcal{E}|^2}{2\epsilon_0  V_0}\int\limits_{t_0}^{t+\tau}\dd\tau_1\int\limits_{t_0}^t\dd\tau_2\,e^{-i\omega_l(\tau_2-\tau_1)}\notag\\
	&\qquad\times C_0(\tau_1,t+\tau,t,\tau_2)
\end{align}
with the four-time correlation function whose dynamics are governed solely by the free Hamiltonian $H$
\begin{align}
	C_0(\tau_1,t+\tau,t,\tau_2)&=\left<\qty[\bm{\epsilon}_l^*\cdot\vb{d}_I^{(+)}(\tau_1)]\qty[ \bm{\epsilon}_{\mathrm{D}}\cdot\vb{d}_I^{(-)}(t+\tau)]\right.\notag\\
	&\left.\times\qty[\bm{\epsilon}^*_{\mathrm{D}}\cdot\vb{d}_I^{(+)}(t)]\qty[\bm{\epsilon}_l\cdot\vb{d}_I^{(-)}(\tau_2)]\right>\,,\label{eq:def_C_0}
\end{align}
where $\expval{...}=\Tr[...\rho(t_0)]$. In the derivation we dropped all contributions with $\vb{d}^{(+)}_I$ to the right or $\vb{d}^{(-)}_I$ to the left acting on $\rho(t_0)$, as well as counter-rotating terms, as discussed previously. For this reason we also only need to go to the linear, not the quadratic order, in Eq.~\eqref{eq:dyn_dip}, to obtain the correlation function in second order, since the zeroth-order contributions always vanish, i.e., combining the dynamics in zeroth order with the dynamics in second order in the correlation function yields a vanishing contribution.

Finally, we consider switching the laser on adiabatically by including a factor $e^{\eta t}$ into $\vb{E}_l(t)$ from Eq.~\eqref{eq:H_l}, taking the limit $\eta\rightarrow 0^+$ at the end of calculations. This allows us to take the limit $t_0\rightarrow-\infty$. Substituting $\tau_2\rightarrow t-\tau_2$ and $\tau_1\rightarrow t+\tau-\tau_1$ then leads to
\begin{align}
	C(t+\tau,t)&=\frac{\omega_l|\mathcal{E}|^2}{2\epsilon_0  V_0}\int\limits_{0}^{\infty}\!\!\dd\tau_1\int\limits_{0}^{\infty}\!\!\dd\tau_2\,e^{-i\omega_l(\tau_1-\tau_2-\tau)}e^{-0^+(\tau_1+\tau_2)} \notag\\
	&\qquad \times C_0(t+\tau-\tau_1,t+\tau,t,t-\tau_2)\,.\label{eq:C_app}
\end{align}
Assuming that the initial density matrix before optical excitation is given by a thermal state at temperature $T$
\begin{equation}\label{eq:rho_thermal}
	\rho(t_0\rightarrow-\infty)=\frac{e^{-\beta H}}{\Tr\qty(e^{-\beta H})}\,,\qquad \beta=1/(k_{\mathrm{B}}T)
\end{equation}
we find that the previous result for the correlation function $C(t+\tau,t)$ becomes independent of $t$ since the free time evolution operator in Eq.~\eqref{eq:U_free} commutes with the initial thermal density matrix. We can thus write
\begin{align}
	C_0&(t+\tau-\tau_1,t+\tau,t,t-\tau_2)=C_0(\tau-\tau_1,\tau,0,-\tau_2)\notag\\
	&=\left<e^{iH\tau}e^{-iH\tau_1}\qty(\bm{\epsilon}_l^*\cdot\vb{d}^{(+)})e^{iH\tau_1}\qty( \bm{\epsilon}_{\mathrm{D}}\cdot\vb{d}^{(-)})e^{-iH\tau}\right.\notag\\
	&\qquad\left.\times\qty(\bm{\epsilon}^*_{\mathrm{D}}\cdot\vb{d}^{(+)})e^{-iH\tau_2}\qty(\bm{\epsilon}_l\cdot\vb{d}^{(-)})e^{iH\tau_2}\right>\,.\label{eq:C_0_1}
\end{align}
When we insert this result into Eq.~\eqref{eq:C_app}, we can identify the polarizability operator from Eq.~\eqref{eq:polariz} (integral over $\tau_2$), as well as its adjoint (integral over $\tau_1$), such that finally the stationary correlation function can be rewritten as presented in Eq.~\eqref{eq:C_stat}.

By making use of the cyclic property of the trace, we can bring Eq.~\eqref{eq:C_0_1} into the form
\begin{align}
	C_0&(\tau-t_1,\tau,0,-t_2)=\Tr\Bigg\lbrace\bm{\epsilon}_{\mathrm{D}}\cdot\vb{d}^{(-)}\mathcal{U}(\tau,\tau-t_1)\notag\\
	&\times\Bigg[\mathcal{U}(\tau-t_1,0)\Big[\bm{\epsilon}^*_{\mathrm{D}}\cdot\vb{d}^{(+)}\mathcal{U}(0,-t_2)\notag\\
	&\times\big[\bm{\epsilon}_l\cdot\vb{d}^{(-)}\mathcal{U}(-t_2,-\infty)[\rho(-\infty)]\big]\Big]\bm{\epsilon}_l^*\cdot\vb{d}^{(+)}\Bigg]\Bigg\rbrace\label{eq:C_0_ordered}
\end{align}
with $\rho(-\infty)$ being the initial thermal state in Eq.~\eqref{eq:rho_thermal}, and
\begin{equation}\label{eq:evol_unitary}
	\mathcal{U}(t,t_0)\qty[A]=U(t,t_0)A U^{\dagger}(t,t_0)
\end{equation}
being the time evolution super-operator acting on density matrices and other Hilbert space operators. Equation~\eqref{eq:C_0_ordered} describes a sequence of free time evolutions via $\mathcal{U}$ and dipole transitions via $\vb{d}^{(\pm)}$ at times $-t_2$, $0$, $\tau-t_1$ and $\tau$. This four-time correlation function is completely time-ordered for $-t_2\leq 0\leq \tau-t_1\leq \tau$, i.e., for $t_2\geq 0$ and $\tau\geq t_1\geq 0$.
\subsection{Isolation of the Raman contribution}\label{app:isolation}
We can isolate the part that contains the Raman scattering contributions by writing the correlation function contained in the scattering spectrum $S(\Omega)$ in Eq.~\eqref{eq:def_S} as (see App.~\ref{app:lin_response})
\begin{align}
	\expval{e^{iH\tau}P^{\dagger}e^{-iH\tau}P}&=\int\limits_0^{\infty}\dd t_1\,\int\limits_0^{\infty}\dd t_2\, e^{-i\omega_l(t_1-t_2)-0^+(t_1+t_2)}\notag\\
	&\qquad\times C_0(\tau-t_1,\tau,0,-t_2)
\end{align}
with the four-time correlation function $C_0$ in Eq.~\eqref{eq:C_0_ordered}.
The integral over $t_1$ contains two separate contributions, one for which the correlation function $C_0$ is time-ordered, i.e., with $t_1\leq \tau$, yielding
\begin{align}
	\expval{e^{iH\tau}P^{\dagger}e^{-iH\tau}P}_1&=\int\limits_0^{\tau}\dd t_1\,\int\limits_0^{\infty}\dd t_2\, e^{-i\omega_l(t_1-t_2)-0^+(t_1+t_2)}\notag\\
	&\qquad\times C_0(\tau-t_1,\tau,0,-t_2),\label{eq:P_corr_1}
\end{align}
and another contribution with $t_1>\tau$ yielding
\begin{align}
	\expval{e^{iH\tau}P^{\dagger}e^{-iH\tau}P}_2&=e^{-i\omega_l\tau}\int\limits_0^{\infty}\!\dd t_1\int\limits_0^{\infty}\!\dd t_2\, e^{-i\omega_l(t_1-t_2)-0^+(t_1+t_2)}\notag\\
	&\qquad\times C_0(-t_1,\tau,0,-t_2)\,,\label{eq:P_corr_2}
\end{align}
where we performed the substitution $\tau-t_1\rightarrow -t_1$. The $\tau$-dependence of this second contribution is therefore determined by the laser frequency $\omega_l$ and the dynamics of the dipole operator $\vb{d}_I^{(-)}(\tau)$ with respect to the Hamiltonian $H$ of the sample [see Eq.~\eqref{eq:def_C_0}]. Naturally, the latter does not contain any further contributions from the laser driving. This implies that the corresponding contribution to the scattering spectrum $S_2(\Omega)$, obtained by replacing the full correlation function in Eq.~\eqref{eq:def_S} with the one in Eq.~\eqref{eq:P_corr_2}, can be written as $S_2(\Omega=\Omega'-\omega_l)$ with the frequencies $\Omega'$ stemming from the free dynamics of the sample described by $C_0$ in Eq.~\eqref{eq:P_corr_2}. In relation to the detected spectrum in Eq.~\eqref{eq:I_id_final_S} this implies that $S_2$ contributes with peaks positioned at frequencies $\omega=\Omega+\omega_l=\Omega'-\omega_l+\omega_l=\Omega'$, i.e., peaks that do not shift in frequency when varying the laser frequency. These stem from incoherent spontaneous emission processes after optical excitation of the sample and the coherent Raman scattering contributions are therefore fully contained in the time-ordered contribution from Eq.~\eqref{eq:P_corr_1}~\cite{groll2023read} with the four-time correlation function in Eq.~\eqref{eq:C_0_ordered} being time-ordered for $t_1, t_2,\tau, \tau-t_1\geq0$.

\section{Diagrammatic evaluation of the scattering amplitude}\label{app:diagrams}
For the sake of consistency of the presented material, in the following we derive the necessary Feynman rules used in the diagrammatic evaluation of scattering amplitudes~\cite{fetter2003quantum,haug2008quantum}.
\subsection{Feynman rules on the single exciton subspace}
We are generally interested in matrix elements of the retarded Green's function between single exciton states, i.e.,
\begin{equation}\label{eq:matrix_elem_G_full}
	\bra{\qty{n}}\bra{\vb{K}',\alpha'}G(E)\ket{\vb{K},\alpha}\ket{\qty{m}}\,.
\end{equation}
Here, $\ket{\vb{K},\alpha}=X^{\dagger}_{\alpha}(\vb{K})\ket{0_x}$ is an arbitrary single exciton state with momentum $\vb{K}$ and additional quantum numbers denoted by $\alpha$. Considering the series expansion in Eq.~\eqref{eq:G_series}, we find that such a matrix element consists of sums of different contributions that themselves consist of products of matrix elements of the free Green's function $G_0$ and the interaction $V$.

Starting with the free Green's function $G_0$, we obtain
\begin{align}
	\bra{\qty{n}}&\bra{\vb{K}',\alpha'}G_0(E)\ket{\vb{K},\alpha}\ket{\qty{m}}=\frac{\delta_{\qty{n}\qty{m}}\delta_{\alpha\alpha'}\delta_{\vb{K}\vb{K}'}}{E-E_{\alpha}(\vb{K})-E_{\qty{n}}+i0^+}\notag\\
	&=\delta_{\qty{n}\qty{m}}\delta_{\alpha\alpha'}\delta_{\vb{K}\vb{K}'}\mathcal{G}^{(0)}_{\alpha}(E-E_{\qty{n}},\vb{K})\,,\label{eq:propagator_free_deriv}
\end{align}
where only identical phonon Fock states $\ket{\qty{n}}=\ket{\qty{m}}$ lead to non-vanishing contributions due to the conservation of the phonon number in absence of exciton-phonon scattering $V=0$. Here,
\begin{equation}
	\mathcal{G}^{(0)}_{\alpha}(E,\vb{K})=\frac{1}{E-E_{\alpha}(\vb{K})+i0^+}\label{eq:propagator_free_app}
\end{equation}
is the free (single) exciton propagator, i.e., the eigenvalue of the non-interacting Green's function $G_0(E)$ corresponding to the single exciton state $\ket{\vb{K},\alpha}$ (together with the phonon vacuum). Note that the presence of additional phonons simply shifts the pole of the free exciton propagator in Eq.~\eqref{eq:propagator_free_deriv} by the additional phonon energy $E_{\qty{n}}$ present in the system [see Eq.~\eqref{eq:E_n}].

Considering now the matrix elements of the exciton-phonon interaction, we obtain
\begin{align}
	\bra{\qty{n}}&\bra{\vb{K}',\alpha'}V\ket{\vb{K},\alpha}\ket{\qty{m}}\notag\\
	&=\sum_{j\vb{q}}\mathcal{V}^{\vb{K}\alpha}_{\vb{K}'\alpha'}(j,\vb{q})\bra{\qty{n}}\qty[b_{j}^{}(\vb{q})+b_{j}^{\dagger}(-\vb{q})]\ket{\qty{m}}\notag\\
	&=\sum_{j\vb{q}}\mathcal{V}^{\vb{K}\alpha}_{\vb{K}'\alpha'}(j,\vb{q})\Big[\sqrt{m_j(\vb{q})}\delta_{\qty{n},\qty{m}-1_j(\vb{q})}\notag\\
	&\qquad+\sqrt{m_j(-\vb{q})+1}\delta_{\qty{n},\qty{m}+1_j(-\vb{q})}\Big]\label{eq:V_elem}
\end{align}
with the interaction vertex
\begin{equation}
	\mathcal{V}^{\vb{K}\alpha}_{\vb{K}'\alpha'}(j,\vb{q})=\bra{\vb{K}',\alpha'}V_j(\vb{q})\ket{\vb{K},\alpha}\,\label{eq:vertex_def_app}
\end{equation}
and $\delta_{\qty{n},\qty{m}\pm 1_j(\vb{q})}$ being zero, except if the Fock states $\ket{\qty{n}}$ and $\ket{\qty{m}}$ have identical occupation numbers apart from mode $(j,\vb{q})$, where $n_j(\vb{q})=m_j(\vb{q})\pm 1$. This implies that at most a single mode $(j,\vb{q})$ from the sum in Eq.~\eqref{eq:V_elem} can contribute to the matrix element and the Fock states have to be identical except for that specific mode. Note that for spatially homogeneous systems the interaction vertex has to obey momentum conservation (see App.~\ref{sec:def_pot})
\begin{equation}\label{eq:vertex_momentum}
	\mathcal{V}^{\vb{K}\alpha}_{\vb{K}'\alpha'}(j,\vb{q})\sim \delta_{\vb{K}',\vb{K}+\vb{q}}\,.
\end{equation}

To determine matrix elements of the interacting Green's function on the single exciton subspace, i.e., Eq.~\eqref{eq:matrix_elem_G_full}, we can formally insert identities
\begin{equation}\label{eq:identity}
	1=\sum_{\vb{K},\alpha,\qty{n}}\ket{\vb{K},\alpha}\ket{\qty{n}}\bra{\qty{n}}\bra{\vb{K},\alpha}
\end{equation}
to the left and right of Eq.~\eqref{eq:G_series}, as well as inbetween the operators $V$ and $G_0$ and find that the matrix elements of $G$ consist of the previously determined building blocks [Eqs.~\eqref{eq:propagator_free_deriv},~\eqref{eq:V_elem}]. We can then represent any matrix element of $G$ as given in Eq.~\eqref{eq:matrix_elem_G_full} diagrammatically according to the following Feynman rules:
\begin{itemize}
	\item[(i)] The initial state $\ket{\vb{K},\alpha}\ket{\qty{m}}$ is drawn on the left-hand side of the diagram with the end-point of a thin straight blue line labeled by $\vb{K},\alpha$ representing the exciton and the end-points of thin whirly red lines labeled by $j,\vb{q}$ representing the phonons contained in the multi-mode Fock state $\ket{\qty{m}}$.
	\item[(ii)] The final state $\ket{\vb{K}',\alpha'}\ket{\qty{n}}$ is drawn on the right-hand side of the diagram as end-points of labeled lines analog to (i).
	\item[(iii)] Absorption of a phonon from mode $(j,\vb{q})$ is depicted as a black dot (vertex) with a corresponding in-going phonon line and an in-going and out-going exciton line. Such a vertex contributes with a factor $\mathcal{V}^{\vb{K}\alpha}_{\vb{K}'\alpha'}(j,\vb{q})\sqrt{n_j(\vb{q})}$ for the case of an in-going $(\vb{K},\alpha)$ exciton and an out-going $(\vb{K}',\alpha')$ exciton. Here, the factor $\sqrt{n_j(\vb{q})}$ is determined by the number of phonons $n_j(\vb{q})$, i.e., number of whirly red lines, corresponding to the mode $(j,\vb{q})$ before the absorption vertex.
	\item[(iv)] Emission of a phonon in mode $(j,\vb{q})$ analog to (iii) is depicted by a vertex with an out-going instead of an in-going phonon line and contributes with a factor $\mathcal{V}^{\vb{K}\alpha}_{\vb{K}'\alpha'}(j,-\vb{q})\sqrt{n_j(\vb{q})+1}$ with $n_j(\vb{q})$ being the number of phonons present in that mode before the vertex. Since the exciton-phonon interaction $V$ is hermitian, we can also write this factor as $\mathcal{V}_{\vb{K}\alpha}^{\vb{K}'\alpha'}(j,\vb{q})^*\sqrt{n_j(\vb{q})+1}$.
	\item[(v)] Free propagation of the system between interaction vertices or start/end-points of the diagram  $\bra{\qty{m}}\bra{\vb{K},\alpha}G_0(E)\ket{\vb{K},\alpha}\ket{\qty{m}}$ is represented by the labeled lines described in (i) and contributes with a factor $\mathcal{G}^{(0)}_{\alpha}(E-E_{\qty{m}},\vb{K})$.
	\item[(vi)] Since the matrix elements are represented in terms of phonon Fock states, diagrams that are identical under exchange of phonon labels, i.e., equivalent diagrams, have to be considered only once.
	\item[(vii)] Finally we have to construct all non-equivalent diagrams  contributing to the desired order in the exciton-phonon coupling. We then add all non-vanishing diagrams and sum over all internal quantum numbers labeling the exciton and phonon lines, i.e., quantum numbers of lines not connected to the initial or final state of the diagram, due to the insertion of the identity operator from Eq.~\eqref{eq:identity} into Eq.~\eqref{eq:G_series}.
\end{itemize}
\begin{figure}
	\centering
	\includegraphics[width=\linewidth]{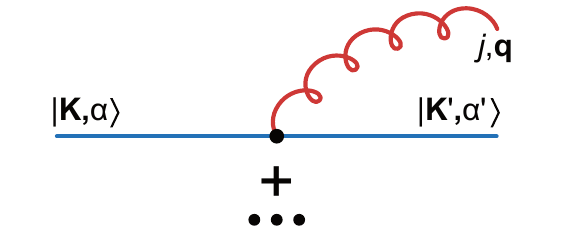}
	\caption{Diagrammatic representation of a phonon emission process to first order in the exciton-phonon coupling, which yields Eq.~\eqref{eq:example_diagram}.}
	\label{fig:example}
\end{figure}

As an illustrative example of these rules, in Fig.~\ref{fig:example} we present the first non-vanishing diagram contributing to the matrix element $\bra{\qty{0}}b_j(\vb{q})\bra{\vb{K}',\alpha'}G(E)\ket{\vb{K},\alpha}\ket{\qty{0}}$ describing emission of a single phonon leading to exciton scattering $\vb{K},\alpha\rightarrow\vb{K}',\alpha'$. Applying the rules, we thus obtain
\begin{align}\label{eq:example_diagram}
	\bra{\qty{0}}&b_j(\vb{q})\bra{\vb{K}',\alpha'}G(E)\ket{\vb{K},\alpha}\ket{\qty{0}}\notag\\
	&=\mathcal{G}^{(0)}_{\alpha}(E,\vb{K})\mathcal{V}^{\vb{K}\alpha}_{\vb{K}'\alpha'}(j,-\vb{q})\mathcal{G}^{(0)}_{\alpha'}(E-\Omega_j(\vb{q}),\vb{K}')\,.
\end{align}
\subsection{Exciton self-energy}
\begin{figure}
	\centering
	\includegraphics[width=\linewidth]{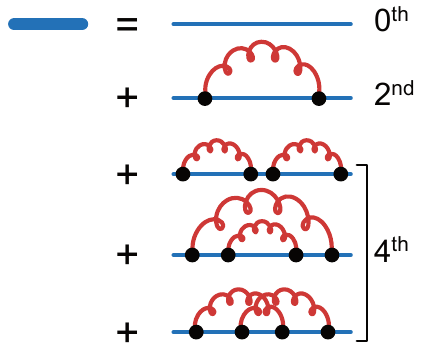}
	\caption{Diagrammatic representation (omitting exciton and phonon labels) of the interacting exciton propagator in Eq.~\eqref{eq:propagator_interacting_app} up to fourth order in the exciton-phonon coupling $V$.}
	\label{fig:propagatorexpansion}
\end{figure}
Scattering with phonons generally modifies the exciton energies, i.e., the poles of exciton propagators such as in Eq.~\eqref{eq:propagator_free_app}. This effect is described by defining the (interacting) exciton propagator (at vanishing temperature)
\begin{equation}\label{eq:propagator_interacting_app}
	\mathcal{G}_{\alpha}(E,\vb{K})=\bra{\qty{0}}\bra{\vb{K},\alpha}G(E)\ket{\vb{K},\alpha}\ket{\qty{0}}\,.
\end{equation}
\begin{figure}
	\centering
	\includegraphics[width=\linewidth]{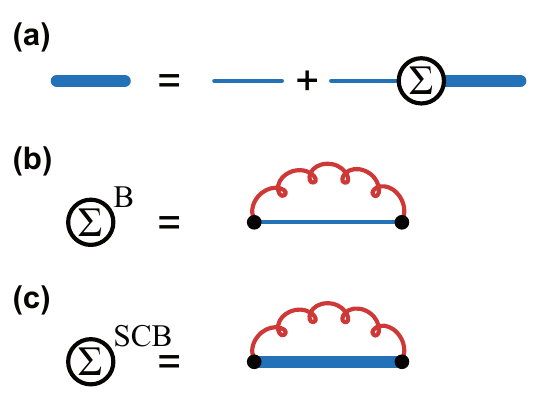}
	\caption{(a) Diagrammatic representation (omitting exciton and phonon labels) of the Dyson series with the self-energy $\Sigma$. (b) Self energy in the Born approximation. (c) Self energy in the self-consistent Born approximation, employed throughout the presented calculations.}
	\label{fig:dyson}
\end{figure}
Using the series expansion in Eq.~\eqref{eq:G_series}, we can represent the exciton propagator diagrammatically as in Fig.~\ref{fig:propagatorexpansion}. Here we use a thick blue line to distinguish the interacting exciton propagator from the free exciton propagator. To deal with this infinite series of diagrams, a myriad of partial resummation schemes exists in the literature~\cite{fetter2003quantum,haug2008quantum}. The key here is to notice that we can rewrite any such diagram in the form presented in Fig.~\ref{fig:dyson}(a), i.e., the Dyson series, where $\Sigma$ denotes the exciton self-energy stemming from exciton-phonon scattering. Choosing the self-energy depicted in Fig.~\ref{fig:dyson}(b) corresponds to the Born approximation (B), which correctly reproduces the first three diagrams of Fig.~\ref{fig:propagatorexpansion}, as well as additional terms in higher orders. An even better approximation is the self-consistent Born approximation (SCB) depicted in Fig.~\ref{fig:dyson}(c), which correctly reproduces the first four diagrams of Fig.~\ref{fig:propagatorexpansion}, as well as additional terms in higher orders. However it does not include any diagrams with crossing phonon lines, such as the last diagram in Fig.~\ref{fig:propagatorexpansion}. In the following we will work in the self-consistent Born approximation, such that the self-energy is given by
\begin{equation}
	\Sigma_{\alpha}(E,\vb{K})=\sum_{j\vb{q}\alpha'\vb{K}'}\qty|\mathcal{V}_{\vb{K}\alpha}^{\vb{K}'\alpha'}(j,\vb{q})|^2 \mathcal{G}_{\alpha'}(E-\Omega_j(\vb{q}),\vb{K}')\,,\label{eq:self_energy_new}
\end{equation}
where we evaluated the diagram in Fig.~\ref{fig:dyson}(c) using the additional rule that we replace $\mathcal{G}^{(0)}$ by $\mathcal{G}$ for the fully interacting (thick) exciton lines. One can generally evaluate the Dyson series in Fig.~\ref{fig:dyson}(a) by inserting the right-hand side into itself iteratively and using the geometric series, leading to
\begin{equation}
	\mathcal{G}_{\alpha}(E,\vb{K})=\frac{1}{E-E_{\alpha}(\vb{K})-\Sigma_{\alpha}(E,\vb{K})+i0^+}\,.
\end{equation}
The exciton energies are thus renormalized by the real part of the self-energy and the resonances, i.e., poles of the Green's function, are broadened by its imaginary part. These effects describe polaron shifts and phonon-induced dephasing of the exciton, respectively.

If we consider a series of diagrams up to a given order containing only free exciton propagators as in Fig.~\ref{fig:example}, we notice that higher order contributions can at least be partially reproduced by replacing the free propagators $\mathcal{G}^{(0)}$ by fully interacting propagators $\mathcal{G}$. This will be done in all considered diagrams in the main text. However one has to be careful not to accidentally double count certain diagrams.

		\section{Derivation of the multi-valley exciton Hamiltonian}\label{app:Hamilton}
		Here we present a full derivation of the exciton-phonon Hamiltonian of a neutral TMDC monolayer. We essentially follow the derivation in Ref.~\cite{groll2026phonon}, generalized to multiple electron and hole valleys.
	\subsection{Electronic multi-valley model}
	We consider the electron-hole Hamiltonian~\cite{katsch2018theory,groll2026phonon}
	\begin{align}\label{eq:H_e_multiband}
		H_{\mathrm{e}}&=\sum_{\vb{k}\lambda}\epsilon^{\mathrm{e}}_{\lambda}(\vb{k})c_{\lambda}^{\dagger}(\vb{k})c_{\lambda}^{}(\vb{k})+\sum_{\vb{k}\lambda'}\epsilon^{\mathrm{h}}_{\lambda'}(\vb{k})d_{\lambda'}^{\dagger}(\vb{k})d_{\lambda'}^{}(\vb{k})\notag\\
		&\quad-\sum_{\vb{k}\vb{k}'\vb{q}\lambda\lambda'}v(\vb{q})c_{\lambda}^{\dagger}(\vb{k})d_{\lambda'}^{\dagger}(\vb{k}')d_{\lambda'}^{}(\vb{k}'-\vb{q})c_{\lambda}^{}(\vb{k}+\vb{q})
	\end{align}
	with the fermionic annihilation (creation) operators $c_{\lambda}^{(\dagger)}(\vb{k})$ and $d_{\lambda'}^{(\dagger)}(\vb{k})$ of electrons in valley $\lambda$ and holes in valley $\lambda'$ with momentum $\vb{k}$, respectively. In the following we will focus on weak optical excitation of a neutral TMDC monolayer, such that we can ignore electron-electron and hole-hole interaction. Here, $v(\vb{q})$ denotes the Coulomb matrix element corresponding to a momentum transfer $\vb{q}$ and in the following we employ the effective mass approximation for the electron and hole energies
	\begin{subequations}
	\begin{align}
		\epsilon_{\lambda}^{\mathrm{e}}(\vb{k})&=E_{\lambda}^{\mathrm{e}}+\frac{(\vb{k}-\vb{k}_\lambda)^2}{2m^{\mathrm{e}}_{\lambda}}\,,\\ \epsilon_{\lambda'}^{\mathrm{h}}(\vb{k})&=-E_{\lambda'}^{\mathrm{h}}+\frac{(\vb{k}+\vb{k}_{\lambda'})^2}{2m^{\mathrm{h}}_{\lambda'}}\,.
	\end{align}
\end{subequations}
	Here, $E_{\lambda}^{\mathrm{e}}$ is the energy of the electronic conduction band minimum at valley $\lambda$, while $E_{\lambda'}^{\mathrm{h}}$ is the energy of the valence band maximum at valley $\lambda'$. The effective electron/hole masses in the different valleys are denoted by $m^{\mathrm{e/h}}_{\lambda^{(\prime)}}$ and the momentum space positions of these valleys are described by $\vb{k}_{\lambda^{(\prime)}}$. Note that we assume that the Coulomb interaction does not change the valley index of electrons and holes, since the different valleys are sufficiently far away in momentum space and the Coulomb matrix element is dominated by contributions around a momentum transfer of $\vb{q}=\vb{0}$. We also neglect the electron-hole exchange interaction, for which the matrix element vanishes in the limit $\vb{q}\rightarrow\vb{0}$~\cite{wu2015exciton}.
	\subsection{Excitons and the Wannier equation}
	Based on the Hamiltonian in Eq.~\eqref{eq:H_e_multiband}, we can define creation operators for intra-valley ($\lambda'=\lambda$) and inter-valley ($\lambda'\neq\lambda$) excitons with total momentum $\overline{\vb{Q}}_{\lambda\lambda'}=\vb{Q}+\vb{k}_{\lambda}-\vb{k}_{\lambda'}$ according to~\cite{brem2020phonon, groll2026phonon}
	\begin{equation}\label{eq:exc_creation}
		X^{(n)\dagger}_{\lambda\lambda'}(\vb{Q})=\sum_{\vb{k}}\phi_{\lambda\lambda'}^{(n)}(\vb{k})c^{\dagger}_{\lambda}(\mu^{\mathrm{e}}_{\lambda\lambda'}\vb{Q}+\vb{k}+\vb{k}_{\lambda})d^{\dagger}_{\lambda'}(\mu^{\mathrm{h}}_{\lambda\lambda'}\vb{Q}-\vb{k}-\vb{k}_{\lambda'})
	\end{equation}
	with $\mu^{\mathrm{e}}_{\lambda\lambda'}=m^{\mathrm{e}}_{\lambda}/M_{\lambda\lambda'}$ and $\mu^{\mathrm{h}}_{\lambda\lambda'}=m^{\mathrm{h}}_{\lambda'}/M_{\lambda\lambda'}$ denoting the mass fractions of electron and hole with respect to the total mass $M_{\lambda\lambda'}=m^{\mathrm{e}}_{\lambda}+m^{\mathrm{h}}_{\lambda'}$ of the electron-hole pair, i.e., the exciton. The exciton wavefunctions in $\vb{k}$-space satisfy the Wannier equation
	\begin{equation}\label{eq:Wannier_2d}
		\sum_{\vb{k}'}\qty[\frac{\vb{k}^2}{2\mu_{\lambda\lambda'}}\delta_{\vb{k}\vb{k}'}-v(\vb{k}'-\vb{k})]\phi_{\lambda\lambda'}^{(n)}(\vb{k}')=B^{(n)}_{\lambda\lambda'}\phi_{\lambda\lambda'}^{(n)}(\vb{k})\,,
	\end{equation}
	with the reduced mass $\mu_{\lambda\lambda'}=m^{\mathrm{e}}_{\lambda}m^{\mathrm{h}}_{\lambda'}/M_{\lambda\lambda'}$, which can be derived from the eigenvalue equation
	\begin{align}\label{eq:wannier_ev_eq}
		H_{\mathrm{e}}X^{(n)\dagger}_{\lambda\lambda'}(\vb{Q})\ket{0}=E^{(n)}_{\lambda\lambda'}(\vb{Q})X^{(n)\dagger}_{\lambda\lambda'}(\vb{Q})\ket{0}
	\end{align}
	using the orthonormality of the electron-hole pair states $c_{\lambda}^{\dagger}(\vb{k})d_{\lambda'}^{\dagger}(\vb{k}')\ket{0}$. The exciton energies thus have to satisfy
	\begin{equation}
		E^{(n)}_{\lambda\lambda'}(\vb{Q})=E_{\lambda}^{\mathrm{e}}-E_{\lambda'}^{\mathrm{h}}+\frac{\vb{Q}^2}{2M_{\lambda\lambda'}}+B^{(n)}_{\lambda\lambda'}\,.
	\end{equation}
	Here, $B^{(n)}_{\lambda\lambda'}$ is the binding energy of the $n$-th exciton state satisfying the Wannier equation~\eqref{eq:Wannier_2d}.
	\subsection{Electronic Hamiltonian in the exciton picture}\label{sec:low-density}
	For undoped semiconductors and weak optical excitation, electrons and holes only ever appear in pairs and we can consider the limit of small electron/hole densities, i.e., exciton densities. In this case we can simplify the electronic Hamiltonian in Eq.~\eqref{eq:H_e_multiband} substantially. For any product of electron operators of the form $c_j^{\dagger}c_i^{}$ with quantum numbers $i$ and $j$ we can then write~\cite{axt1994dynamics,katsch2018theory,groll2026phonon}
	\begin{equation}\label{eq:low-density_approx}
		c_j^{\dagger}c_i^{}\approx c_j^{\dagger}N_{\mathrm{h}}c_i^{}
	\end{equation}
	with the hole number operator
	\begin{equation}
		N_{\mathrm{h}}=\sum_k d_k^{\dagger}d_k^{}\,,
	\end{equation}
	where $k$ are the possible hole quantum numbers. This is due to the fact that acting on the vacuum state, both sides of Eq.~\eqref{eq:low-density_approx} vanish anyway and acting on any single-electron state, there will be a hole present for optically excited undoped semiconductors, such that $N_{\mathrm{h}}$ simply yields unit value. On states with more than one electron equality does not hold, since then $N_{\mathrm{h}}$ would yield a factor of at least $2$, making Eq.~\eqref{eq:low-density_approx} an approximation valid at small electron/hole densities. In an analog fashion we can also write
	\begin{equation}
		d_j^{\dagger}d_i^{}\approx N_{\mathrm{e}}	d_j^{\dagger}d_i^{}=\sum_k c_k^{\dagger}d_j^{\dagger}d_i^{} c_k^{}\,,\qquad N_{\mathrm{e}}=\sum_k c_k^{\dagger}c_k^{}\,.
	\end{equation}
	With these approximations we can rewrite the Hamiltonian in Eq.~\eqref{eq:H_e_multiband} as
	\begin{align}
		H_{\mathrm{e}}&=\sum_{\vb{k}\vb{k}'\vb{q}\lambda\lambda'}\qty{\qty[\epsilon^{\mathrm{e}}_{\lambda}(\vb{k})+\epsilon^{\mathrm{h}}_{\lambda'}(\vb{k}')]\delta_{\vb{q}\vb{0}}-v(\vb{q})}\notag\\
		&\qquad\times c_{\lambda}^{\dagger}(\vb{k})d_{\lambda'}^{\dagger}(\vb{k}')d_{\lambda'}^{}(\vb{k}'-\vb{q})c_{\lambda}^{}(\vb{k}+\vb{q})\,.\label{eq:H_e_multiband_low_density}
	\end{align}
	We can furthermore use the completeness of the exciton wavefunctions, being solutions to the Wannier equation~\eqref{eq:Wannier_2d} to invert the definition of the exciton creation operator in Eq.~\eqref{eq:exc_creation} and obtain
	\begin{align}
		\sum_n &\phi^{(n)*}_{\lambda\lambda'}(\vb{k})X^{(n)\dagger}_{\lambda\lambda'}(\vb{Q})\notag\\&=c^{\dagger}_{\lambda}(\mu^{\mathrm{e}}_{\lambda\lambda'}\vb{Q}+\vb{k}+\vb{k}_{\lambda})d^{\dagger}_{\lambda'}(\mu^{\mathrm{h}}_{\lambda\lambda'}\vb{Q}-\vb{k}-\vb{k}_{\lambda'})\,.\label{eq:exc_op_invert}
	\end{align}
	With this relation we can rewrite Eq.~\eqref{eq:H_e_multiband_low_density} and apply the Wannier equation in form of the eigenvalue equation~\eqref{eq:wannier_ev_eq} to obtain
	\begin{align}\label{eq:H_exc}
		H_{\mathrm{e}}&=\sum_{n\vb{Q}\lambda\lambda'}	E^{(n)}_{\lambda\lambda'}(\vb{Q})X_{\lambda\lambda'}^{(n)\dagger}(\vb{Q})X_{\lambda\lambda'}^{(n)}(\vb{Q})\,,
	\end{align}
	where we furthermore used the orthonormality of the exciton wavefunctions. Note that in the low density limit considered here, the exciton operators in Eq.~\eqref{eq:exc_creation} satisfy bosonic commutation relations~\cite{katsch2018theory}
	\begin{equation}
		\comm{X_{\lambda\lambda'}^{(n)}(\vb{Q})}{X_{\lambda''\lambda'''}^{(n')\dagger}(\vb{Q}')}=\delta_{nn'}\delta_{\vb{Q}\vb{Q}'}\delta_{\lambda\lambda''}\delta_{\lambda'\lambda'''}\,,
	\end{equation}
	such that the form of the exciton Hamiltonian in Eq.~\eqref{eq:H_exc} is consistent with the eigenvalue equation~\eqref{eq:wannier_ev_eq}.
	\subsection{Exciton-phonon coupling}
	In addition to the excitonic structure discussed previously, we consider exciton-phonon coupling. Starting in the electron-hole picture, we take into account phonons in mode $j$ with wavenumber $\vb{q}$ scattering with electrons and holes, respectively, including possible inter-valley scattering, leading to the interaction Hamiltonian~\cite{krummheuer2002theory,selig2016excitonic,lengers2020theory,groll2026phonon}
	\begin{align}
		H_{\mathrm{ep}}&=\sum_{j\vb{k}\vb{q}\lambda\lambda'}\Big[g_{j\lambda\lambda'}^{(\mathrm{e})}(\vb{q})c_{\lambda}^{\dagger}(\vb{k}+\vb{q})c^{}_{\lambda'}(\vb{k})\label{eq:H_ep}\\
		&\qquad-g_{j\lambda\lambda'}^{(\mathrm{h})}(\vb{q})d_{\lambda}^{\dagger}(\vb{k}+\vb{q})d_{\lambda'}^{}(\vb{k})\Big]\qty[b_{j}^{}(\vb{q})+b_{j}^{\dagger}(-\vb{q})]\,.\notag
	\end{align}
	Here, $g^{(\mathrm{e/h} )}_{j\lambda\lambda'}(\vb{q})$ denote the scattering matrix elements for electrons/holes from valley $\lambda'$ to valley $\lambda$ via a phonon in mode $j$ with wavevector $\vb{q}$. We assume these to be independent of $\vb{k}$, which is justified when employing the deformation potential approximation~\cite{jin2014intrinsic, selig2016excitonic, lengers2020theory}. Note that the electron, hole and phonon wavevectors are defined as lying within the first Brillouin zone, i.e., one might need to add reciprocal lattice vectors to $\vb{k}+\vb{q}$ accordingly. The phonon annihilation (creation) operators $b_{j}^{(\dagger)}(\vb{q})$ fulfill the usual bosonic commutation relations and the free phonon Hamiltonian describing phonons with dispersion relation $\Omega_{j}(\vb{q})$ is given by
	\begin{equation}
		H_{\mathrm{p}}=\sum_{j\vb{q}}\Omega_{j}(\vb{q})b_{j}^{\dagger}(\vb{q})b_{j}^{}(\vb{q})\,.
	\end{equation}
	Using the low-density approximation discussed in Sec.~\ref{sec:low-density}, we can rewrite the interaction Hamiltonian in Eq.~\eqref{eq:H_ep} in terms of exciton operators. To do so, we make use of Eqs.~\eqref{eq:low-density_approx} and \eqref{eq:exc_op_invert} leading to
	\begin{align}
		c_{\lambda}^{\dagger}&(\vb{k}+\vb{q})c^{}_{\lambda'}(\vb{k})\approx\sum_{\vb{K}\lambda''}c_{\lambda}^{\dagger}(\vb{k}+\vb{q})d^{\dagger}_{\lambda''}(\vb{K})d^{}_{\lambda''}(\vb{K})c_{\lambda'}^{}(\vb{k})
		\notag\\
		&=\sum_{\vb{K}\lambda''}\left\lbrace\sum_n \phi_{\lambda\lambda''}^{(n)*}\qty[\mu^{\mathrm{h}}_{\lambda\lambda''}(\vb{k}+\vb{q}-\vb{k}_\lambda)-\mu^{\mathrm{e}}_{\lambda\lambda''}(\vb{K}+\vb{k}_{\lambda''})]\right.\notag\\
		&\qquad\quad\qquad\times\left.X^{(n)\dagger}_{\lambda\lambda''}( \vb{k}+\vb{q}+\vb{K}-\vb{k}_\lambda+\vb{k}_{\lambda''})\right\rbrace\notag\\
		&\qquad\times \left\lbrace\sum_n \phi_{\lambda'\lambda''}^{(n)}\qty[\mu^{\mathrm{h}}_{\lambda'\lambda''}(\vb{k}-\vb{k}_{\lambda'})-\mu^{\mathrm{e}}_{\lambda'\lambda''}(\vb{K}+\vb{k}_{\lambda''})]\right.\notag\\
		&\qquad\quad\qquad\qquad\times\left.X^{(n)}_{\lambda'\lambda''}( \vb{k}+\vb{K}-\vb{k}_{\lambda'}+\vb{k}_{\lambda''})\right\rbrace\,.
	\end{align}
	Performing first the substitutions $\vb{k}\rightarrow \vb{k}+\vb{k}_{\lambda'}$ and $\vb{K}\rightarrow \vb{K}-\vb{k}_{\lambda''}$ and then the substitution $\vb{k}+\vb{K}\rightarrow\vb{k}$ leads to
	\begin{align}
		\sum_{\vb{k}}&c_{\lambda}^{\dagger}(\vb{k}+\vb{q})c^{}_{\lambda'}(\vb{k})\\
		&=\sum_{\vb{k}\vb{K}\lambda''nn'} \phi^{(n)*}_{\lambda\lambda''}\qty[\mu^{\mathrm{h}}_{\lambda\lambda''}(\vb{k}+\vb{q}-\vb{k}_\lambda+\vb{k}_{\lambda'})-\vb{K}]\notag\\
		&\quad\times\phi^{(n')}_{\lambda'\lambda''}\qty(\mu^{\mathrm{h}}_{\lambda'\lambda''}\vb{k}-\vb{K})X^{(n)\dagger}_{\lambda\lambda''}( \vb{k}+\vb{q}-\vb{k}_\lambda+\vb{k}_{\lambda'})X^{(n')}_{\lambda'\lambda''}(\vb{k})\,.\notag\,
	\end{align}
	Doing the same calculation for the holes yields
	\begin{align}
		\sum_{\vb{k}}&d_{\lambda}^{\dagger}(\vb{k}+\vb{q})d_{\lambda'}^{}(\vb{k})\\
		&=\sum_{\vb{k}\vb{K}\lambda''nn'} \phi_{\lambda''\lambda}^{(n)*}\qty[\vb{K}-\mu^{\mathrm{e}}_{\lambda''\lambda}(\vb{k}+\vb{q}-\vb{k}_{\lambda'}+\vb{k}_{\lambda})]\notag\\
		&\quad\times\phi_{\lambda''\lambda'}^{(n')}\qty(\vb{K}-\mu^{\mathrm{e}}_{\lambda''\lambda'}\vb{k})X^{(n)\dagger}_{\lambda''\lambda}(\vb{k}+\vb{q}+\vb{k}_\lambda-\vb{k}_{\lambda'})X^{(n')}_{\lambda''\lambda'}(\vb{k})\notag\,.
	\end{align}
	In both of these expressions we identify the exciton form factors for electron and hole scattering processes~\cite{krummheuer2002theory,lengers2020theory,groll2026phonon}
	\begin{equation}\label{eq:form_factor}
		F^{(nn')}_{\lambda_{\mathrm{e}}\lambda_{\mathrm{h}};\lambda_{\mathrm{e}}'\lambda_{\mathrm{h}}'}(\vb{q})=\sum_{\vb{K}}\phi_{\lambda_{\mathrm{e}}\lambda_{\mathrm{h}}}^{(n)*}\qty(\vb{K}-\vb{q})\phi_{\lambda_{\mathrm{e}}'\lambda_{\mathrm{h}}'}^{(n')}\qty(\vb{K})
	\end{equation}
	and we can insert the previous expressions into Eq.~\eqref{eq:H_ep} to obtain the exciton-phonon interaction Hamiltonian in the exciton picture
	\begin{align}\label{eq:H_ep_ex_pic}
		H_{\mathrm{ep}}\!&=\!\!\sum_{j\vb{k}\vb{q}\lambda_{\mathrm{e}}\lambda_{\mathrm{e}}'\lambda_{\mathrm{h}}nn'}\qty[b_{j}^{}(\vb{q})+b_{j}^{\dagger}(-\vb{q})]g_{j\lambda_{\mathrm{e}}\lambda_{\mathrm{e}}'}^{(\mathrm{e})}(\vb{q})\notag\\
		&\!\!\!\quad \times X^{(n)\dagger}_{\lambda_{\mathrm{e}}\lambda_{\mathrm{h}}}( \vb{k}+\vb{q}-\vb{k}_{\lambda_{\mathrm{e}}}+\vb{k}_{\lambda_{\mathrm{e}}'})X^{(n')}_{\lambda_{\mathrm{e}}'\lambda_{\mathrm{h}}}(\vb{k})\notag\\
		&\!\!\!\quad \times  F^{(nn')}_{\lambda_{\mathrm{e}}\lambda_{\mathrm{h}};\lambda_{\mathrm{e}}'\lambda_{\mathrm{h}}}\qty[-\mu^{\mathrm{h}}_{\lambda_{\mathrm{e}}\lambda_{\mathrm{h}}}(\vb{k}+\vb{q}-\vb{k}_{\lambda_{\mathrm{e}}}+\vb{k}_{\lambda_{\mathrm{e}}'})+\mu^{\mathrm{h}}_{\lambda_{\mathrm{e}}'\lambda_{\mathrm{h}}}\vb{k}]\notag\\
		&-\!\!\sum_{j\vb{k}\vb{q}\lambda_{\mathrm{h}}\lambda_{\mathrm{h}}'\lambda_{\mathrm{e}}nn'}\qty[b_{j}^{}(\vb{q})+b_{j}^{\dagger}(-\vb{q})]g_{j\lambda_{\mathrm{h}}\lambda_{\mathrm{h}}'}^{(\mathrm{h})}(\vb{q})\notag\\
		&\!\!\!\quad \times X^{(n)\dagger}_{\lambda_{\mathrm{e}}\lambda_{\mathrm{h}}}(\vb{k}+\vb{q}+\vb{k}_{\lambda_{\mathrm{h}}}-\vb{k}_{\lambda_{\mathrm{h}}'})X^{(n')}_{\lambda_{\mathrm{e}}\lambda_{\mathrm{h}}'}(\vb{k})\notag\\
		&\!\!\!\quad\times F^{(nn')}_{\lambda_{\mathrm{e}}\lambda_{\mathrm{h}};\lambda_{\mathrm{e}}\lambda_{\mathrm{h}}'}\qty[\mu^{\mathrm{e}}_{\lambda_{\mathrm{e}}\lambda_{\mathrm{h}}}(\vb{k}+\vb{q}-\vb{k}_{\lambda_{\mathrm{h}}'}+\vb{k}_{\lambda_{\mathrm{h}}})-\mu^{\mathrm{e}}_{\lambda_{\mathrm{e}}\lambda_{\mathrm{h}}'}\vb{k}].
	\end{align}
This Hamiltonian is of the general form assumed in Eq.~\eqref{eq:H_ep_general}, which was used for the definition of the exciton-phonon vertices in Eqs.~\eqref{eq:vertex_def} and ~\eqref{eq:vertex_def_app}. The form of the vertices in our model is discussed further in the context of Eq.~\eqref{eq:vertex_main_new} in the main text [Eq.~\eqref{eq:vertex} in App.~\ref{sec:def_pot}].
\section{Details of the TMDC monolayer modeling}\label{app:model_params}
In the following we present a detailed description of the numerical modeling of the excitonic properties, as well as the exciton-phonon coupling for the considered TMDC monolayers. The discussion is accompanied by a presentation of the considered model parameters.
\subsection{Modeling anisotropic screening}\label{app:screening}
We want to determine the quasi-static Coulomb field generated by a charge distribution $\rho$ which is placed in an anisotropic dielectric material~\cite{cudazzo2011dielectric,van2018coulomb,brem2020microscopic}. Two of Maxwell's equations are relevant for this situation:
\begin{equation}
	\vb{\nabla}\cdot \vb{D}=\rho\,,\qquad \vb{\nabla}\times\vb{E}=0\,.
\end{equation}
The second equation allows us to express the electric field via the Coulomb potential
\begin{equation}
	\vb{E}=-\vb{\nabla}\Phi\,,
\end{equation}
while in the first equation we can write the displacement field as
\begin{equation}
	\vb{D}(\vb{r})=\epsilon_0 \underline{\epsilon}(\vb{r})\cdot\vb{E}(\vb{r})\,,
\end{equation}
where $\underline{\epsilon}$ is the dielectric tensor of the macroscopic dielectric environment. This leads us to the generalized anisotropic Poisson's equation for the Coulomb potential generated by the charge density $\rho$
\begin{equation}
	\nabla\cdot\underline{\epsilon}\cdot\nabla \Phi=-\frac{\rho}{\epsilon_0}\,.
\end{equation}
In the following we consider a situation where the dielectric environment consists of homogeneous sheets of a certain height, stacked on top of each other in $z$-direction. In the corresponding Cartesian basis spanned by $\vb{e}_{i}$, $i=x,y,z$ the dielectric tensor is then of the form
\begin{equation}
	\underline{\epsilon}(\vb{r})=\text{diag}(\epsilon_{||}(z),\epsilon_{||}(z), \epsilon_{\bot}(z))
\end{equation}
with $\epsilon_{||}$ denoting the in-plane and $\epsilon_{\bot}$ the out-of plane dielectric constant, respectively. This leads us to the following form of Poisson's equation
\begin{equation}
	\epsilon_{||}(z)\nabla_{\vb{r}_{||}}^2 \Phi(\vb{r}_{||},z)+\pdv{z}\epsilon_{\bot}(z)\pdv{z}\Phi(\vb{r}_{||},z)=-\frac{\rho(\vb{r}_{||},z)}{\epsilon_0}\,,
\end{equation}
where $\vb{r}_{||}$ is the component of $\vb{r}$ orthogonal to $\vb{e}_z$, i.e., in-plane. Performing a two-dimensional Fourier transformation
\begin{equation}
	f(\vb{q},z)=N\int\dd^2\vb{r}_{||} e^{-i\vb{q}\cdot\vb{r}_{||}}f(\vb{r}_{||},z)\,,
\end{equation}
with $N$ being an arbitrary normalization factor, we obtain
\begin{equation}\label{eq:main}
	\pdv{z}\epsilon_{\bot}(z)\pdv{z}\Phi(\vb{q},z)-q^2\epsilon_{||}(z)\Phi(\vb{q},z)=-\frac{\rho(\vb{q},z)}{\epsilon_0}\,.
\end{equation}
In the following we assume that the charge density is localized at $z=0$, i.e., 
\begin{equation}
	\rho(\vb{q},z)=\rho(\vb{q})\delta(z)\,.
\end{equation}
Furthermore we consider a situation, where $z=0$ denotes the center of a monolayer material ($m$) of thickness $d$, which is placed on top of a substrate (-) for $z<-d/2$ and capped with another material (+) on top for $z>d/2$ (this can potentially be air or vacuum). The dielectric constants then obey
\begin{equation}
	\epsilon_{i}(z)=\left\lbrace\begin{matrix}\epsilon_{i}^+&z>d/2\\\epsilon_{i}^m&|z|<d/2\\ \epsilon_{i}^-&z<-d/2\end{matrix}\right.\,,\qquad i=||,\bot\,.
\end{equation}
We can now solve Eq.~\eqref{eq:main} by first considering the continuity properties of the potential and its spatial derivative. Integrating over an infinitesimal length around $z=h$ yields
\begin{equation}\label{eq:deriv_cont}
	\eval{\epsilon_{\bot}(z)\pdv{z}\Phi(\vb{q},z)}_{h-0^+}^{h+0^+}=-\frac{\rho(\vb{q})}{\epsilon_0}\int\limits_{h-0^+}^{h+0^+}\dd z\,\delta(z)\,.
\end{equation}
where we assumed that the potential itself has to be continuous, otherwise the electric field that generates a force on charged particles could become infinitely large. We can now distinguish two cases: (i) $|h|>0$ shows that $\epsilon_{\bot}(z)\pdv{z}\Phi(\vb{q},z)$ is continuous for $|z|>0$ and (ii) $h=0$ shows that it is discontinuous for $z=0$ and obeys
\begin{equation}\label{eq:discon}
	\eval{\epsilon_{\bot}(z)\pdv{z}\Phi(\vb{q},z)}_{0^-}^{0^+}=-\frac{\rho(\vb{q})}{\epsilon_0}\,.
\end{equation}
Furthermore both of these conditions are consistent with the assumed continuity of $\Phi(\vb{q},z)$. Solving Eq.~\eqref{eq:main} outside of $z=0$, considering that the potential has to vanish for $|z|\rightarrow\infty$, we obtain four distinct solutions
\begin{align}
	\Phi(\vb{q},z)=\left\lbrace\begin{matrix}A_1 e^{-\alpha_+ qz}&z>d/2\\A_2 e^{-\alpha_m qz}+B_2e^{\alpha_m qz}&0<z<d/2\\A_3 e^{-\alpha_m qz}+B_3e^{\alpha_m qz}&-d/2<z<0\\B_4 e^{\alpha_-qz}&z<-d/2\end{matrix}\right.
\end{align}
with
\begin{equation}
	\alpha_i=\sqrt{\epsilon_{||}^i/\epsilon_{\bot}^i}\,,\qquad i=+,-,m\,.
\end{equation}
Considering the continuity of $\epsilon_{\bot}(z)\pdv{z}\Phi(\vb{q},z)$ and $\Phi(\vb{q},z)$, as well as the discontinuity from Eq.~\eqref{eq:discon}, we obtain six equations determining the six prefactors $A_i$ and $B_i$ from the ansatz, yielding the potential
\begin{align}
	\Phi(\vb{q},z=0)&=\frac{\rho(\vb{q})}{2q\epsilon_0\epsilon(q)}
\end{align}
with the stationary dielectric function
\begin{equation}\label{eq:eps_q_res}
	\epsilon(q)=\frac{\kappa_m}{2}\qty(\frac{1+\frac{\kappa_+-\kappa_m}{\kappa_++\kappa_m}e^{-\alpha_m qd}}{1-\frac{\kappa_+-\kappa_m}{\kappa_++\kappa_m}e^{-\alpha_m qd}}+\frac{1+\frac{\kappa_--\kappa_m}{\kappa_-+\kappa_m}e^{-\alpha_m qd}}{1-\frac{\kappa_--\kappa_m}{\kappa_-+\kappa_m}e^{-\alpha_m qd}})\,,
\end{equation}
where
\begin{equation}
	\kappa_i=\sqrt{\epsilon_{||}^i\epsilon_{\bot}^i}\,,\qquad i=+,-,m\,.
\end{equation}
The widely used Keldysh potential is obtained in the long wavelength limit~\cite{cudazzo2011dielectric,berkelbach2013theory,chernikov2014exciton,selig2016excitonic,lengers2020theory}
\begin{align}
	\epsilon(q)&=\epsilon(0)+\epsilon'(0)q+\mathcal{O}(q^2)\notag\\
	&=\frac{\kappa_++\kappa_-}{2}+\frac{1}{4}\kappa_m\alpha_m\qty(2-\frac{\kappa_+^2+\kappa_-^2}{\kappa_m^2}) q d+\mathcal{O}(q^2)
\end{align}
assuming $\kappa_{\pm}\ll\kappa_m$, such that
\begin{equation}
	\epsilon(q)\approx \epsilon_s \qty(1+r_0 q)
\end{equation}
with
\begin{equation}
	\epsilon_s =\frac{\kappa_++\kappa_-}{2}
\end{equation}
and
\begin{equation}
	r_0=\frac{d\kappa_m\alpha_m}{2\epsilon_s}=\frac{d\epsilon_{||}^m}{2\epsilon_s}\,.
\end{equation}
For identical screening parameters above and below the material $\kappa_+=\kappa_-$, Eq.~\eqref{eq:eps_q_res} can be rewritten in the compact form~\cite{brem2019intrinsic,brem2020phonon}
\begin{equation}\label{eq:eps_q_final}
	\epsilon(q)=\kappa_m\tanh\qty{\frac{1}{2}\qty[\alpha_m q d-\ln\qty(\frac{\kappa_m-\kappa_+}{\kappa_m+\kappa_+})]}\,,
\end{equation}
which is the screening model used in the following to describe hBN-encapsulated TMDC monolayers. The relevant parameters are summarized in Tab.~\ref{tab:screening}.
\begin{table}
	\caption{Screening parameters for hBN-encapsulated TMDC monolayers, described by Eq.~\eqref{eq:eps_q_final}.}
	\begin{tabular}{c|c|c|c|c|c}
		& MoSe$_2$ & WSe$_2$ & WS$_2$ & hBN & Ref.\\
		\hline
		\hline
		$\epsilon_{||}^m$ &16.8 &15.3 &14 &- &\cite{laturia2018dielectric}\\
		$\epsilon_{\bot}^m$ & 7.2 &7.4 & 6.1 &- &\cite{laturia2018dielectric}\\
		$d$ (nm) & 0.649 &0.648 &0.612 &- &\cite{laturia2018dielectric} \\
		$\kappa_{\pm}$ & -& -& -& 4.5&\cite{geick1966normal,brem2019intrinsic}
	\end{tabular}\label{tab:screening}
\end{table}
\subsection{Solving the Wannier equation}\label{app:Wannier_numerical}
Here we describe the numerical solution of the Wannier equation~\eqref{eq:Wannier_2d}. We model the Coulomb matrix element $v(\vb{q})=v(q)$ with $q=|\vb{q}|$ via the anisotropic screening from Eq.~\eqref{eq:eps_q_final}
\begin{subequations}
\begin{align}
	v(q)&=\frac{v_{\mathrm{2D}}(q)}{\epsilon(q)}\,,\\
	v_{\mathrm{2D}}(q)&=\frac{e^2}{2A\epsilon_0q}\,,
\end{align}
\end{subequations}
where $A$ denotes the normalization area, and $-e$ the electron charge. We can separate this potential into two components
\begin{equation}
	v(q)=\frac{v_{\mathrm{2D}}(q)}{\epsilon(0)}-\frac{v_{\mathrm{2D}}(q)}{\epsilon(0)}\frac{\epsilon(q)-\epsilon(0)}{\epsilon(q)}\,.\label{eq:keldysh_split}
\end{equation}
The first term diverges for $q=0$, as it is simply the Coulomb potential with static screening $\epsilon_s=\epsilon(0)$. However the second term does not diverge, since $\epsilon(q)-\epsilon(0)=\mathcal{O}(q)$. This separation allows for an efficient numerical solution of the Wannier equation~\eqref{eq:Wannier_2d}.

Focusing on the radially symmetric 1s-excitons with $\phi_{\lambda_{\mathrm{e}}\lambda_{\mathrm{h}}}^{(1\mathrm{s})}(\vb{k})\equiv\phi_{\lambda_{\mathrm{e}}\lambda_{\mathrm{h}}}(\vb{k})=\phi_{\lambda_{\mathrm{e}}\lambda_{\mathrm{h}}}(|\vb{k}|)$, dropping the index $(n=1\mathrm{s})$ in the following, we can perform the integral over the polar angle $\varphi'$ in  Eq.~\eqref{eq:Wannier_2d}, obtaining an effectively one-dimensional Wannier equation depending only on the absolute values $k=|\vb{k}|$ of the momenta
\begin{equation}\label{eq:wannier_1d}
	\sum_{k'}\qty[\frac{k^2}{2\mu_{\lambda_{\mathrm{e}}\lambda_{\mathrm{h}}}}\delta_{kk'}-\bar{v}(k,k')]\phi^{}_{\lambda_{\mathrm{e}}\lambda_{\mathrm{h}}}(k')=B^{}_{\lambda_{\mathrm{e}}\lambda_{\mathrm{h}}}\phi^{}_{\lambda_{\mathrm{e}}\lambda_{\mathrm{h}}}(k)
\end{equation}
with
\begin{align}
	\bar{v}(k,k')&=\frac{L}{2\pi}\int\limits_{0}^{2\pi}\dd\varphi'\, k' v\qty(|\vb{k}-\vb{k}'|)\notag\\
	&=\bar{v}_{\mathrm{2D}}(k,k')+\Delta \bar{v}(k,k')\,.
\end{align}
Here, $L=\sqrt{A}$ is the normalization length and we separated the total potential into a two-dimensional Coulomb contribution and a remainder due to the screening according to the separation in Eq.~\eqref{eq:keldysh_split}, leading to
\begin{align}
	\bar{v}_{\mathrm{2D}}(k\neq k')&=\frac{k'e^2}{\pi\epsilon_0\epsilon_sL}\frac{1}{\sqrt{2kk'}}\frac{1}{\sqrt{\eta}}K\qty(-2/\eta)\,,\\ \eta&=\frac{(k-k')^2}{2kk'}\notag
\end{align}
with $K(m)$ being the complete elliptic integral of first kind with parameter $m$~\cite{abramowitz1964handbook} and
\begin{align}\label{eq:Delta_V}
	\Delta \bar{v}(k,k')&=-\frac{L}{2\pi}\int\limits_{0}^{2\pi}\dd\varphi' k' \frac{v_{\mathrm{2D}}(|\vb{k}-\vb{k}'|)}{\epsilon(0)}\frac{\epsilon(|\vb{k}-\vb{k}'|)-\epsilon(0)}{\epsilon(|\vb{k}-\vb{k}'|)}\,.
\end{align}
The latter integral is calculated numerically. The expression for $\bar{v}_{\mathrm{2D}}(k,k')$ has a pole at $k=k'$, which can be regularized, using instead
\begin{align}
	\bar{v}^{\delta}_{\mathrm{2D}}(k,k')&=\frac{k'e^2}{\pi\epsilon_0\epsilon_sL}\frac{1}{\sqrt{2kk'}}\frac{1}{\sqrt{\eta(\delta)}}K\qty[-2/\eta(\delta)]\,,\\ \eta(\delta)&=\frac{(k-k')^2}{2kk'}+\delta\notag
\end{align}
and letting $\delta\rightarrow 0^+$ until the isotropic solutions of the Wannier equation~\eqref{eq:Wannier_2d}, i.e., the solutions of Eq.~\eqref{eq:wannier_1d}, converge. The electron band parameters relevant to the solution of the Wannier equation are summarized in Tab.~\ref{tab:bands}. The results of the numerical solution are given in Tab.~\ref{tab:wannier}.

\begin{table}
	\caption{Electron band parameters taken from Ref.~\cite{kormanyos2015k}. Electron and hole masses are given in units of the free electron mass $m_0$. Effective masses in the Q-valley are rendered isotropic by performing the average $2/m^{\mathrm{e}}_{\mathrm{Q}}=1/m^{\mathrm{e}}_{\mathrm{Q},x}+1/m^{\mathrm{e}}_{\mathrm{Q},y}$ over the two principal axes~\cite{selig2018dark}.}
	\begin{tabular}{c|c|c|c}
		& MoSe$_2$ & WSe$_2$ & WS$_2$\\
		\hline
		\hline
		$m^{\mathrm{e}}_{\mathrm{K}}$ ($m_0$) &0.50 &0.29 &0.27 \\
		$m^{\mathrm{e}}_{\mathrm{Q}}$ ($m_0$)&0.71 &0.56 &0.62 \\
		$m^{\mathrm{e}}_{\mathrm{K'}}$ ($m_0$)&0.58 &0.40 &0.36 \\
		$m^{\mathrm{h}}_{\mathrm{K}}$ ($m_0$)&0.60 &0.36 &0.36 \\
		$\Delta^{\mathrm{e}}_{\mathrm{KQ}}$ (meV) & -163 &5 &-27 \\
		$\Delta^{\mathrm{e}}_{\mathrm{KK'}}$ (meV) & -20 &37 &31
	\end{tabular}\label{tab:bands}
\end{table}

\subsection{Exciton wave functions and form factors}\label{app:form_factors}
After solving Eq.~\eqref{eq:wannier_1d} numerically, we fit the resulting 1s wavefunction, using a hydrogenic wavefunction in real space
\begin{equation}
	\psi_{\lambda_{\mathrm{e}}\lambda_{\mathrm{h}}}(\vb{r})=\sqrt{\frac{2}{\pi \sigma_{\lambda_{\mathrm{e}}\lambda_{\mathrm{h}}}^2}}e^{-\frac{|\vb{r}|}{\sigma_{\lambda_{\mathrm{e}}\lambda_{\mathrm{h}}}}}\,,
\end{equation}
which corresponds to 
\begin{equation}
	\phi_{\lambda_{\mathrm{e}}\lambda_{\mathrm{h}}}(\vb{k})=\sqrt{\frac{2}{A\pi \sigma_{\lambda_{\mathrm{e}}\lambda_{\mathrm{h}}}^2}}\frac{2\pi \sigma_{\lambda_{\mathrm{e}}\lambda_{\mathrm{h}}}^2}{\qty(|\vb{k}|^2\sigma_{\lambda_{\mathrm{e}}\lambda_{\mathrm{h}}}^2+1)^{3/2}}
\end{equation}
in reciprocal space. Here, the width $\sigma_{\lambda_{\mathrm{e}}\lambda_{\mathrm{h}}}$ describes the exciton Bohr radius and the normalization in reciprocal space is chosen, such that the exciton form factor
\begin{equation}\label{eq:form_factor_lorentz}
	F^{}_{\lambda_{\mathrm{e}}\lambda_{\mathrm{h}};\lambda_{\mathrm{e}}'\lambda_{\mathrm{h}}'}(\vb{q})=\frac{4\sigma_{\lambda_{\mathrm{e}}\lambda_{\mathrm{h}}}\sigma_{\lambda_{\mathrm{e}}'\lambda_{\mathrm{h}}'}}{\qty(\sigma_{\lambda_{\mathrm{e}}\lambda_{\mathrm{h}}}+\sigma_{\lambda_{\mathrm{e}}'\lambda_{\mathrm{h}}'})^2}\frac{1}{\qty[\qty(|\vb{q}|\frac{\sigma_{\lambda_{\mathrm{e}}\lambda_{\mathrm{h}}}\sigma_{\lambda_{\mathrm{e}}'\lambda_{\mathrm{h}}'}}{\sigma_{\lambda_{\mathrm{e}}\lambda_{\mathrm{h}}}+\sigma_{\lambda_{\mathrm{e}}'\lambda_{\mathrm{h}}'}})^2+1]^{3/2}}
\end{equation}
is normalized to unity for $\vb{q}=\vb{0}$ and $\lambda_{\mathrm{e}}=\lambda_{\mathrm{e}}'$, as well as $\lambda_{\mathrm{h}}=\lambda_{\mathrm{h}}'$. The advantage of using these wavefunctions instead of the explicit numerical solutions, is that we will be able to perform polar angle integrals analytically, such that the numerical demand on the self-consistent self-energy calculations is reduced drastically (see App.~\ref{app:self_energy}). At the same time we find that the fits are of excellent quality, consistent with the findings in Ref.~\cite{henriques2021analytical}.

For the determination of the radiative decay rate of the bright KK exciton, which within a $\vb{k}\cdot\vb{p}$ treatment is given by~\cite{xiao2012coupled,brem2019intrinsic}
\begin{equation}
	\gamma_{\mathrm{r}}=\frac{8\pi}{\sqrt{\kappa_{\pm}}}\alpha \frac{\hbar^2|\psi_{\mathrm{KK}}(\vb{r}=\vb{0})|^2}{M_{\mathrm{KK}}}\frac{E_{\mathrm{KK}}(Q=0)-B_{\mathrm{KK}}}{E_{\mathrm{KK}}(Q=0)}
\end{equation}
with $\alpha$ being the fine structure constant, we however still use the numerically exact result for $|\psi_{\mathrm{KK}}(\vb{0})|$, as it is impacted by the reciprocal space wavefunction $\phi_{\mathrm{KK}}$ at large momenta, for which the hydrogen wavefunction fit is not particularly sensitive.

The fitted form factor parameters, as well as radiative decay rates, are summarized in Tab.~\ref{tab:wannier}.

\begin{table}
	\caption{Parameters obtained by solving the Wannier equation~\eqref{eq:wannier_1d}. For the numerical solution of Eq.~\eqref{eq:Delta_V} we use $10^3$ steps on the interval $\varphi\in (0,\pi]$. We furthermore discretize $k$ and $k'$ in Eq.~\eqref{eq:wannier_1d} into $4\cdot 10^3$ equidistant values on the interval $[0,10/a_{\lambda_{\mathrm{e}}\lambda_{\mathrm{h}}}]$, where the effective Bohr radius is determined via $a_{\lambda_{\mathrm{e}}\lambda_{\mathrm{h}}}=\epsilon(q=0)\hbar /(\mu_{\lambda_{\mathrm{e}}\lambda_{\mathrm{h}}}\alpha c)$ with $c$ being the speed of light and $\alpha$ being the fine structure constant. The bright exciton energy $E_{\mathrm{KK}}(Q=0)$, being a free parameter, is chosen in accordance with experimental data~\cite{meshulam2026temperature}.}
	\begin{tabular}{c|c|c|c}
		& MoSe$_2$ & WSe$_2$ & WS$_2$\\
		\hline
		\hline
		$B_{\mathrm{KK}}$ (meV) &-230.2 &-171.6 &-187.7 \\
		$B_{\mathrm{QK}}$ (meV)&-258.5 &-210.3 &-246.8 \\
		$B_{\mathrm{K' K}}$ (meV)&-242.3 &-191.2 &-209.1 \\
		$\sigma_{\mathrm{KK}}$ (nm)&0.91 &1.35 &1.29 \\
		$\sigma_{\mathrm{QK}}$ (nm) &0.78 &1.05 &0.93 \\
		$\sigma_{\mathrm{K' K}}$ (nm) & 0.85 &1.18 &1.13 \\
		$|\psi_{\mathrm{KK}}(\vb{r}=\vb{0})|^2$ (nm$^{-2}$) &0.549 &0.233 &0.258 \\
		$\gamma_{\mathrm{r}}$ (meV) &3.8 & 2.6 & 3.0\\
		$E_{\mathrm{KK}}(Q=0)$ (eV) &1.64 & 1.74& 2.06\\
	\end{tabular}\label{tab:wannier}
\end{table}

\subsection{Scattering matrix elements and deformation potential approximation}\label{sec:def_pot}
The interaction Hamiltonian in Eq.~\eqref{eq:H_ep_ex_pic} describes both intra- and inter-valley scattering with $\lambda_{\mathrm{e/h}}=\lambda_{\mathrm{e/h}}'$ and $\lambda_{\mathrm{e/h}}\neq\lambda_{\mathrm{e/h}}'$, respectively.  Absorption of a phonon from mode $(j,\vb{q})$, changing the single exciton state from $\ket{\vb{Q}\lambda_{\mathrm{e}}\lambda_{\mathrm{h}}}$ to $\ket{\vb{Q}'\lambda_{\mathrm{e}}'\lambda_{\mathrm{h}}'}$, is described by the matrix element [see Eq.~\eqref{eq:vertex_def}]
\begin{equation}
	\mathcal{V}^{\vb{Q}\lambda_{\mathrm{e}}\lambda_{\mathrm{h}}}_{\vb{Q}'\lambda_{\mathrm{e}}'\lambda_{\mathrm{h}}'}(j,\vb{q})=\bra{\vb{Q}'\lambda_{\mathrm{e}}'\lambda_{\mathrm{h}}'}V_{j}(\vb{q})\ket{\vb{Q}\lambda_{\mathrm{e}}\lambda_{\mathrm{h}}}\label{eq:vertex}
\end{equation}
while the reversed process is described by its complex conjugate. The intra-valley scattering contribution is given by
\begin{align}\label{eq:vertex_intra_main}
	\mathcal{V}^{\vb{Q}\lambda_{\mathrm{e}}\lambda_{\mathrm{h}}}_{\vb{Q}'\lambda_{\mathrm{e}}\lambda_{\mathrm{h}}}(j,\vb{q})=\delta_{\vb{Q}',\vb{Q}+\vb{q}}	g_{j\lambda_{\mathrm{e}}\lambda_{\mathrm{h}}}^{(x)}(\vb{q})
\end{align}
with the intra-valley exciton-phonon coupling constant
\begin{align}
	g_{j\lambda_{\mathrm{e}}\lambda_{\mathrm{h}}}^{(x)}(\vb{q})&=g_{j\lambda_{\mathrm{e}}\lambda_{\mathrm{e}}}^{(\mathrm{e})}(\vb{q})F_{\lambda_{\mathrm{e}}\lambda_{\mathrm{h}};\lambda_{\mathrm{e}}\lambda_{\mathrm{h}}}\qty(-\mu^{\mathrm{h}}_{\lambda_{\mathrm{e}}\lambda_{\mathrm{h}}}\vb{q})\notag\\&\qquad-g_{j\lambda_{\mathrm{h}}\lambda_{\mathrm{h}}}^{(\mathrm{h})}(\vb{q})F_{\lambda_{\mathrm{e}}\lambda_{\mathrm{h}};\lambda_{\mathrm{e}}\lambda_{\mathrm{h}}}\qty(\mu^{\mathrm{e}}_{\lambda_{\mathrm{e}}\lambda_{\mathrm{h}}}\vb{q})\,.\label{eq:g_x}
\end{align}
For inter-valley scattering we can distinguish between (i) electron-phonon scattering and (ii) hole-phonon scattering. The corresponding matrix elements are  
\begin{align}
	&\eval{\mathcal{V}^{\vb{Q}\lambda_{\mathrm{e}}\lambda_{\mathrm{h}}}_{\vb{Q}'\lambda_{\mathrm{e}}'\lambda_{\mathrm{h}}}(j,\vb{q}+\vb{k}_{\lambda_{\mathrm{e}}'}-\vb{k}_{\lambda_{\mathrm{e}}})}_{\lambda_{\mathrm{e}}'\neq \lambda_{\mathrm{e}}}\!\!\!=\delta_{\vb{Q}',\vb{Q}+\vb{q}}g_{j\lambda_{\mathrm{e}}'\lambda_{\mathrm{e}}}^{(\mathrm{e})}(\vb{q}+\vb{k}_{\lambda_{\mathrm{e}}'}-\vb{k}_{\lambda_{\mathrm{e}}})\notag\\
	&\qquad\times	F_{\lambda_{\mathrm{e}}'\lambda_{\mathrm{h}};\lambda_{\mathrm{e}}\lambda_{\mathrm{h}}}\qty[-\mu^{\mathrm{h}}_{\lambda_{\mathrm{e}}'\lambda_{\mathrm{h}}}(\vb{Q}+\vb{q})+\mu^{\mathrm{h}}_{\lambda_{\mathrm{e}}\lambda_{\mathrm{h}}}\vb{Q}]\label{eq:vertex_electron_main}
\end{align}
for (i) electron-phonon inter-valley scattering with $\lambda_{\mathrm{e}}'\neq \lambda_{\mathrm{e}}$, and 
\begin{align}
	&\eval{\mathcal{V}^{\vb{Q}\lambda_{\mathrm{e}}\lambda_{\mathrm{h}}}_{\vb{Q}'\lambda_{\mathrm{e}}\lambda_{\mathrm{h}}'}(j,\vb{q}+\vb{k}_{\lambda_{\mathrm{h}}}-\vb{k}_{\lambda_{\mathrm{h}}'})}_{\lambda_{\mathrm{h}}'\neq \lambda_{\mathrm{h}}}\!\!\!=-\delta_{\vb{Q}',\vb{Q}+\vb{q}}g_{j\lambda_{\mathrm{h}}'\lambda_{\mathrm{h}}}^{(\mathrm{h})}(\vb{q}+\vb{k}_{\lambda_{\mathrm{h}}}-\vb{k}_{\lambda_{\mathrm{h}}'})\notag\\
	&\qquad\times F_{\lambda_{\mathrm{e}}\lambda_{\mathrm{h}}';\lambda_{\mathrm{e}}\lambda_{\mathrm{h}}}\qty[\mu^{\mathrm{e}}_{\lambda_{\mathrm{e}}\lambda_{\mathrm{h}}'}(\vb{Q}+\vb{q})-\mu^{\mathrm{e}}_{\lambda_{\mathrm{e}}\lambda_{\mathrm{h}}}\vb{Q}]\label{eq:vertex_hole_main}\,
\end{align}
for (ii) hole-phonon inter-valley scattering with $\lambda_{\mathrm{h}}'\neq \lambda_{\mathrm{h}}$. In all three instances [Eqs.~\eqref{eq:vertex_intra_main}, \eqref{eq:vertex_electron_main}, and \eqref{eq:vertex_hole_main}] the momentum argument of the phonons is shifted such that $\vb{q}$ reflects the phonon momentum relative to the momentum difference of electrons/holes between the final and initial valleys. 

\begin{table}
	\caption{Phonon frequencies and effective average deformation potential parameters from Ref.~\cite{jin2014intrinsic}. The sign of the acoustic deformation potential in the K-valley valence band is chosen to reflect the opposite shifts of conduction and valence bands in TMDCs under deformation~\cite{he2013experimental,rasmussen2015computational,lengers2020theory, brem2020microscopic}. The acoustic deformation potential for scattering with phonons at the $\Gamma$-point is furthermore scaled by a factor of $1/\sqrt{2}$ with respect to Ref.~\cite{jin2014intrinsic} to reflect the negligible influence of piezo coupling in the charge neutral excitons~\cite{kaasbjerg2012phonon,lengers2020theory}. The sign of the optical deformation potential is kept identical between valence and conduction band reflecting the dominating influence of the Fröhlich coupling~\cite{brem2020microscopic}. Choosing the opposite sign for the valence band would result in resonance Raman profiles incompatible with experimental observation (see discussion in the context of Fig.~\ref{fig:resonance_MoSe2}). Note that there are three degenerate Q-valleys. For notational simplicity we denote all of these by the label Q, however the deformation potentials for intra-valley scattering can be distinguished from those for inter-valley scattering by the involved phonon momentum. The unit cell density is calculated via $\rho_{\mathrm{m}}=(2M_{\mathrm{C}}+M_{\mathrm{M}})(a_0^2\sqrt{3}/2)^{-1}$ with the lattice constant $a_0$ and the masses $M_{\mathrm{C}}$ and $M_{\mathrm{M}}$ of the chalcogen and transition metal atom, respectively~\cite{rasmussen2015computational,lengers2020theory}.}
	\begin{tabular}{c|c|c|c}
		& MoSe$_2$ & WSe$_2$ & WS$_2$\\
		\hline
		\hline
		$c_{\mathrm{s}}$ (km/s) &4.1 &3.3 &4.3 \\
		$\Omega_{\mathrm{opt},\Gamma}$ (meV) & 34.3 &30.7 &46.8 \\
		$\Omega_{\mathrm{ac},\mathrm{K}}$ (meV) &18.3 &16.8 &20.5 \\
		$\Omega_{\mathrm{opt},\mathrm{K}}$ (meV) &32.8 &29.7 &45 \\
		$\Omega_{\mathrm{ac},\mathrm{M}}$ (meV) &18 &15.8 &19.6 \\
		$\Omega_{\mathrm{opt},\mathrm{M}}$ (meV) &33.7 &30 &45.8 \\
		$\Omega_{\mathrm{ac},\mathrm{Q}}$ (meV) &15.1 &13 &17.7 \\
		$\Omega_{\mathrm{opt},\mathrm{Q}}$ (meV) &33.7 &30 &45.9 \\
		$\Omega_{\mathrm{A}_1}$ (meV) &30.3 &30.8 &51.8 \\
		$\sqrt{2}D^{(\mathrm{e})}_{\mathrm{ac},\Gamma,\mathrm{KK}}$ (eV) & 3.4 & 3.2 & 3.2\\
		$\sqrt{2}D^{(\mathrm{h})}_{\mathrm{ac},\Gamma,\mathrm{KK}}$ (eV) & -2.8 & -2.1 & -1.7\\
		$\sqrt{2}D^{(\mathrm{e})}_{\mathrm{ac},\Gamma,\mathrm{QQ}}$ (eV) & 3.1 & 1.9 & 1.8\\
		$D^{(\mathrm{e})}_{\mathrm{ac},\mathrm{K},\mathrm{KK'}}$ (eV/nm) & 18 & 13 & 12\\
		$D^{(\mathrm{e})}_{\mathrm{ac},\mathrm{Q},\mathrm{KQ}}$ (eV/nm) & 15 & 16 & 14\\
		$D^{(\mathrm{e})}_{\mathrm{ac},\mathrm{M},\mathrm{K' Q}}$ (eV/nm) & 49 & 41 & 40\\
		$D^{(\mathrm{e})}_{\mathrm{ac},\mathrm{M},\mathrm{Q Q}}$ (eV/nm) & 22 & 18 & 15\\
		$D^{(\mathrm{e})}_{\mathrm{opt},\Gamma,\mathrm{KK}}$ (eV/nm) & 52 & 23 & 31\\
		$D^{(\mathrm{h})}_{\mathrm{opt},\Gamma,\mathrm{KK}}$ (eV/nm) & 49 & 31 & 23\\
		$D^{(\mathrm{e})}_{\mathrm{opt},\Gamma,\mathrm{Q Q}}$ (eV/nm) & 78 & 27 & 34\\
		$D^{(\mathrm{e})}_{\mathrm{opt},\mathrm{K},\mathrm{KK'}}$ (eV/nm) & 21 & 8 & 11\\
		$D^{(\mathrm{e})}_{\mathrm{opt},\mathrm{Q},\mathrm{KQ}}$ (eV/nm) & 30 & 10 & 13\\
		$D^{(\mathrm{e})}_{\mathrm{opt},\mathrm{M},\mathrm{K' Q}}$ (eV/nm) & 83 & 28 & 46\\
		$D^{(\mathrm{e})}_{\mathrm{opt},\mathrm{M},\mathrm{Q Q}}$ (eV/nm) & 59 & 16 & 19\\
		$\rho_{\mathrm{m}}$ ($10^{-7}$ g/cm$^2$) & 4.42 & 5.95 & 4.67
	\end{tabular}\label{tab:phon}
\end{table}

The exciton form factors $F^{}_{\lambda_{\mathrm{e}}\lambda_{\mathrm{h}};\lambda_{\mathrm{e}}'\lambda_{\mathrm{h}}'}(\vb{k})$ from Eqs.~\eqref{eq:form_factor_main} are localized around $\vb{k}=\vb{0}$ and decay on a scale, called $1/a_{\mathrm{B}}$ for the moment, determined by the inverse of the exciton Bohr radii. Since typical exciton Bohr radii in TMDCs are on the order of $a_{\mathrm{B}}\sim 1$~nm (see Tab.~\ref{tab:wannier}), while the different valleys are separated by wavevectors $|\vb{k}_{\lambda}-\vb{k}_{\lambda'}|>0.5\Angstrom^{-1}$ for $\lambda\neq \lambda'$~\cite{deilmann2019finite}, we find for $\mu^{\mathrm{e/h}}_{\kappa\kappa'}=\mathcal{O}(1)$ that only phonon momenta with  $|\vb{q}|\sim a_{\mathrm{B}}^{-1}\ll |\vb{k}_{\lambda}-\vb{k}_{\lambda'\neq\lambda}|$ are relevant for the scattering processes described by the three classes of scattering matrix elements given above [Eqs.~\eqref{eq:vertex_intra_main}, \eqref{eq:vertex_electron_main}, and \eqref{eq:vertex_hole_main}]. This implies that, instead of classifying the phonon modes by a branch index $j$ and a wavevector $\vb{q}$ within the first Brillouin zone, we can classify them by a branch index $j$, a momentum center $\vb{k}_{\kappa}-\vb{k}_{\lambda}$ at a high symmetry point, denoted by $\overline{\kappa\lambda}$, and an arbitrary wavevector $\vb{q}$, whose sufficient smallness is guaranteed by the presence of the exciton form factors. In this notation the $\Gamma$-point of the phonons is given by $\Gamma=\overline{\lambda\lambda}$ for arbitrary valleys $\lambda$. Intra-valley scattering is then exclusively mediated by phonons located around the $\Gamma$-point~\cite{jin2014intrinsic}. 

In the following we will work in the deformation potential approximation and collect the acoustic and optical phonon modes around each high symmetry point into a single effective branch~\cite{jin2014intrinsic,selig2016excitonic,selig2018dark,lengers2020theory}. At the $\Gamma$ point we then have the acoustic phonon dispersion
\begin{equation}\label{eq:disp_LA_Gamma}
	\Omega_{\mathrm{ac},\Gamma}(\vb{q})=c_{\mathrm{s}}|\vb{q}|
\end{equation}
with the speed of sound $c_{\mathrm{s}}$, and the optical phonon dispersion
\begin{equation}
	\Omega_{\mathrm{opt},\Gamma}(\vb{q})=\Omega_{\mathrm{opt},\Gamma},
\end{equation}
treated within the Debye and Einstein model, respectively. At high symmetry points different from the $\Gamma$ point we furthermore apply the Einstein model
\begin{equation}
	\Omega_{j,\overline{\kappa\lambda}\neq\Gamma}(\vb{q})=\Omega_{j,\overline{\kappa\lambda}\neq\Gamma}
\end{equation}
for both acoustic and optical modes. Within the deformation potential approximation the coupling constants for electrons and holes are given by
\begin{subequations}\label{eq:def_pot}
	\begin{align}
		g^{(\mathrm{e})}_{j\lambda\kappa}(\vb{q}+\vb{k}_{\lambda}-\vb{k}_{\kappa})&=g^{(\mathrm{e})}_{j\overline{\lambda\kappa}\lambda\kappa}(\vb{q})=\frac{\Delta V^{(\mathrm{e})}_{j\overline{\lambda\kappa}\lambda\kappa}(\vb{q})}{\sqrt{2A\rho_{\mathrm{m}}\Omega_{j\overline{\lambda\kappa}}(\vb{q})}}\,,\\
		g^{(\mathrm{h})}_{j\lambda\kappa}(\vb{q}-\vb{k}_{\lambda}+\vb{k}_{\kappa})&=g^{(\mathrm{h})}_{j\overline{\kappa\lambda}\lambda\kappa}(\vb{q})=\frac{\Delta V^{(\mathrm{h})}_{j\overline{\kappa\lambda}\lambda\kappa}(\vb{q})}{\sqrt{2A\rho_{\mathrm{m}}\Omega_{j\overline{\kappa\lambda}}(\vb{q})}}\,.
	\end{align}
\end{subequations}
Here, $\rho_{\mathrm{m}}$ denotes the two-dimensional mass density of the TMDC monolayer and $A$ is a normalization area. For intra-valley acoustic phonon scattering the deformation potential is given by
\begin{equation}\label{eq:def_pot_ac_low}
	\Delta V^{(\mathrm{e/h})}_{\mathrm{ac},\Gamma,\lambda\lambda}(\vb{q})=D^{(\mathrm{e/h})}_{\mathrm{ac},\Gamma,\lambda\lambda}|\vb{q}|
\end{equation}
while for all other cases, i.e., $\overline{\lambda\kappa}\neq\Gamma$ or $\overline{\lambda\kappa}=\Gamma$ but $j=\mathrm{opt}$,  we have
\begin{equation}\label{eq:def_pot_high}
	\Delta V^{(\mathrm{e})}_{j,\overline{\lambda\kappa},\lambda\kappa}(\vb{q})=D^{(\mathrm{e})}_{j,\overline{\lambda\kappa},\lambda\kappa}\,,\qquad \Delta V^{(\mathrm{h})}_{j,\overline{\kappa\lambda},\lambda\kappa}(\vb{q})=D^{(\mathrm{h})}_{j,\overline{\kappa\lambda},\lambda\kappa}
\end{equation}
with $D^{(\mathrm{e/h})}_{j,\overline{\lambda\kappa}/\overline{\kappa\lambda},\lambda\kappa}$ being the corresponding deformation potential constants~\cite{jin2014intrinsic}. The phonon and deformation potential parameters considered in this paper are summarized in Tab.~\ref{tab:phon}.

\section{Self energy calculations}\label{app:self_energy}
\subsection{Including temperature-dependent linewidths in acoustic phonon intra-valley scattering}\label{app:master_eq}
In the diagrammatic Green's function approach at vanishing temperature finite linewidths of the excitonic energy levels emerged via partial resummation of diagrams as depicted in Fig.~\ref{fig:propagatorexpansion}. Here we will provide an alternative approach to obtaining such linewidths by employing a master equation approach and compare the results. This will allow us to (partially) incorporate effects due to finite temperatures.

To this aim we assume that there is a clear distinction between high frequency, e.g., Raman modes in which one might be interested in, and low frequency phonon modes which are thermally occupied but not directly measured in the Raman signal of interest. The phonon Hamiltonian can then be written as
\begin{equation}
	H_{\mathrm{p}}=\sum_{\sigma=\pm}H_{\mathrm{p}}^{\sigma}=\sum_{\sigma=\pm}\sum_{j_{\sigma}\vb{q}}\Omega_{j_{\sigma}}(\vb{q})b_{j_{\sigma}}^{\dagger}(\vb{q})b_{j_{\sigma}}^{}(\vb{q})\,.
\end{equation}
Here, $\sigma=+$ describes phonon branches (or parts thereof) in the considered model which are not thermally occupied, e.g., optical phonons or acoustic phonons around high symmetry points different from the $\Gamma$-point at $\vb{q}=\vb{0}$, i.e., $\Omega_{j_{+}\vb{q}}\gg k_{\mathrm{B}}T$. The potentially thermally occupied low frequency modes, e.g., acoustic phonons around the $\Gamma$-point, are collectively denoted by $\sigma=-$. Performing the same separation also in $H_{\mathrm{ep}}=\sum_\sigma H_{\mathrm{ep}}^{\sigma}$, we will treat the impact of the low frequency phonons in an effective fashion using open quantum system master equation techniques~\cite{breuer2002theory}. To this aim we define the system (S), bath (B) and system-bath interaction (SB) Hamiltonians as
\begin{subequations}
	\begin{align}
		H_{\mathrm{S}}&=H_{\mathrm{e}}+H_{\mathrm{ep}}^++H_{\mathrm{p}}^+\,,\\
		H_{\mathrm{B}}&=H_{\mathrm{p}}^-\,,\\
		H_{\mathrm{SB}}&=H_{\mathrm{ep}}^-\,.
	\end{align}
\end{subequations}
Treating the system-bath interaction in a perturbative fashion and assuming a Born approximation of the density matrix $\rho(t)\approx\rho_{\mathrm{S}}(t)\otimes\rho_{\mathrm{B}}$ with a $\rho_{\mathrm{B}}$ being a thermal bath state [already satisfied initially by Eq.~\eqref{eq:rho}], the reduced density matrix of the system $\rho_{\mathrm{S}}$ consisting of excitons and high frequency phonons then obeys the Markovian master equation~\cite{breuer2002theory}
\begin{subequations}\label{eq:TCL}
	\begin{align}
		\dv{t}\rho_{\mathrm{S}}(t)=&-i\comm{H_{\mathrm{S}}}{\rho_{\mathrm{S}}(t)}\\
		&-\int\limits_0^{\infty}\dd\tau\, \Tr_{\mathrm{B}}\qty{\comm{H_{\mathrm{SB}}}{\comm{H_{\mathrm{SB}}^I(-\tau)}{\rho_{\mathrm{S}}(t)\rho_{\mathrm{B}}}}}\notag\,,\\
		H_{\mathrm{SB}}^I(-\tau)&=e^{-i\qty(H_{\mathrm{e}}+H_{\mathrm{p}}^-)\tau} H_{\mathrm{SB}}e^{i\qty(H_{\mathrm{e}}+H_{\mathrm{p}}^-)\tau}
	\end{align}
\end{subequations}
with $\Tr_{\mathrm{B}}$ denoting the trace over the bath states. In Eq.~\eqref{eq:TCL} we included only contributions of second order in the exciton phonon coupling, by neglecting $H_{\mathrm{ep}}^+$ from the exponentials appearing in the interaction picture operator $H_{\mathrm{SB}}^I$. Note that $H_{\mathrm{p}}^+$ commutes with $H_{\mathrm{SB}}$, since the latter describes only the direct coupling of excitons with low frequency phonons, while the former describes the free dynamics of the high frequency phonons. Therefore it also does not appear in the expression for $H_{\mathrm{SB}}^I$.

In the assumed Markov limit we can apply the quantum regression theorem to evaluate time-ordered correlation functions such as Eq.~\eqref{eq:C_0_ordered}~\cite{breuer2002theory,groll2023read}. Note that this correlation function contains only dipole operators, which according to Eq.~\eqref{eq:dipole_number} lower or raise the number of excitons. Focusing on the excitonic content, the correlation function in Eq.~\eqref{eq:C_0_ordered} can thus be evaluated as follows: 

Initially at $t=-\infty$ the density matrix is given by Eq.~\eqref{eq:rho}, such that the excitonic part is of the form $\ket{0_x}\bra{0_x}$, i.e., the system contains no excitons. Since the free time evolution due to the Hamiltonian in Eq.~\eqref{eq:H_generic} conserves the number of excitons, the excitonic part of the density matrix remains in this form until at time $t=-t_2$ the dipole operator $\vb{d}^{(-)}$ acts on the left, changing the excitonic part of the effective density matrix to $\ket{1_x}\bra{0_x}$ with $\ket{1_x}$ being any state from the subspace containing only single excitons, i.e., with $N_x\ket{1_x}=1\ket{1_x}$. The excitonic structure again remains like this until at time $t=0$ the operator $\vb{d}^{(+)}$ acts on the left, bringing the excitonic part of the effective density matrix back to $\ket{0_x}\bra{0_x}$. At time $t=\tau-t_1$ the excitonic content changes to $\ket{0_x}\bra{1_x}$ via action of $\vb{d}^{(+)}$ on the right, which is changed back to $\ket{0_x}\bra{0_x}$ at $t=\tau$ via action of $\vb{d}^{(-)}$ on the right using the cyclic property of the trace. In essence we can summarize as follows: For the determination of the Raman signal, which is encapsulated by the correlation function in Eq.~\eqref{eq:C_0_ordered}, only excitonic density matrix elements of the form $\ket{0_x}\bra{0_x}$, $\ket{0_x}\bra{1_x}$ and $\ket{1_x}\bra{0_x}$ are relevant. 

This motivates us to evaluate Eqs.~\eqref{eq:TCL} on density matrix elements of the form $\rho_{\mathrm{S}}(t)=\ket{n_x}\bra{0_x}\rho_{\mathrm{p}}^+(t)$ with $\rho_{\mathrm{p}}^+$ encapsulating the state of the high frequency phonon modes and $\ket{n_x}$ being any $n$-exciton state with $n=0,1$. Since the total Hamiltonian $H$ conserves the number of excitons, the master equation projects density matrix elements of this form onto themselves and acts as a von-Neumann equation with an effective non-Hermitian Hamiltonian $\tilde{H}_{\mathrm{S}}$
\begin{subequations}\label{eq:TCL_eff_1}
	\begin{align}
		\dv{t}\rho_{\mathrm{S}}(t)=&-i\comm{\tilde{H}_{\mathrm{S}}}{\rho_{\mathrm{S}}(t)}\label{eq:TCL_eff_form}\,,\\
		\tilde{H}_{\mathrm{S}}=&H_{\mathrm{S}}-i\int\limits_0^{\infty}\dd\tau\,\Tr_{\mathrm{B}}\qty[H_{\mathrm{SB}}H_{\mathrm{SB}}^I(-\tau)\rho_{\mathrm{B}}]\,,\\
		\rho_{\mathrm{S}}(t)&=\ket{n_x}\bra{0_x}\rho_{\mathrm{p}}^+(t) \text{ with  }n=0,1\,.
	\end{align}
\end{subequations}
Note that density matrix elements of the form $\rho_{\mathrm{S}}(t)=\ket{0_x}\bra{n_x}\rho_{\mathrm{p}}^+(t)$ with $n=0,1$ obey the same equation but with $\tilde{H}_{\mathrm{S}}\rightarrow \tilde{H}_{\mathrm{S}}^{\dagger}$. We can thus summarize the time evolution of the relevant density matrix elements as follows
\begin{subequations}\label{eq:evol_non_unitary}
	\begin{align}
		\rho_{\mathrm{S}}(t)&=e^{-i\tilde{H}_{\mathrm{S}}(t-t_0)}\rho_{\mathrm{S}}(t_0)e^{i\tilde{H}_{\mathrm{S}}^{\dagger}(t-t_0)}\,,\\
		\rho_{\mathrm{S}}(t)&=\ket{n_x}\bra{0_x}\rho_{\mathrm{p}}^+(t)\notag\\
		&\qquad\text{ or }\ket{0_x}\bra{n_x}\rho_{\mathrm{p}}^+(t) \text{ with  }n=0,1\,,
	\end{align}
\end{subequations}
which is the non-unitary equivalent to Eq.~\eqref{eq:evol_unitary}.

Considering that the low-frequency modes are given by acoustic phonons around the $\Gamma$-point with dispersion relation $\Omega_{\mathrm{ac},\Gamma}(\vb{q})=c_{\mathrm{s}}|\vb{q}|$ given in Eq.~\eqref{eq:disp_LA_Gamma}, the system-bath interaction can be written as
\begin{align}\label{eq:H_SB}
	H_{\mathrm{SB}}\!&=\!\!\sum_{\vb{k}\vb{q}\lambda_{\mathrm{e}}\lambda_{\mathrm{h}}}	g_{\mathrm{ac},\Gamma,\lambda_{\mathrm{e}}\lambda_{\mathrm{h}}}^{(x)}(\vb{q})\qty[b_{\mathrm{ac},\Gamma}^{}(\vb{q})+b_{\mathrm{ac},\Gamma}^{\dagger}(-\vb{q})]\\
	&\qquad\times\notag X^{\dagger}_{\lambda_{\mathrm{e}}\lambda_{\mathrm{h}}}(\vb{k}+\vb{q})X^{}_{\lambda_{\mathrm{e}}\lambda_{\mathrm{h}}}(\vb{k})
\end{align}
with the intra-valley exciton-phonon coupling constants from Eq.~\eqref{eq:g_x}. Here we employed the relabeling of phonon modes around high symmetry points in accordance with App.~\ref{sec:def_pot}. With these considerations we obtain
\begin{align}\label{eq:H_eff_1}
	\tilde{H}_{\mathrm{S}}-H_{\mathrm{S}}&=-i\sum_{\vb{q}\vb{k}\vb{k}'\lambda_{\mathrm{e}}\lambda_{\mathrm{h}}\lambda_{\mathrm{e}}'\lambda_{\mathrm{h}}'}g_{\mathrm{ac},\Gamma,\lambda_{\mathrm{e}}\lambda_{\mathrm{h}}}^{(x)*}(\vb{q})X^{\dagger}_{\lambda_{\mathrm{e}}\lambda_{\mathrm{h}}}(\vb{k})X^{}_{\lambda_{\mathrm{e}}\lambda_{\mathrm{h}}}(\vb{k}+\vb{q})\notag\\
	&\qquad\times g_{\mathrm{ac},\Gamma,\lambda_{\mathrm{e}}'\lambda_{\mathrm{h}}'}^{(x)}(\vb{q})X^{\dagger}_{\lambda_{\mathrm{e}}'\lambda_{\mathrm{h}}'}(\vb{k}'+\vb{q})X^{}_{\lambda_{\mathrm{e}}'\lambda_{\mathrm{h}}'}(\vb{k}')\notag\\
	&\qquad\times\sum_{\eta=\pm(1)}\int\limits_0^{\infty}\dd\tau\, N^{(-\eta)}_{\mathrm{ac},\Gamma}(\eta\vb{q})\notag\\
	&\qquad\times e^{i\qty[\eta\Omega_{\mathrm{ac},\Gamma}(\eta\vb{q})-E_{\lambda_{\mathrm{e}}'\lambda_{\mathrm{h}}'}(\vb{k}'+\vb{q})+E_{\lambda_{\mathrm{e}}'\lambda_{\mathrm{h}}'}(\vb{k}')]\tau}
\end{align}
with 
\begin{equation}
	N^{(\eta)}_{\mathrm{ac},\Gamma}(\vb{q})=\expval{b_{\mathrm{ac},\Gamma}^{\dagger}(\vb{q})b_{\mathrm{ac},\Gamma}(\vb{q})}+\frac{1}{2}(1+\eta)
\end{equation}
being determined by the thermal acoustic phonon occupation
\begin{equation}
	\expval{b_{\mathrm{ac},\Gamma}^{\dagger}(\vb{q})b_{\mathrm{ac},\Gamma}(\vb{q})}=\qty{\exp[\beta \Omega_{\mathrm{ac},\Gamma}(\vb{q})]-1}^{-1}
\end{equation}
and $\eta=\pm(1)$ describing the impact of phonon emission/absorption processes, respectively. We can normal-order the exciton operators and make use of the fact that in Eq.~\eqref{eq:TCL_eff_1} they act on states containing at most one exciton, such that we can effectively write
\begin{equation}\label{eq:H_eff_final}
	\tilde{H}_{\mathrm{S}}-H_{\mathrm{S}}\simeq\sum_{\vb{k}\lambda_{\mathrm{e}}\lambda_{\mathrm{h}}}\Sigma_{\lambda_{\mathrm{e}}\lambda_{\mathrm{h}}}^{\mathrm{ac}\Gamma}(\vb{k})X^{\dagger}_{\lambda_{\mathrm{e}}\lambda_{\mathrm{h}}}(\vb{k})X^{}_{\lambda_{\mathrm{e}}\lambda_{\mathrm{h}}}(\vb{k})
\end{equation}
with the self-energy due to acoustic phonon intra-valley scattering
\begin{align}\label{eq:sigma_ac_1}
	&\Sigma_{\lambda_{\mathrm{e}}\lambda_{\mathrm{h}}}^{\mathrm{ac}\Gamma}(\vb{k})=-i\sum_{\vb{q},\eta=\pm(1)}N_{\mathrm{ac},\Gamma}^{(-\eta)}(\eta\vb{q})\qty|g_{\mathrm{ac},\Gamma,\lambda_{\mathrm{e}}\lambda_{\mathrm{h}}}^{(x)}(\vb{q})|^2\\
	&\qquad\qquad\times \int\limits_0^{\infty}\dd\tau\,e^{i\qty[\eta\Omega_{\mathrm{ac},\Gamma}(\eta\vb{q})-E_{\lambda_{\mathrm{e}}\lambda_{\mathrm{h}}}(\vb{k}+\vb{q})+E_{\lambda_{\mathrm{e}}\lambda_{\mathrm{h}}}(\vb{k})+i0^+]\tau}\notag\\
	&\qquad=\sum_{\vb{q},\eta=\pm(1)}\frac{N_{\mathrm{ac},\Gamma}^{(-\eta)}(\eta\vb{q})\qty|g_{\mathrm{ac},\Gamma,\lambda_{\mathrm{e}}\lambda_{\mathrm{h}}}^{(x)}(\vb{q})|^2}{\eta\Omega_{\mathrm{ac},\Gamma}(\eta\vb{q})-E_{\lambda_{\mathrm{e}}\lambda_{\mathrm{h}}}(\vb{k}+\vb{q})+E_{\lambda_{\mathrm{e}}\lambda_{\mathrm{h}}}(\vb{k})+i0^+}\,.\notag
\end{align}
This result corresponds to the contribution to the self-energy $\Sigma_{\lambda_{\mathrm{e}}\lambda_{\mathrm{h}}}(E,\vb{k})$ in Eq.~\eqref{eq:self_energy_new_new} due to acoustic phonon intra-valley scattering, however in the usual Born approximation (not self-consistent) with $\mathcal{G}\rightarrow\mathcal{G}^{(0)}$, as well as in quasi-particle approximation $E=E_{\lambda_{\mathrm{e}}\lambda_{\mathrm{h}}}(\vb{k})$~\cite{fetter2003quantum, haug2008quantum}, while being generalized to non-vanishing temperatures. 

Considering that the Raman signal is contained in the time-ordered correlation function in Eq.~\eqref{eq:C_0_ordered} we can apply the quantum regression theorem in the Markov limit, allowing us to effectively replace Eq.~\eqref{eq:evol_unitary} with Eqs.~\eqref{eq:evol_non_unitary} in the calculation of Eq.~\eqref{eq:C_0_ordered}. This in turn implies that we can simply replace the full Hamiltonian $H$ by the effective Hamiltonian $\tilde{H}_{\mathrm{S}}$ in the calculation of the scattering amplitude in Eq.~\eqref{eq:scattering_amplitude}. This replacement will lead to errors due to the presence of the non-time-ordered contribution in Eq.~\eqref{eq:P_corr_2}, which also contains exciton states of the form $\ket{1_x}\bra{1_x}$. However since this will not contribute to the true Raman signal, i.e., the part of the detected spectrum that moves with the laser frequency $\omega_l$ [as discussed in the context of Eq.~\eqref{eq:P_corr_2}], this error will not affect the coherent Raman part of the scattering spectrum.

This replacement effectively removes the explicit appearance of the low frequency acoustic phonons from the diagrams in Fig.~\ref{fig:stokes}, i.e., from the summation over all modes in Eq.~\eqref{eq:A_R_final}, but includes their temperature-dependent impact on the excitons via
\begin{equation}\label{eq:self_energy_low_impact}
	E_{\lambda_{\mathrm{e}}\lambda_{\mathrm{h}}}(\vb{k})\rightarrow	\tilde{E}_{\lambda_{\mathrm{e}}\lambda_{\mathrm{h}}}(\vb{k})= E_{\lambda_{\mathrm{e}}\lambda_{\mathrm{h}}}(\vb{k})+\Sigma_{\lambda_{\mathrm{e}}\lambda_{\mathrm{h}}}^{\mathrm{ac}\Gamma}(\vb{k})\,.
\end{equation}
\subsection{Low-frequency phonon contributions to the exciton self-energy}\label{app:self_energy_low}
Using the deformation potential model and acoustic phonon dispersion relation given in App.~\ref{sec:def_pot}, we can rewrite Eq.~\eqref{eq:sigma_ac_1} as
\begin{align}
	\Sigma_{\lambda_{\mathrm{e}}\lambda_{\mathrm{h}}}^{\mathrm{ac}\Gamma}(\vb{k})&=-i\sum_{\vb{q},\eta=\pm(1)}\qty[\frac{1}{e^{\beta c_{\mathrm{s}}q}-1}+\frac{1}{2}(1-\eta)]\qty|g_{\mathrm{ac},\Gamma,\lambda_{\mathrm{e}}\lambda_{\mathrm{h}}}^{(x)}(q)|^2\notag\\
	&\times \int\limits_0^{\infty}\dd\tau\,e^{i\qty[\eta c_{\mathrm{s}}q -\frac{q^2}{2M_{\lambda_{\mathrm{e}}\lambda_{\mathrm{h}}}}-\frac{qk}{M_{\lambda_{\mathrm{e}}\lambda_{\mathrm{h}}}}\cos(\varphi)+i0^+]\tau}
\end{align}
with $q=|\vb{q}|$, $k=|\vb{k}|$, and $\vb{k}\cdot\vb{q}=kq\cos(\varphi)$. Here, the exciton-phonon coupling [see Eqs.~\eqref{eq:g_x}, \eqref{eq:def_pot} and \eqref{eq:def_pot_ac_low}]
\begin{align}
	g_{\mathrm{ac},\Gamma,\lambda_{\mathrm{e}}\lambda_{\mathrm{h}}}^{(x)}(\vb{q})&=g_{\mathrm{ac},\Gamma,\lambda_{\mathrm{e}}\lambda_{\mathrm{h}}}^{(x)}(q)\notag\\
	&=\sqrt{\frac{q}{2A\rho_{\mathrm{m}}c_{\mathrm{s}}}}\Big[D^{(\mathrm{e})}_{\mathrm{ac},\Gamma,\lambda_{\mathrm{e}}\lambda_{\mathrm{e}}}F_{\lambda_{\mathrm{e}}\lambda_{\mathrm{h}};\lambda_{\mathrm{e}}\lambda_{\mathrm{h}}}\qty(\mu^{\mathrm{h}}_{\lambda_{\mathrm{e}}\lambda_{\mathrm{h}}}q)\notag\\
	&\qquad-D^{(\mathrm{h})}_{\mathrm{ac},\Gamma,\lambda_{\mathrm{h}}\lambda_{\mathrm{h}}}F_{\lambda_{\mathrm{e}}\lambda_{\mathrm{h}};\lambda_{\mathrm{e}}\lambda_{\mathrm{h}}}\qty(\mu^{\mathrm{e}}_{\lambda_{\mathrm{e}}\lambda_{\mathrm{h}}}q)\Big]\,.\label{eq:g_x_low}
\end{align}
only depends on $q$, since the form factors in Eqs.~\eqref{eq:form_factor_main} are isotropic for 1s excitons [see Eq.~\eqref{eq:form_factor_isotropy}]. Rewriting the sum over $\vb{q}$ as an integral over the absolute value $q$ and the polar angle $\varphi$ yields
\begin{align}\label{eq:sigma_ac_2}
	&\Sigma_{\lambda_{\mathrm{e}}\lambda_{\mathrm{h}}}^{\mathrm{ac}\Gamma}(\vb{k})=-\frac{iA}{2\pi}\sum_{\eta=\pm(1)}\int\limits_0^{\infty}\dd q\, q\qty(\frac{1}{e^{\beta c_{\mathrm{s}}q}-1}+\frac{1-\eta}{2})\\
	&\times \qty|g_{\mathrm{ac},\Gamma,\lambda_{\mathrm{e}}\lambda_{\mathrm{h}}}^{(x)}(q)|^2\int\limits_0^{\infty}\dd\tau\,e^{i\qty(\eta c_{\mathrm{s}}q -\frac{q^2}{2M_{\lambda_{\mathrm{e}}\lambda_{\mathrm{h}}}}+i0^+)\tau}J_0\qty(\frac{qk}{M_{\lambda_{\mathrm{e}}\lambda_{\mathrm{h}}}}\tau)\,,\notag
\end{align}
where the Bessel function $J_0$ stems from the polar angle integral~\cite{abramowitz1964handbook}. The $\tau$-integral now corresponds to a Laplace transform of the Bessel function, yielding
\begin{align}
	\Sigma_{\lambda_{\mathrm{e}}\lambda_{\mathrm{h}}}^{\mathrm{ac}\Gamma}(\vb{k})&=-\frac{iA}{2\pi}\sum_{\eta=\pm(1)}\int\limits_0^{\infty}\dd q\, q\qty(\frac{1}{e^{\beta c_{\mathrm{s}}q}-1}+\frac{1-\eta}{2})\notag\\
	&\times \frac{\qty|g_{\mathrm{ac},\Gamma,\lambda_{\mathrm{e}}\lambda_{\mathrm{h}}}^{(x)}(q)|^2}{\sqrt{\qty(\frac{qk}{M_{\lambda_{\mathrm{e}}\lambda_{\mathrm{h}}}})^2-\qty(\eta c_{\mathrm{s}}q -\frac{q^2}{2M_{\lambda_{\mathrm{e}}\lambda_{\mathrm{h}}}}+i0^+)^2}}\,.\label{eq:sigma_ac_final}
\end{align}
The integral over $q$ can be evaluated numerically by introducing a small finite broadening $\epsilon$, replacing $i0^+$ by $i\epsilon$ in the previous equations and then letting $\epsilon\rightarrow 0^+$ until the result converges.

\subsection{High-frequency phonon contributions to the exciton self-energy}\label{app:self_energy_high}
With the classification of intra/inter-valley scattering from App.~\ref{sec:def_pot} at hand we can separate the contributions to the exciton self-energy in Eq.~\eqref{eq:self_energy_new_new} due to high-frequency phonon scattering via
\begin{subequations}\label{eq:self_energy_valleys}
	\begin{align}
		\Sigma_{\lambda_{\mathrm{e}}\lambda_{\mathrm{h}}}^+(E,\vb{k})&=\sum_{j_+\lambda_{\mathrm{e}}'\lambda_{\mathrm{h}}'}\Sigma_{\lambda_{\mathrm{e}}\lambda_{\mathrm{h}}}^{j_+\lambda_{\mathrm{e}}'\lambda_{\mathrm{h}}'}(E,\vb{k})\,,\\
		\Sigma_{\lambda_{\mathrm{e}}\lambda_{\mathrm{h}}}^{j_+\lambda_{\mathrm{e}}'\lambda_{\mathrm{h}}'}(E,\vb{k})&=\sum_{\vb{k}'\vb{q}}\qty|\mathcal{V}_{\vb{k}\lambda_{\mathrm{e}}\lambda_{\mathrm{h}}}^{\vb{k}'\lambda_{\mathrm{e}}'\lambda_{\mathrm{h}}'}(j_+,\vb{q})|^2\notag\\
		&\qquad\times\mathcal{G}_{\lambda_{\mathrm{e}}'\lambda_{\mathrm{h}}'}\qty(E-\Omega_{j_+}(\vb{q}),\vb{k}')\,,
	\end{align}
\end{subequations}
where the label $j_+$ indicates that we exclude the low-frequency acoustic phonons at the $\Gamma$ point, whose impact has already been dealt with [see Eqs.~\eqref{eq:sigma_ac_1} and \eqref{eq:self_energy_low_impact}]. Note that according to the discussion in the context of Eq.~\eqref{eq:self_energy_low_impact}, this impact has to be included here by replacing $E_{\lambda_{\mathrm{e}}'\lambda_{\mathrm{h}}'}(\vb{k}')$ with $\tilde{E}_{\lambda_{\mathrm{e}}'\lambda_{\mathrm{h}}'}(\vb{k}')$ appearing in the single exciton Green's function from Eq.~\eqref{eq:G_full}. We can here either have intra-valley scattering with $\lambda_{\mathrm{e}}=\lambda_{\mathrm{e}}'$ and $\lambda_{\mathrm{h}}=\lambda_{\mathrm{h}}'$ or inter-valley scattering with either $\lambda_{\mathrm{e}}\neq\lambda_{\mathrm{e}}'$ or $\lambda_{\mathrm{h}}\neq\lambda_{\mathrm{h}}'$. However there is no scattering process with $\lambda_{\mathrm{e}}\neq\lambda_{\mathrm{e}}'$ and $\lambda_{\mathrm{h}}\neq\lambda_{\mathrm{h}}'$.

We can evaluate the self-energy contributions further by employing momentum conservation, which holds at every vertex, and by shifting the argument $\vb{q}$ according to Eqs.~\eqref{eq:vertex_intra_main}, \eqref{eq:vertex_electron_main} and \eqref{eq:vertex_hole_main}. This leads to
\begin{align}\label{eq:sigma_general_2}
	\Sigma_{\lambda_{\mathrm{e}}\lambda_{\mathrm{h}}}^{j_+\lambda_{\mathrm{e}}'\lambda_{\mathrm{h}}'}(E,\vb{k})&=\sum_{\vb{q}}\qty|\mathcal{V}_{\vb{k}\lambda_{\mathrm{e}}\lambda_{\mathrm{h}}}^{(\vb{k}-\vb{q})\lambda_{\mathrm{e}}'\lambda_{\mathrm{h}}'}(j_+,\vb{q}+\vb{k}_{\lambda_{\mathrm{e}}}-\vb{k}_{\lambda_{\mathrm{e}}'}-\vb{k}_{\lambda_{\mathrm{h}}}+\vb{k}_{\lambda_{\mathrm{h}}'})|^2\notag\\
	&\qquad\times\mathcal{G}_{\lambda_{\mathrm{e}}'\lambda_{\mathrm{h}}'}\qty(E-\Omega_{\lambda_{\mathrm{e}}\lambda_{\mathrm{h}}}^{j_+\lambda_{\mathrm{e}}'\lambda_{\mathrm{h}}'},\vb{k}-\vb{q})\,.
\end{align}
Here, we made use of the Einstein model for the high frequency modes as discussed in App.~\ref{sec:def_pot} and employed the shorthand notation
\begin{equation}
	\Omega_{\lambda_{\mathrm{e}}\lambda_{\mathrm{h}}}^{j_+\lambda_{\mathrm{e}}'\lambda_{\mathrm{h}}'}=\Omega_{j_+}(\vb{q}+\vb{k}_{\lambda_{\mathrm{e}}}-\vb{k}_{\lambda_{\mathrm{e}}'}-\vb{k}_{\lambda_{\mathrm{h}}}+\vb{k}_{\lambda_{\mathrm{h}}'})\,,
\end{equation}
which implies, e.g., for electron inter-valley scattering
\begin{equation}
	\Omega_{\lambda_{\mathrm{e}}\lambda_{\mathrm{h}}}^{j_+\lambda_{\mathrm{e}}'\lambda_{\mathrm{h}}}=\Omega_{j_+\overline{\lambda_{\mathrm{e}}\lambda_{\mathrm{e}}'}}\,.
\end{equation}
We can make further progress by substituting $\vb{k}-\vb{q}\rightarrow\vb{q}$ in Eq.~\eqref{eq:sigma_general_2}, which makes the single exciton Green's functions therein independent of the polar angle $\mathcal{G}_{\lambda_{\mathrm{e}}\lambda_{\mathrm{h}}}\qty(E,\vb{q})=\mathcal{G}_{\lambda_{\mathrm{e}}\lambda_{\mathrm{h}}}\qty(E,q)$, since the exciton energies are isotropic $\tilde{E}_{\lambda_{\mathrm{e}}\lambda_{\mathrm{h}}}(\vb{q})=\tilde{E}_{\lambda_{\mathrm{e}}\lambda_{\mathrm{h}}}(q)$, even when including the impact of low frequency acoustic phonon scattering [see Eqs.~\eqref{eq:disp_ex}, \eqref{eq:self_energy_low_impact} and \eqref{eq:sigma_ac_final}]. We thus obtain the isotropic exciton self-energy contributions
\begin{equation}
	\Sigma_{\lambda_{\mathrm{e}}\lambda_{\mathrm{h}}}^{j^+\lambda_{\mathrm{e}}'\lambda_{\mathrm{h}}'}(E,k)=\int\limits_0^{\infty}\dd q\, q \mathcal{K}_{\lambda_{\mathrm{e}}\lambda_{\mathrm{h}}}^{j^+\lambda_{\mathrm{e}}'\lambda_{\mathrm{h}}'}(k,q)\mathcal{G}_{\lambda_{\mathrm{e}}'\lambda_{\mathrm{h}}'}\qty(E-\Omega_{\lambda_{\mathrm{e}}\lambda_{\mathrm{h}}}^{j^+\lambda_{\mathrm{e}}'\lambda_{\mathrm{h}}'},q)\label{eq:sigma_final}
\end{equation}
with the integration kernel
\begin{align}
	&\mathcal{K}_{\lambda_{\mathrm{e}}\lambda_{\mathrm{h}}}^{j^+\lambda_{\mathrm{e}}'\lambda_{\mathrm{h}}'}(k,q)=\\
	&\quad\frac{A}{(2\pi)^2}\int\limits_0^{2\pi}\dd\varphi\,\qty|\mathcal{V}_{\vb{k}\lambda_{\mathrm{e}}\lambda_{\mathrm{h}}}^{\vb{q}\lambda_{\mathrm{e}}'\lambda_{\mathrm{h}}'}(j_+,\vb{k}-\vb{q}+\vb{k}_{\lambda_{\mathrm{e}}}-\vb{k}_{\lambda_{\mathrm{e}}'}-\vb{k}_{\lambda_{\mathrm{h}}}+\vb{k}_{\lambda_{\mathrm{h}}'})|^2\notag
\end{align}
stemming from the integration over the polar angle between $\vb{k}$ and $\vb{q}$, i.e., $\vb{k}\cdot\vb{q}=kq\cos(\varphi)$. Note that the momentum dependence of the vertices $\mathcal{V}$ in the previous equation stems solely from the exciton form factors [see Eqs.~\eqref{eq:vertex_intra_main}, \eqref{eq:vertex_electron_main} and \eqref{eq:vertex_hole_main}], while the electron/hole-phonon coupling is momentum independent for the high frequency phonon modes in deformation potential approximation as discussed in App.~\ref{sec:def_pot}. This implies 
\begin{align}\label{eq:kernel_e}
	\eval{\mathcal{K}_{\lambda_{\mathrm{e}}\lambda_{\mathrm{h}}}^{j^+\lambda_{\mathrm{e}}'\lambda_{\mathrm{h}}}(k,q)}_{\lambda_{\mathrm{e}}\neq\lambda_{\mathrm{e}}'}&=\frac{\qty|D^{(\mathrm{e})}_{j^+,\overline{\lambda_{\mathrm{e}}\lambda_{\mathrm{e}}'},\lambda_{\mathrm{e}}\lambda_{\mathrm{e}}'}|^2}{8\pi^2 \rho_{\mathrm{m}}\Omega_{j^+\overline{\lambda_{\mathrm{e}}\lambda_{\mathrm{e}}'}}}\\
	&\qquad\times\int\limits_0^{2\pi}\dd\varphi\,\qty|F_{\lambda_{\mathrm{e}}\lambda_{\mathrm{h}};\lambda_{\mathrm{e}}'\lambda_{\mathrm{h}}}(-\mu^{\mathrm{h}}_{\lambda_{\mathrm{e}}\lambda_{\mathrm{h}}}\vb{k}+\mu^{\mathrm{h}}_{\lambda_{\mathrm{e}}'\lambda_{\mathrm{h}}}\vb{q})|^2\notag\,,
\end{align}
for electron inter-valley scattering,
\begin{align}\label{eq:kernel_h}
	\eval{\mathcal{K}_{\lambda_{\mathrm{e}}\lambda_{\mathrm{h}}}^{j^+\lambda_{\mathrm{e}}\lambda_{\mathrm{h}}'}(k,q)}_{\lambda_{\mathrm{h}}\neq\lambda_{\mathrm{h}}'}&=\frac{\qty|D^{(\mathrm{h})}_{j^+,\overline{\lambda_{\mathrm{h}}'\lambda_{\mathrm{h}}},\lambda_{\mathrm{h}}\lambda_{\mathrm{h}}'}|^2}{8\pi^2 \rho_{\mathrm{m}}\Omega_{j^+\overline{\lambda_{\mathrm{h}}'\lambda_{\mathrm{h}}}}}\\
	&\qquad\times\int\limits_0^{2\pi}\dd\varphi\,\qty|F_{\lambda_{\mathrm{e}}\lambda_{\mathrm{h}};\lambda_{\mathrm{e}}\lambda_{\mathrm{h}}'}(\mu^{\mathrm{e}}_{\lambda_{\mathrm{e}}\lambda_{\mathrm{h}}}\vb{k}-\mu^{\mathrm{e}}_{\lambda_{\mathrm{e}}\lambda_{\mathrm{h}}'}\vb{q})|^2\notag\,,
\end{align}
for hole inter-valley scattering, and
\begin{align}
	&\mathcal{K}_{\lambda_{\mathrm{e}}\lambda_{\mathrm{h}}}^{\mathrm{opt}\lambda_{\mathrm{e}}\lambda_{\mathrm{h}}}(k,q)=\eval{\mathcal{K}_{\lambda_{\mathrm{e}}\lambda_{\mathrm{h}}}^{\mathrm{opt}\lambda_{\mathrm{e}}'\lambda_{\mathrm{h}}}(k,q)}_{\lambda_{\mathrm{e}}\neq\lambda_{\mathrm{e}}'}^{\lambda_{\mathrm{e}}'\rightarrow\lambda_{\mathrm{e}}}+\eval{\mathcal{K}_{\lambda_{\mathrm{e}}\lambda_{\mathrm{h}}}^{\mathrm{opt}\lambda_{\mathrm{e}}\lambda_{\mathrm{h}}'}(k,q)}_{\lambda_{\mathrm{h}}\neq\lambda_{\mathrm{h}}'}^{\lambda_{\mathrm{h}}'\rightarrow\lambda_{\mathrm{h}}}\notag\\
	&-\text{Re}\Bigg\lbrace\frac{D^{(\mathrm{e})}_{\mathrm{opt},\Gamma,\lambda_{\mathrm{e}}\lambda_{\mathrm{e}}}D^{(\mathrm{h})*}_{\mathrm{opt},\Gamma,\lambda_{\mathrm{h}}\lambda_{\mathrm{h}}}}{4\pi^2\rho_{\mathrm{m}}\Omega_{\mathrm{opt},\Gamma}}\int\limits_{0}^{2\pi}\dd\varphi\, F_{\lambda_{\mathrm{e}}\lambda_{\mathrm{h}};\lambda_{\mathrm{e}}\lambda_{\mathrm{h}}}[\mu^{\mathrm{h}}_{\lambda_{\mathrm{e}}\lambda_{\mathrm{h}}}(\vb{q}-\vb{k})]\Bigg.\notag\\
	&\qquad\qquad\qquad\times \Bigg.F^*_{\lambda_{\mathrm{e}}\lambda_{\mathrm{h}};\lambda_{\mathrm{e}}\lambda_{\mathrm{h}}}[\mu^{\mathrm{e}}_{\lambda_{\mathrm{e}}\lambda_{\mathrm{h}}}(\vb{k}-\vb{q})]\Bigg\rbrace\label{eq:kernel_intra}
\end{align}
for optical phonon intra-valley scattering. Here, the expressions with $\lambda_{\mathrm{e/h}}'\rightarrow\lambda_{\mathrm{e/h}}$ from the first line correspond to the right-hand-side of the previous two equations~\eqref{eq:kernel_e} and \eqref{eq:kernel_h}, evaluated at $\lambda_{\mathrm{e/h}}'=\lambda_{\mathrm{e/h}}$, respectively.

Approximating the form factors by Lorentzians, as in Eq.~\eqref{eq:form_factor_lorentz}, we can make further progress in evaluating the integration kernel $\mathcal{K}_{\lambda_{\mathrm{e}}\lambda_{\mathrm{h}}}^{j^+\lambda_{\mathrm{e}}'\lambda_{\mathrm{h}}'}$, since then
\begin{align}
	&\int\limits_0^{2\pi}\dd\varphi\,\qty|F_{\lambda_{\mathrm{e}}\lambda_{\mathrm{h}};\lambda_{\mathrm{e}}'\lambda_{\mathrm{h}}'}(\alpha\vb{k}-\beta\vb{q})|^2\\
	&\quad=\frac{16(\sigma_{\lambda_{\mathrm{e}}\lambda_{\mathrm{h}}}\sigma_{\lambda_{\mathrm{e}}'\lambda_{\mathrm{h}}'})^2}{\qty(\sigma_{\lambda_{\mathrm{e}}\lambda_{\mathrm{h}}}+\sigma_{\lambda_{\mathrm{e}}'\lambda_{\mathrm{h}}'})^4}\int\limits_{0}^{2\pi}\dd\varphi\,\frac{1}{\qty[\qty(|\alpha\vb{k}-\beta\vb{q}|\frac{\sigma_{\lambda_{\mathrm{e}}\lambda_{\mathrm{h}}}\sigma_{\lambda_{\mathrm{e}}'\lambda_{\mathrm{h}}'}}{\sigma_{\lambda_{\mathrm{e}}\lambda_{\mathrm{h}}}+\sigma_{\lambda_{\mathrm{e}}'\lambda_{\mathrm{h}}'}})^2+1]^{3}}\notag
\end{align}
can be calculated analytically with the help of
\begin{equation}\label{eq:helper_integral}
	\int\limits_0^{2\pi}\dd x\, \frac{1}{[a+b\cos(x)]^3}=\pi \frac{2a^2+b^2}{(a^2-b^2)^{5/2}}\,,\qquad a>b\,,
\end{equation}
which can be determined by performing the second derivative with respect to $a$ of [$t=\tan(x/2)$]
\begin{align}
	\int\limits_{-\pi}^{\pi}\dd x\, \frac{1}{a+b\cos(x)}&=\int\limits_{-\infty}^{\infty}\frac{2\dd t\,}{1+t^2} \frac{1}{a+b\frac{1-t^2}{1+t^2}}\notag\\
	&=\int\limits_{-\infty}^{\infty}\dd t\,\frac{2}{a+b+t^2(a-b)}\notag\\
	&=\eval{2\frac{\tan^{-1}\qty(\frac{\sqrt{a-b}}{\sqrt{a+b}} t)}{\sqrt{a^2-b^2}}}_{-\infty}^{\infty}\notag\\
	&=\frac{2\pi}{\sqrt{a^2-b^2}}\,.
\end{align}
With this we find analytical expressions for all polar-angle integrals, needed to determine the integration kernel, except for the mixed integral in Eq.~\eqref{eq:kernel_intra}, i.e., for optical phonon intra-valley scattering. Due to the large optical phonon energies, as well as the fact that initial and final exciton state in the scattering process belong to the same combination of valleys, the impact of this process is negligible compared to inter-valley scattering. For this reason we approximate both $\mu_{\lambda_{\mathrm{e}}\lambda_{\mathrm{h}}}^{\mathrm{e/h}}$ in Eq.~\eqref{eq:kernel_intra} by the same quantity $\bar{\mu}_{\lambda_{\mathrm{e}}\lambda_{\mathrm{h}}}$, which we choose as their average 
\begin{equation}
	\mu_{\lambda_{\mathrm{e}}\lambda_{\mathrm{h}}}^{\mathrm{e/h}}\rightarrow\bar{\mu}_{\lambda_{\mathrm{e}}\lambda_{\mathrm{h}}}\rightarrow \frac{\mu_{\lambda_{\mathrm{e}}\lambda_{\mathrm{h}}}^{\mathrm{e}}+\mu_{\lambda_{\mathrm{e}}\lambda_{\mathrm{h}}}^{\mathrm{h}}}{2}=\frac{1}{2}
\end{equation}
for the final calculations. However we checked numerically that the results of the numerical simulations for the self-energies as well as the Raman scattering amplitudes do not change in any relevant fashion, when varying $\bar{\mu}_{\lambda_{\mathrm{e}}\lambda_{\mathrm{h}}}$ around $1/2$. With this replacement $\mu_{\lambda_{\mathrm{e}}\lambda_{\mathrm{h}}}^{\mathrm{e/h}}\rightarrow \bar{\mu}_{\lambda_{\mathrm{e}}\lambda_{\mathrm{h}}}$ we can also evaluate Eq.~\eqref{eq:kernel_intra} using Eq.~\eqref{eq:helper_integral}.

\subsection{Self consistent numerical calculation of self-energies}\label{app:self_consistent}
To obtain the self-energies in a self-consistent fashion, we start by evaluating the impact of the low frequency acoustic modes via numerical integration of Eq.~\eqref{eq:sigma_ac_final}, which yields the corresponding isotropic self-energy contribution $\Sigma_{\lambda_{\mathrm{e}}\lambda_{\mathrm{h}}}^{\mathrm{ac}\Gamma}(\vb{k})=\Sigma_{\lambda_{\mathrm{e}}\lambda_{\mathrm{h}}}^{\mathrm{ac}\Gamma}(k)$. This renormalizes the exciton energies according to Eq.~\eqref{eq:self_energy_low_impact}, such that the single exciton propagators appearing in Eq.~\eqref{eq:sigma_final} are of the form
\begin{equation}\label{eq:propagator_isotropic}
	\mathcal{G}_{\lambda_{\mathrm{e}}\lambda_{\mathrm{h}}}(E,q)=\frac{1}{E-\tilde{E}_{\lambda_{\mathrm{e}}\lambda_{\mathrm{h}}}(q)-\Sigma^+_{\lambda_{\mathrm{e}}\lambda_{\mathrm{h}}}(E,q)+i0^+}
\end{equation} 
according to the distinction between high and low frequency phonon modes discussed in previous subsections. Here, the isotropic self-energy stemming from scattering with high frequency phonons $\Sigma^+_{\lambda_{\mathrm{e}}\lambda_{\mathrm{h}}}(E,q)$ is in turn determined by Eqs.~\eqref{eq:self_energy_valleys} and~\eqref{eq:sigma_final}.

To solve this problem self-consistently, in the first step we use Eq.~\eqref{eq:propagator_isotropic} with $\Sigma^+=0$ to determine the initial estimate for the self-energy contributions to $\Sigma^+$ via Eq.~\eqref{eq:sigma_final}. These are in turn inserted into Eq.~\eqref{eq:propagator_isotropic} to find a better estimate by repeating this process until convergence is reached, i.e., until the self-energy $\Sigma^+_{\lambda_{\mathrm{e}}\lambda_{\mathrm{h}}}(E,q)$ is stable. In each step we perform the integrals in Eq.~\eqref{eq:sigma_final} numerically.

To reduce the numerical cost of the procedure, we employ the quasi-particle approximation by evaluating the self-energy $\Sigma^+$ in Eq.~\eqref{eq:propagator_isotropic} only at the pole~\cite{fetter2003quantum,haug2008quantum}
\begin{equation}
	E=\text{Re}\qty[\tilde{E}_{\lambda_{\mathrm{e}}\lambda_{\mathrm{h}}}(q)+\Sigma_{\lambda_{\mathrm{e}}\lambda_{\mathrm{h}}}^{+}(q)]\,,
\end{equation}
where $\Sigma_{\lambda_{\mathrm{e}}\lambda_{\mathrm{h}}}^{+}(q)$ denotes the self-energy determined in the previous iteration, however evaluated at the same kind of pole determined with the self-energies from the iteration before that.

Finally in each iteration we redefine the poles according to
\begin{equation}
	\text{Re}\qty[\tilde{E}_{\lambda_{\mathrm{e}}\lambda_{\mathrm{h}}}(q)+\Sigma_{\lambda_{\mathrm{e}}\lambda_{\mathrm{h}}}^{+}(q)]\rightarrow E_{\lambda_{\mathrm{e}}\lambda_{\mathrm{h}}}(q)\,,
\end{equation}
such that we effectively change the bare exciton energies in the Hamiltonian in Eq.~\eqref{eq:H_exc} in each iteration in such a way that the polaron-shifted energies constitute the parabolas in Eq.~\eqref{eq:disp_ex}. This implies that the renormalized single exciton propagator finally reads
\begin{equation}\label{eq:propagator_renorm}
	\mathcal{G}_{\lambda_{\mathrm{e}}\lambda_{\mathrm{h}}}(E,q)=\frac{1}{E-E_{\lambda_{\mathrm{e}}\lambda_{\mathrm{h}}}(q)-i\text{Im}[\Sigma^+_{\lambda_{\mathrm{e}}\lambda_{\mathrm{h}}}(q)+\Sigma_{\lambda_{\mathrm{e}}\lambda_{\mathrm{h}}}^{\mathrm{ac}\Gamma}(q)]+i0^+}
\end{equation}
with
\begin{equation}
	\Sigma^+_{\lambda_{\mathrm{e}}\lambda_{\mathrm{h}}}(q)=\Sigma^+_{\lambda_{\mathrm{e}}\lambda_{\mathrm{h}}}(E=E_{\lambda_{\mathrm{e}}\lambda_{\mathrm{h}}}(q),q)\,.
\end{equation}
Convergence of the self-energies $\Sigma^+_{\lambda_{\mathrm{e}}\lambda_{\mathrm{h}}}(q)$ then implies convergence of the bare exciton energies, while keeping the renormalized exciton energies fixed.

\section{Resonance Raman profiles for MoSe$_2$ and WS$_2$}\label{app:rrp_add}
\begin{figure}[t!]
	\centering
	\includegraphics[width=\linewidth]{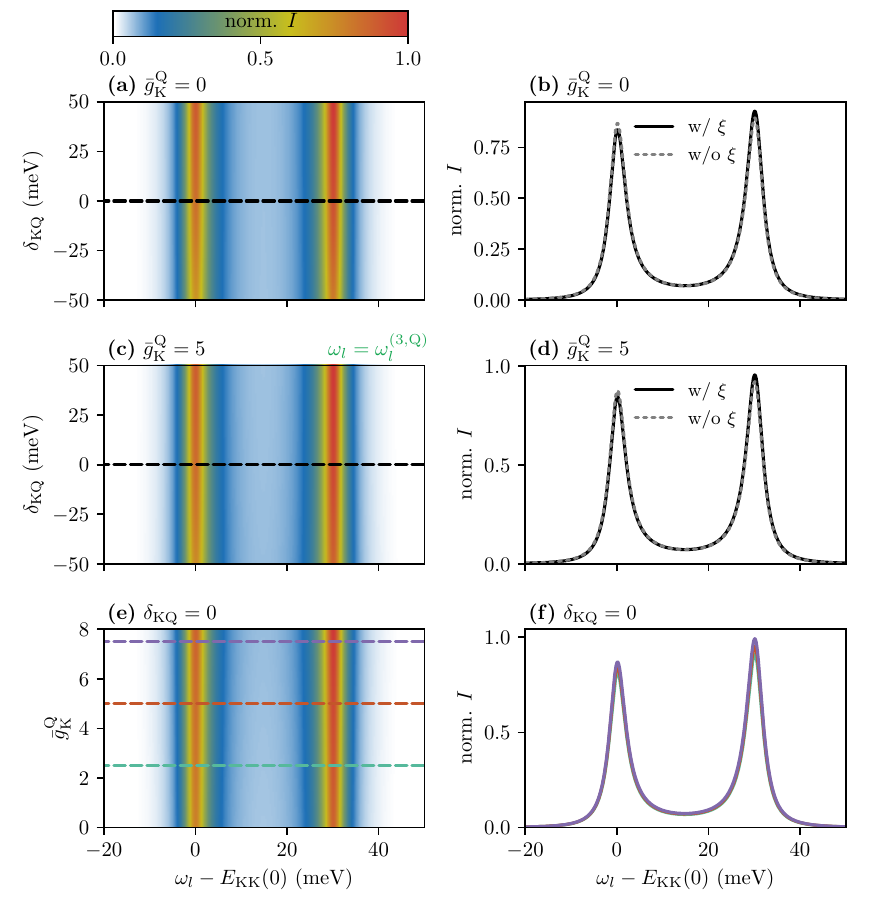}
	\caption{Impact of strain and inter-valley scattering on resonance Raman profiles $I(\omega_l-\Omega_{\mathrm{A}_1})$ of hBN-encapsulated monolayer MoSe$_2$ at $T=4$~K. (a) Strain-dependent resonance Raman profile for suppressed A$_1$-phonon emission from the Q-valley $\bar{g}_{\mathrm{K}}^{\mathrm{Q}}=0$. (b) Cut in (a) as marked by horizontal dashed line, showing the unstrained case. Full simulation including the factor $\xi(\omega_l-\Omega_{\mathrm{A}_1})$ from Eq.~\eqref{eq:phase_space} (solid) vs. neglected factor $\xi(\omega_l-\Omega_{\mathrm{A}_1})\rightarrow \xi(E_{\mathrm{KK}}(0)-\Omega_{\mathrm{A}_1}/2)$ (dashed). (c,d) same as (a,b) for non-vanishing A$_1$-phonon emission from the Q-valley $\bar{g}_{\mathrm{K}}^{\mathrm{Q}}=5$. In contrast to Fig.~\ref{fig:resonance}, there is no green line in (c) marking the energetic position of the third order resonance $\omega_l=E_{\mathrm{KQ}}(0)+\Omega_{\mathrm{A}_1}+\Omega_{\mathrm{ac},\mathrm{Q}}\equiv \omega_l^{(3,\mathrm{Q})}$, since the KQ exciton lies more than 100~meV above the KK exciton. (e,f) Resonance Raman profiles for the unstrained case $\delta_{\mathrm{KQ}}=0$ as a function of the A$_1$-phonon emission strength in the Q-valley $\bar{g}^{\mathrm{Q}}_{\mathrm{K}}$. Lines in (f) correspond to identically colored horizontal cuts in (e).}
	\label{fig:resonance_MoSe2}
\end{figure}
\begin{figure}[h!]
	\centering
	\includegraphics[width=\linewidth]{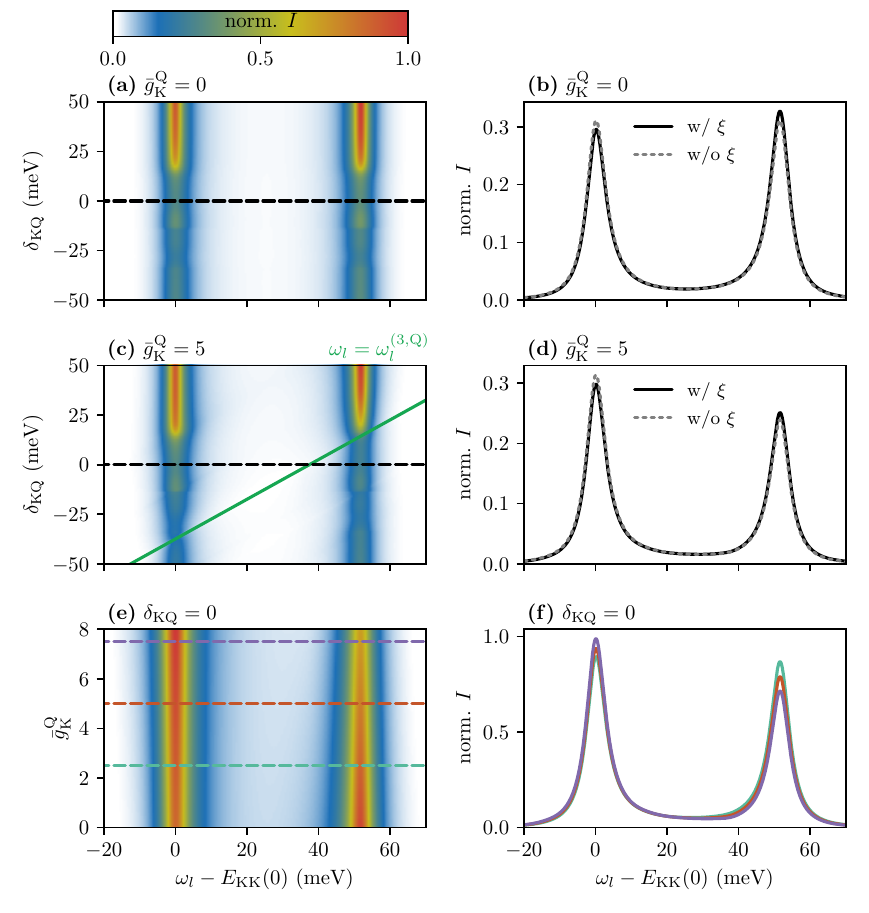}
	\caption{Impact of strain and inter-valley scattering on resonance Raman profiles $I(\omega_l-\Omega_{\mathrm{A}_1})$ of hBN-encapsulated monolayer WS$_2$ at $T=4$~K. (a) Strain-dependent resonance Raman profile for suppressed A$_1$-phonon emission from the Q-valley $\bar{g}_{\mathrm{K}}^{\mathrm{Q}}=0$. (b) Cut in (a) as marked by horizontal dashed line, showing the unstrained case. Full simulation including the factor $\xi(\omega_l-\Omega_{\mathrm{A}_1})$ from Eq.~\eqref{eq:phase_space} (solid) vs. neglected factor $\xi(\omega_l-\Omega_{\mathrm{A}_1})\rightarrow \xi(E_{\mathrm{KK}}(0)-\Omega_{\mathrm{A}_1}/2)$ (dashed). (c,d) same as (a,b) for non-vanishing A$_1$-phonon emission from the Q-valley $\bar{g}_{\mathrm{K}}^{\mathrm{Q}}=5$. The green line in (c) marks the energetic position of the third order resonance $\omega_l=E_{\mathrm{KQ}}(0)+\Omega_{\mathrm{A}_1}+\Omega_{\mathrm{ac},\mathrm{Q}}\equiv \omega_l^{(3,\mathrm{Q})}$. (e,f) Resonance Raman profiles for the unstrained case $\delta_{\mathrm{KQ}}=0$ as a function of the A$_1$-phonon emission strength in the Q-valley $\bar{g}^{\mathrm{Q}}_{\mathrm{K}}$. Lines in (f) correspond to identically colored horizontal cuts in (e).}
	\label{fig:resonance_WS2}
\end{figure}
Figs.~\ref{fig:resonance_MoSe2} and \ref{fig:resonance_WS2} show the same calculations as discussed for WSe$_2$ in the context of Fig.~\ref{fig:resonance} from the main text, however for hBN-encapsulated MoSe$_2$ and WS$_2$ monolayers, respectively. 

Note that the sign of the ab initio deformation potential parameters presented in Tab.~\ref{tab:phon} cannot be inferred from their original publication in Ref.~\cite{jin2014intrinsic}. However choosing the opposite sign for the optical deformation potential for hole intra-valley scattering at the K-point $D^{(\mathrm{h})}_{\mathrm{opt},\Gamma,\mathrm{KK}}\rightarrow -D^{(\mathrm{h})}_{\mathrm{opt},\Gamma,\mathrm{KK}}$ would lead to strongly asymmetric resonance Raman profiles in the case of MoSe$_2$ (not shown), which is inconsistent with experimental data~\cite{mcdonnell2020observation}. This supports the choice of the signs for the deformation potential constants discussed in the caption of Tab.~\ref{tab:phon}.

\clearpage
\bibliography{bib_Raman.bib}

\end{document}